\documentclass[]{aa}
\usepackage[varg]{txfonts}
\usepackage{amsmath}
\usepackage{graphicx}
\usepackage{longtable}
\usepackage{multicol}
\usepackage{multirow}
\usepackage{wrapfig}
\usepackage{adjustbox}
\usepackage[colorinlistoftodos]{todonotes}
\usepackage{natbib,twoopt}
\usepackage[draft]{hyperref} 
\bibpunct{(}{)}{;}{a}{}{,}             
\makeatletter
  \newcommandtwoopt{\citeads}[3][][]{\href{http://adsabs.harvard.edu/abs/#3}%
    {\def\hyper@linkstart##1##2{}%
     \let\hyper@linkend\@empty\citealp[#1][#2]{#3}}}
  \newcommandtwoopt{\citepads}[3][][]{\href{http://adsabs.harvard.edu/abs/#3}%
    {\def\hyper@linkstart##1##2{}%
     \let\hyper@linkend\@empty\citep[#1][#2]{#3}}}
  \newcommandtwoopt{\citetads}[3][][]{\href{http://adsabs.harvard.edu/abs/#3}%
    {\def\hyper@linkstart##1##2{}%
     \let\hyper@linkend\@empty\citet[#1][#2]{#3}}}
  \newcommandtwoopt{\citeyearads}[3][][]%
    {\href{http://adsabs.harvard.edu/abs/#3}
    {\def\hyper@linkstart##1##2{}%
     \let\hyper@linkend\@empty\citeyear[#1][#2]{#3}}}

\makeatother

\usepackage{tikz}
\usetikzlibrary{shapes, arrows, positioning}
\definecolor{dgreen}{rgb}{0, 0.7, 0}
\title{The ALMA-PILS survey: First detection of the unsaturated 3-carbon molecules Propenal (C$_2$H$_3$CHO) and Propylene (C$_3$H$_6$) towards IRAS~16293--2422~B}
\author{S. Manigand$^{\ref{inst-NBI}}$\and 
                        A. Coutens$^{\ref{inst-LAB}}$ \and 
                        J.-C. Loison$^{\ref{inst-ISM}}$ \and 
                        V. Wakelam$^{\ref{inst-LAB}}$ \and 
                        H. Calcutt$^{\ref{inst-Sweden}}$ \and           
                        H.~S.~P. M{\"u}ller$^{\ref{inst-Koln}}$ \and
                        J.~K. J{\o}rgensen$^{\ref{inst-NBI}}$\and
                        V. Taquet$^{\ref{inst-INAF}}$\and
                        S.~F. Wampfler$^{\ref{inst-Bern}}$\and
                        T.~L. Bourke$^{\ref{inst-SKA}}$\and
                        B.~M. Kulterer$^{\ref{inst-Bern}}$\and
                        E.~F. van Dishoeck$^{\ref{inst-Leiden},\,\ref{inst-MPE}}$\and
                        M.~N. Drozdovskaya$^{\ref{inst-Bern}}$\and
                        N.~F.~W. Ligterink$^{\ref{inst-Bern}}$
                        }
\institute{
Niels Bohr Institute \& Centre for Star and Planet Formation, University of Copenhagen, \O ster Voldgade 5–7, DK-1350 Copenhagen K., Denmark\label{inst-NBI} \and
Laboratoire d'Astrophysique de Bordeaux, Univ. Bordeaux, CNRS, B18N, all{\'e}e Geoffroy Saint-Hilaire, 33615 Pessac, France \label{inst-LAB} \and
Institut des Sciences Mol{\'e}culaires (ISM), CNRS, Univ. Bordeaux, 351 cours de la Lib{\'e}ration, 33400, Talence, France \label{inst-ISM} \and
Department of Space, Earth and Environment, Chalmers University of Technology, 41296, Gothenburg, Sweden\label{inst-Sweden} \and
I. Physikalisches Institut, Universit{\"a}t zu K{\"o}ln, Z{\"u}lpicher Str. 77, 50937 K{\"o}ln, Germany\label{inst-Koln} \and
INAF, Osservatorio Astrofisico di Arcetri, Largo E. Fermi 5, 50125 Firenze, Italy\label{inst-INAF} \and
Center for Space and Habitability, University of Bern, Gesellschaftsstrasse 6, 3012 Bern, Switzerland\label{inst-Bern} \and
SKA Organisation, Jodrell Bank, Lower Withington, Macclesfield, Cheshire SK11 9FT, UK\label{inst-SKA} \and
Leiden Observatory, Leiden University, PO Box 9513, 2300 RA Leiden, The Netherlands\label{inst-Leiden}\and
Max-Planck Institut f{\"u}r Extraterrestrische Physik (MPE), Giessenbachstr. 1, 85748 Garching, Germany\label{inst-MPE}
} 
\date{Received: April 8, 2020 / Accepted: July 2, 2020}
\abstract{Complex organic molecules with three carbon atoms are found in the earliest stages of star formation. In particular, propenal (C$_2$H$_3$CHO) is a species of interest due to its implication in the formation of more complex species and even biotic molecules. }
{This study aims to search for the presence of C$_2$H$_3$CHO and other three-carbon species such as propylene (C$_3$H$_6$) in the hot corino region of the low-mass protostellar binary IRAS 16293--2422 to understand their formation pathways.}
{We use ALMA observations in Band 6 and 7 from various surveys to search for the presence of C$_3$H$_6$ and C$_2$H$_3$CHO towards the protostar IRAS 16293--2422 B (IRAS 16293B). The identification of the species and the estimates of the column densities and excitation temperatures are carried out by modeling the observed spectrum under the assumption of local thermodynamical equilibrium. }
{We report the detection of both C$_3$H$_6$ and C$_2$H$_3$CHO towards IRAS 16293B, however, no unblended lines were found towards the other component of the binary system, IRAS 16293A. We derive column density upper limits for C$_3$H$_8$, HCCCHO, \emph{n}-C$_3$H$_7$OH, \emph{i}-C$_3$H$_7$OH, 
C$_3$O, and \emph{cis}-HC(O)CHO towards IRAS 16293B. We then use a three-phase chemical model to simulate the formation of these species in a typical prestellar environment followed by its hydrodynamical collapse until the birth of the central protostar. Different formation paths, such as successive hydrogenation and radical-radical additions on grain surfaces, are tested and compared to the observational results in a number of different simulations, to assess which are the dominant formation mechanisms in the most embedded region of the protostar.  } 
{The simulations reproduce the abundances within one order of magnitude from those observed towards IRAS 16293B, with the best agreement found for a rate of $10^{-12}$ cm$^3$~s$^{-1}$ for the gas-phase reaction C$_3$ + O $\rightarrow$ C$_2$ + CO. 
Successive hydrogenations of C$_3$, HC(O)CHO, and CH$_3$OCHO on grain surfaces are a major and crucial formation route of complex organics molecules, whereas both successive hydrogenation pathways and radical-radical addition reactions contribute to the formation of C$_2$H$_5$CHO.}

\begin{document}

\titlerunning{First detection of C$_2$H$_3$CHO and C$_3$H$_6$ towards IRAS~16293--2422~B}
\authorrunning{S. Manigand et al.}
\maketitle
\bibpunct{(}{)}{;}{a}{}{,} 

\section{Introduction}


Aldehyde molecules, which contain a functional group CHO, play an important role in the formation of complex organic molecules \citep[COMs, molecules containing
six or more atoms with at least one carbon,][]{Herbst-2009}.
The two-carbon to three-carbon chain aldehydes are generally found in the cold dense regions of star formation as well as in the innermost warm region of protostellar envelope, the hot core. 
In particular, propenal (C$_2$H$_3$CHO), also called acrolein, is considered as a prebiotic species due to its formation after the decomposition of sugars \citep{Moldoveanu-2010, Bermudez-2013}
and its role in the synthesis of amino acids via Strecker-type reactions \citep{Strecker-1850, Strecker-1854} as tested in laboratory studies \citep[e.g.][]{vanTrump-1972, Shibasaki-2008, Grefenstette-2017}. On the other hand, on the primordial Earth, C$_2$H$_3$CHO is one of a few species that readily reacts with nucleobases of the ribonucleic acid \citep[RNA,][]{Nelsestuen-1980}, which makes it a possible sink for the nucleobases and an important hindrance to the start of the RNA world \citep[e.g.][]{astrobiology, Neish-2010}.
Propynal (HCCCHO), C$_2$H$_3$CHO and propanal (C$_2$H$_5$CHO) are suspected to be linked through their formation on ice surfaces by successive hydrogenation \citep[e.g.][]{Hudson-1999}. C$_2$H$_3$CHO and C$_2$H$_5$CHO were first detected in the interstellar medium toward the Galactic Center source Sgr~B2(N) \citep{Hollis-2004, Requena-Torres-2008}, while HCCCHO was first detected towards the dark cloud TMC--1 \citep{Irvine-1988, Turner-1991}. 

Three-carbon-chain molecules have also been observed in the ISM and are, in general, more abundant towards protostars characterised by warm carbon-chain chemistry (WCCC), such as L1527 \citep{Sakai-2008}, and cold dense clouds. For example, propylene (C$_3$H$_6$) was detected for the first time towards TMC-1 \citep{Marcelino-2007}. It has so far never been reported in  warmer environments, where COMs are found with high abundances, such as hot cores/corinos, contrary to methyl acetylene (CH$_3$CCH) already found with high abundances in such objects \citep[e.g.][]{VanDishoeck-1995, Cazaux-2003}.

In this paper we report detections of C$_2$H$_3$CHO and C$_3$H$_6$ towards the low-mass Class 0 protostellar binary IRAS 16293--2422 (IRAS 16293 hereafter). This source, located in the $\rho$ Ophiuchus cloud complex  at a distance of 144$\pm$7 pc \citep{Zucker-2019}, is well-known as a reference in astrochemistry because of its molecule-rich envelope and the presence of numerous bright emission lines at millimetre wavelengths \citep[for example,][]{VanDishoeck-1995, Cazaux-2003, Caux-2011}. 
The high sensitivity and the high angular resolution of the Atacama Large (sub)Millimeter Array (ALMA) allowed many new detections around solar-type protostars and in the interstellar medium. 
Using ALMA observations of the Protostellar Interferometric Line Survey \citep[PILS,][]{Jorgensen-2016}, some of these new detections of COMs, such as deuterated formamide (NHDCHO and NH$_2$CDO) and isocyanic acid (HNCO) by \cite{Coutens-2016}, methyl chloride \citep[CH$_3$Cl,][]{Fayolle-2017}, methyl isocyanate \citep[CH$_3$NCO,][]{Ligterink-2017}, cyanamide \citep[NH$_2$CN,][]{Coutens-2018}, methyl isocyanide \citep[CH$_3$NC,][]{Calcutt-2018a}, and nitrous acid \citep[HONO,][]{Coutens-2019} have been reported towards this protostellar binary. 
C$_2$H$_5$CHO \citep{Lykke-2017} and CH$_3$CCH \citep{Calcutt-2019} are also detected in the embedded hot corino towards the ``B component'' (IRAS 16293B) of this source. This demonstrates the possibility of finding the unsaturated precursors of C$_2$H$_5$CHO and provide new constraints on the formation of these species in low-mass protostellar environments. It also suggests the need to develop  models to describe these three-carbon species (C$_3$-species hereafter) using laboratory measurements and theoretical calculations both for the gas phase and grain surfaces \citep{Loison-2014, Loison-2017, Hickson-2016b, Qasim-2019}.

This paper is organised as follows: Section 2 describes the observations and the spectroscopic data used in this study. The results and the analysis are presented in Sect.~3. These results are then compared to a chemical model which is described in Sect.~4. Finally, the comparison between the results and the model are discussed in Sect.~5 and the conclusions summarised in Sect.~6.

\section{Observations}

\begin{table*}[t]
\caption{\label{app-tab-obs}\small List of the observations towards IRAS 16293B used in this study.}
\centering
\begin{tabular}{cccccc}
\hline\hline
Project & Band & Frequency range & Spectral resolution &  Beam size & Sensitivity  \\
 & & (GHz) & (MHz) & (arcsec) & (mJy beam$^{-1}$ channel$^{-1}$) \\
 \hline
\multirow{8}{*}{2012.1.00712.S} & \multirow{8}{*}{Band 6} & 221.77 -- 222.23 & \multirow{8}{*}{0.122} & \multirow{8}{*}{0.5} & \multirow{8}{*}{7 -- 10} \\
  &  & 224.77 -- 225.23 &  & &  \\
  &  & 231.02 -- 231.48 &  &  &  \\
  &  & 232.16 -- 232.62 &  &  &  \\
  &  & 239.42 -- 239.88 &  &  &  \\
  &  & 240.17 -- 240.63 &  &  &  \\
  &  & 247.30 -- 247.76 &  &  &  \\
  &  & 250.26 -- 250.72 &  &  &  \\
\hline
\multirow{3}{*}{2016.1.01150.S}  & \multirow{3}{*}{Band 6} & 233.71 -- 234.18 & \multirow{3}{*}{0.122} & \multirow{3}{*}{0.5} & \multirow{3}{*}{1.2 -- 1.4} \\
  &  & 234.92 -- 235.39 &  &  &  \\
  &  & 235.91 -- 236.84 &  &  &  \\
\hline
2013.1.00278.S & Band 7 & 329.15 -- 362.90 & 0.244 & 0.5 & 7 -- 10 \\
\hline
\end{tabular}
\end{table*}

Observations at 1.3 and 0.8 mm wavelength, carried out with ALMA, corresponding to Band 6 and 7, respectively, towards IRAS 16293 were used in this study. The pointing centre, located between the two protostars of the binary system at $\alpha_\text{J2000} = 16^\text{h}32^\text{m}22{\fs}72$; $\delta_\text{J2000} = -24^\circ28^\prime34{\farcs}3$, was the same for all the observations.
The species that are the main focus of this study are the C$_3$-species C$_2$H$_3$CHO and C$_3$H$_6$, as well as the chemically related species HCCCHO, C$_2$H$_5$CHO, propanol (\textit{n}-C$_3$H$_7$OH) and its \emph{iso} conformer (\textit{i}-C$_3$H$_7$OH), 
tricarbon monoxide (C$_3$O), C$_3$H$_8$, and glyoxal (\emph{cis-}HC(O)CHO). Note that the most abundant conformer of HC(O)CHO, \textit{trans-}HC(O)CHO, has no dipole moment and does not display any pure rotational transitions, thus it cannot be detected at millimetre wavelengths.

\subsection{Band 6 observations}

The ALMA-Band 6 data are a combination of PILS observations in Cycle 1 (project-id: 2012.1.00712.S, PI: J{\o}rgensen, J. K.) and Cycle 4 observations taken from \cite{Taquet-2018} (project-id: 2016.1.01150.S, PI: Taquet, V.). 
The observations cover a total of $\sim$5.5 GHz spread between 221.7 and 250.7 GHz with a frequency resolution of 0.122 MHz, corresponding to a velocity resolution of $\sim$0.16~km~s$^{-1}$. The two datasets have been treated the same way and were restored with the same 0\farcs 5 circular beam as in Band 7. The continuum subtraction and the data calibration are detailed in \cite{Jorgensen-2016} and \cite{Taquet-2018}, for Cycle 1 and Cycle 4 observations, respectively. Calibration uncertainties are better than 5\% and the sensitivity reached is 1 -- 10 mJy~beam$^{-1}$ per channel, depending on the observations. 

\subsection{Band 7 observations}

The Band 7 data are taken from the PILS observations (project-id: 2013.1.00278.S, PI: J{\o}rgensen, J. K.) and cover the full range of 329 to 363 GHz at a frequency resolution of 0.244 MHz, which corresponds to $\sim$0.2 km s$^{-1}$ in this frequency range, and were restored with a circular beam of 0\farcs 5. The continuum subtraction and the data reduction are described in \cite{Jorgensen-2016}. The sensitivity reaches down to 7--10 mJy beam$^{-1}$ per channel and the relative calibration uncertainty across the band is $\sim$5\%. Table \ref{app-tab-obs} summarises the details of the spectral windows used in this study. 

\subsection{Spectroscopic data}

The spectroscopic data used to identify C$_2$H$_3$CHO transitions are taken from the Cologne Database for Molecular Spectroscopy \citep[CDMS,][]{cdms-1, cdms-2}. The CDMS entry is based on \cite{Daly-2015} with additional data from \cite{Winnewisser-1975, Cherniak-1966}. The dipole moment was determined by \cite{Blom-1984}. 

The C$_3$H$_6$ rotational transitions are taken from its CDMS entry, based on \cite{Craig-2016} with numerous additional transition frequencies from \cite{Hirota-1966}, \cite{Pearson-1994} and \cite{Wlodarczak-1994}. The dipole moment was measured by \cite{Lide-1957}. The spectroscopic entry includes the transitions of \emph{A} and \emph{E} torsional substate conformers, which arise from the methyl internal rotation splitting. 

Five spectroscopic conformers of n-C$_3$H$_7$OH exist, distinguished by the orientation of their methyl and OH groups. Ga-n-C$_3$H$_7$OH is the lowest vibrational ground state conformer. The CDMS entry used in this study is largely based on \cite{Kisiel-2010}, with interactions with other conformers studied by \cite{Kahn-2005} and additional data from \cite{Maeda-2006-n, White-1975}. The entry only takes the Ga- conformer into account. The correction factor corresponding to the contribution of the other conformers is 2.75 at $T_\mathrm{ex}$ = 100 K. 

The CDMS entries for the two conformers \emph{gauche}, the energetically lowest, and \emph{anti} of \textit{i}-C$_3$H$_7$OH are based on \cite{Maeda-2006-i, Ulenikov-1991, Hirota-1979}. Rotation-tunneling interactions between the conformers are incorporated in the Hamiltonian using the formulation described in \cite{Christen-2003}. Unlike the n-C$_3$H$_7$OH entry, the i-C$_3$H$_7$OH partition function already includes the contribution of \textit{gauche} and \textit{anti} conformers.

Spectroscopic data of HCCCHO were taken from its CDMS entry, which is based on \cite{McKellar-2008} with additional data from \cite{Costain-1959} and \cite{Winnewisser-1973}. The dipole moment was measured by \cite{Brown-1984}.

C$_3$O spectroscopic data are found in the CDMS. The entry is based on the work of \cite{Bizzocchi-2008}, where they used previous studies and measurements \citep{Klebsch-1985, Tang-1985, Brown-1983}. The dipole moment is reported in the last cited work. The entry does not take the contribution of the torsional/vibrational excited state into account.

C$_3$H$_8$ spectroscopic data are taken from the Jet Propulsion Laboratory database \citep[JPL,][]{jpl_0}. The JPL entry is based on \cite{Drouin-2006}, which is a compilation of extensive own data and data taken from \cite{Lide-1960}, \cite{Bestmann-1985a} and \cite{Bestmann-1985b}. The partition function includes contributions from the first vibrational and torsional excited states. The dipole moment is reported in \cite{Lide-1960}.

\emph{cis}-HC(O)CHO is higher in energy by $1555 \pm 48$ cm$^{-1}$ or $2237 \pm 69$ K than \emph{trans}-HC(O)CHO, however, the latter has no dipole moment thus does not display any rotational transitions. The spectroscopic data are taken from its CDMS entry which is based on the study of \cite{Hubner-1997}, which also includes the measurement of the dipole moment. Due to the enthalpy difference between \emph{cis} and \emph{trans} conformers, the population fraction of \emph{cis}-HC(O)CHO is $1.9\times10^{-10}$ at 100~K. The abundance derived from the molecular emission of this conformer has to be divided by this factor to retrieve the total abundance of HC(O)CHO.

The main spectroscopic parameters of the different species analysed in this study are summarised in Table \ref{app-tab-spectro}.

\section{Observational results}

For this study we follow the majority of the previous PILS papers and focus on an offset position by  0\farcs 5 in the South-West direction from the most compact source of the binary system, IRAS 16293B \citep[e.g.][]{Jorgensen-2016, Calcutt-2018b, Ligterink-2017}. The spectrum at this position shows narrow lines and is less affected by the strong absorption arising from the continuum emission. At this position the molecular lines have a full width at half-maximum (FWHM) of $\sim$1 km s$^{-1}$ and a velocity peak position ($v_{lsr}$) of  $\sim 2.5-2.8$ km s$^{-1}$.  

\subsection{C$_3$-species lines}

\begin{figure}[t]
\centering
\includegraphics[width=0.24\textwidth]{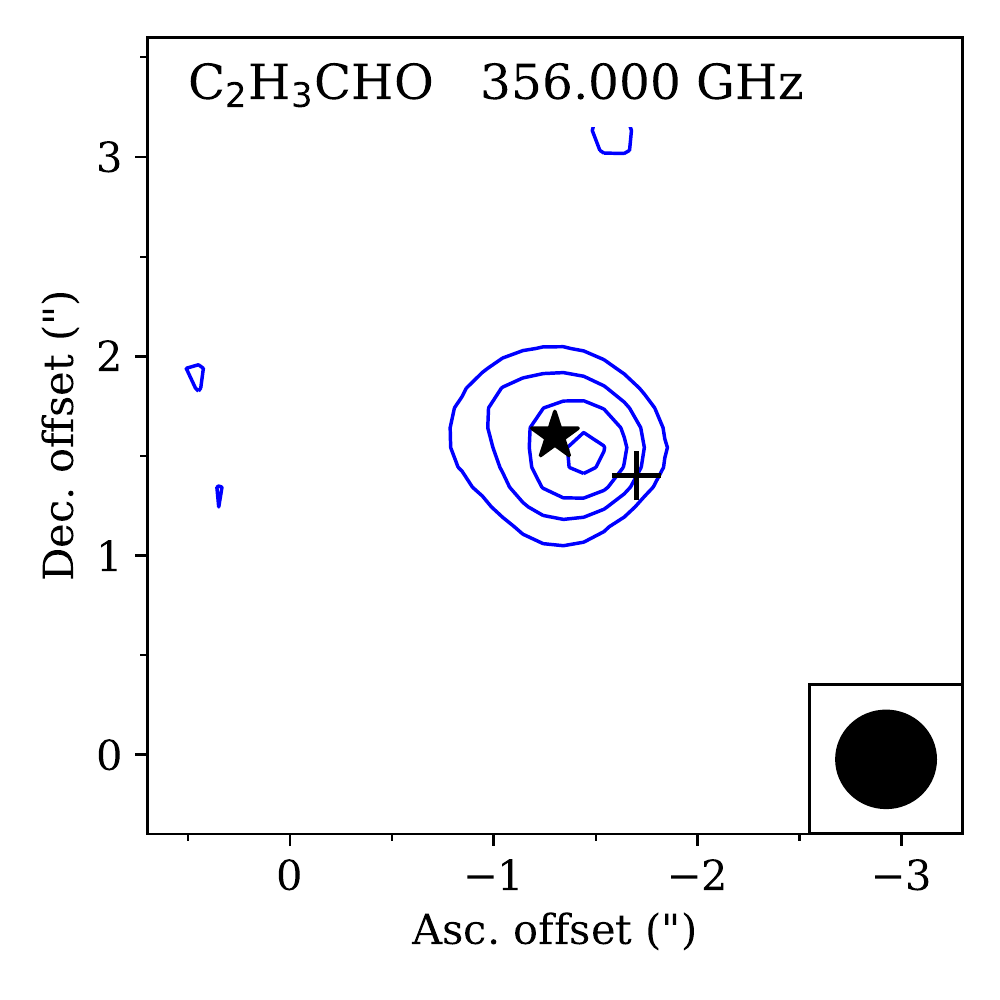}
\includegraphics[width=0.24\textwidth]{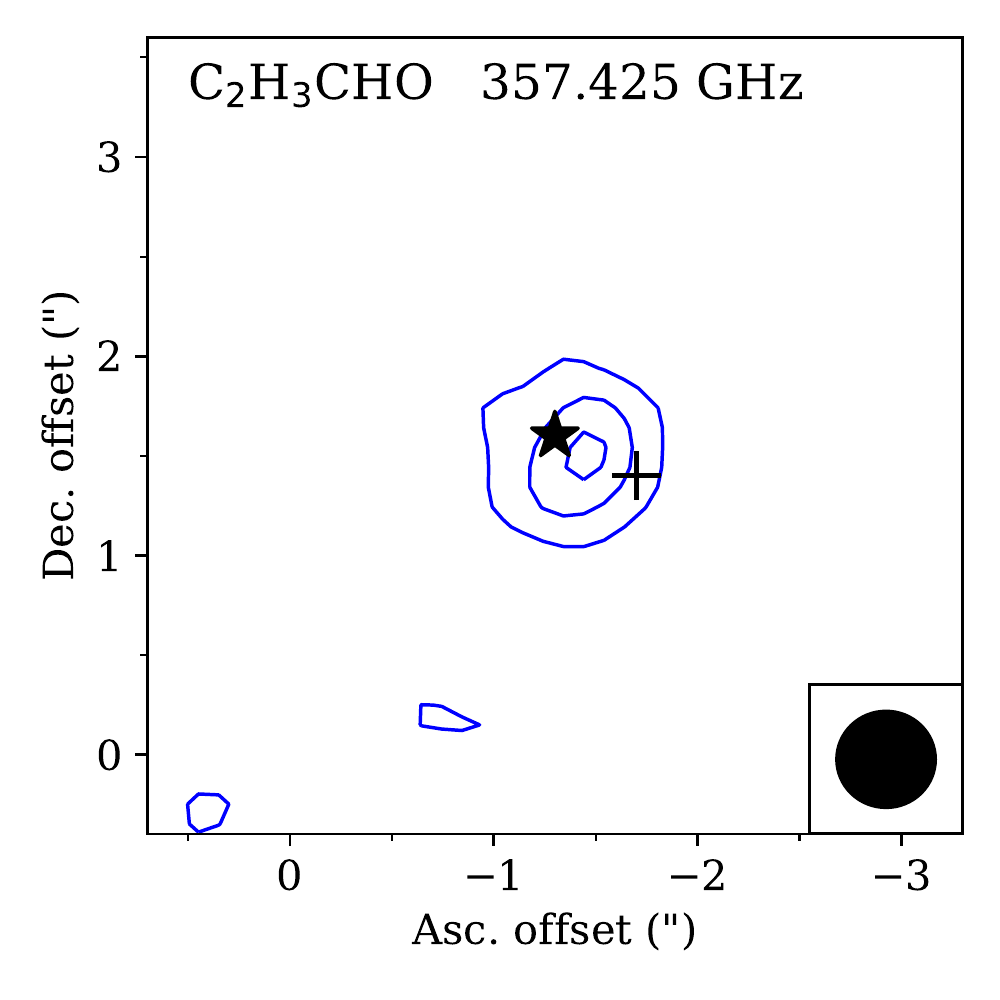}\\
\includegraphics[width=0.24\textwidth]{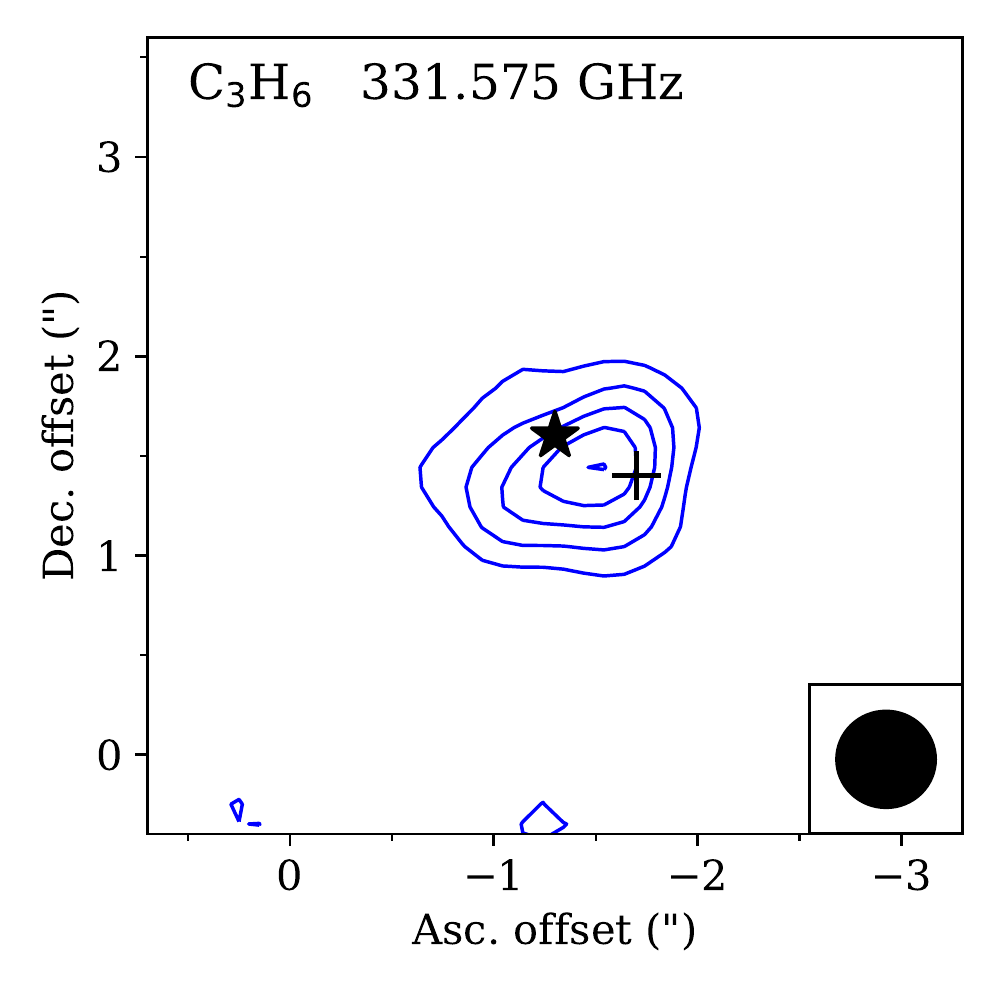}
\includegraphics[width=0.24\textwidth]{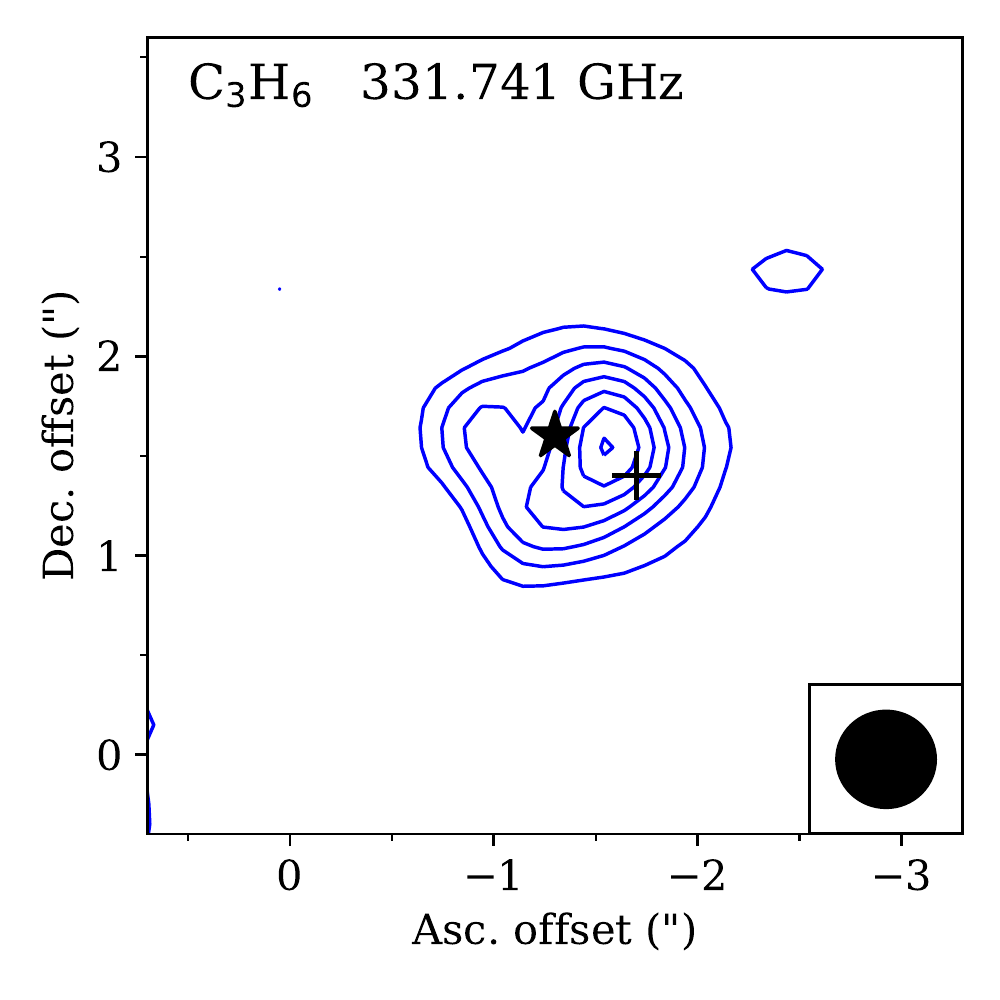}
\caption{\small\label{fig:maps} Integrated intensity maps of the line emission for C$_3$H$_6$ and C$_2$H$_3$CHO, summed over $\pm0.5$ MHz. The locations of IRAS16293B continuum peak and the offset position are marked by the black star and plus sign, respectively. The blue contours start at 4 $\sigma$ and increase by steps of 4 $\sigma$, where $\sigma$ is 5 mJy beam$^{-1}$ km s$^{-1}$ for the integrated intensity. A representative beam of 0\farcs5 is shown in the
lower right corner of each panel.}
\end{figure}

\begin{figure*}[t]
\centering
\includegraphics[width=\textwidth]{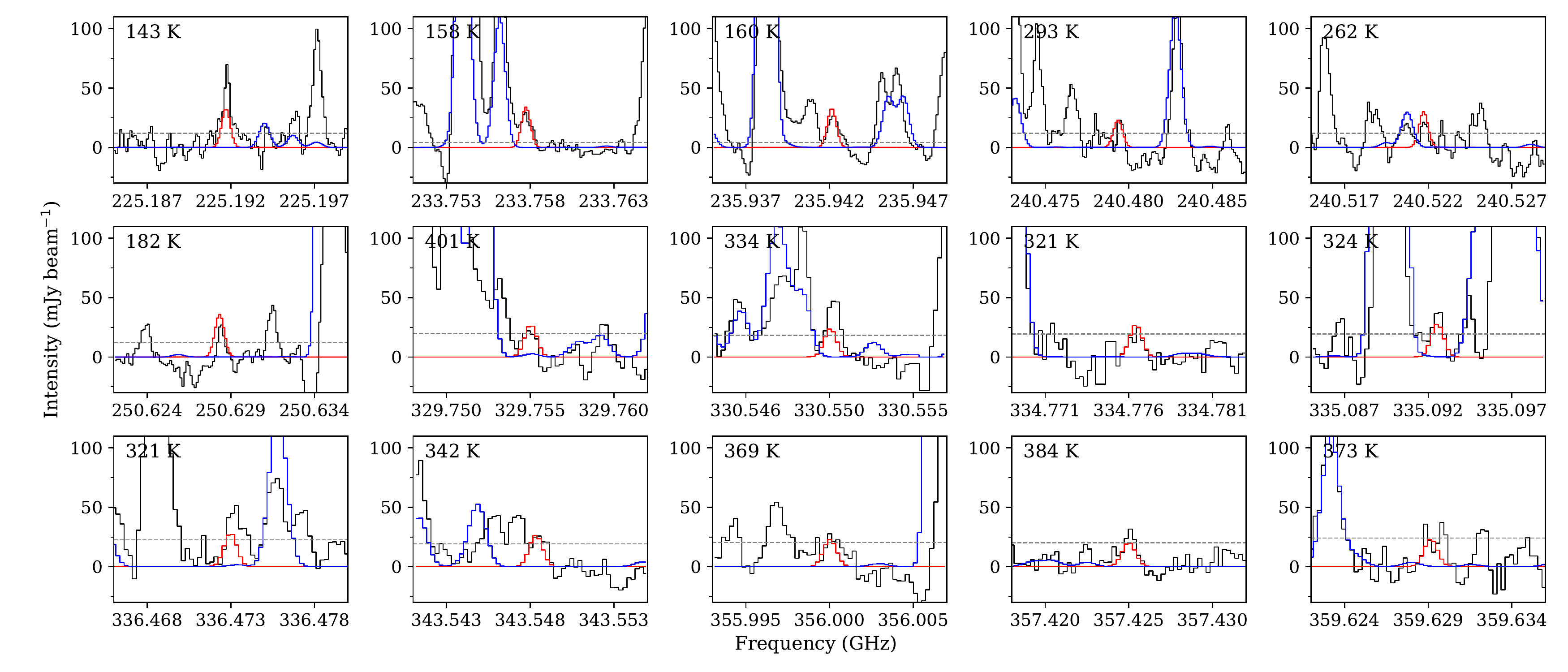}
\caption{\label{fig-fit}Identified lines of C$_2$H$_3$CHO: the synthetic spectrum, in red, is plotted along with the offset position spectrum towards IRAS 16293B. The straight dashed black line represents the detection limit at the intensity of 3$\sigma$. The upper-level energy of the transition is noted in the upper left corner of each panel. The reference spectrum in blue takes the previous species reported in PILS into account \citep{Jorgensen-2016, Jorgensen-2018, Lykke-2017, Coutens-2016, Ligterink-2017, Persson-2018, Calcutt-2018a, Drozdovskaya-2018, Manigand-2019}.}
\end{figure*}

\begin{figure*}[h]\centering
\includegraphics[width=\textwidth]{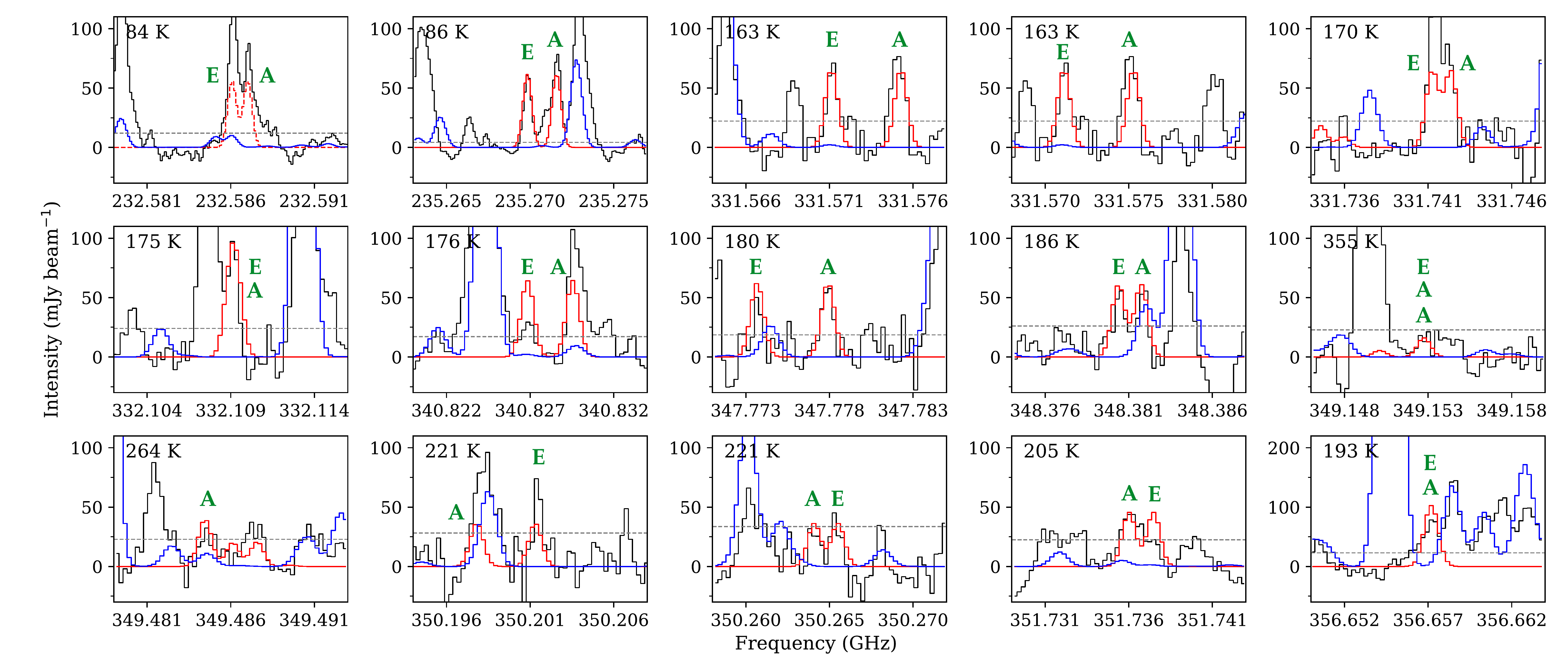}
\caption{\label{fig-fit-propene}Identified lines of C$_3$H$_6$: the synthetic spectrum, in red, is plotted along with the offset position spectrum towards IRAS 16293B. The dashed lines correspond to optically thick lines. The straight dashed black line represents the detection limit at the intensity of 3$\sigma$. The upper-level energy of the transition is noted in the upper left corner of each panel. The torsional state `E' or `A' of each transition is noted in green on the spectra. The reference spectrum in blue takes the previous species reported in PILS into account (see Figure \ref{fig-fit} and references therein).}
\end{figure*}

A total of 12 unblended lines of C$_2$H$_3$CHO have been identified across Bands 6 and 7, with an intensity higher than 3$\sigma$ -- with $\sigma$ calculated as the root mean square (RMS) of the noise spectrum. 38 other lines that match in terms of frequency were found in the range of the observations. However, they show significant blending with other species. All the lines predicted by the local thermodynamic equilibrium (LTE) model are present in the observations. The upper-level energies of the detected transitions lie between 158 and 401 K, which provide constraints on the excitation temperature. Among these transitions, those at frequencies of 225.192, 330.550 and 336.473~GHz have been used in the fit even though they seem to be slightly blended with other species.  
All these transitions are listed in Table \ref{app-linelist-C2H3CHO}. C$_2$H$_3$CHO has been searched for towards IRAS 16293A as well, however, no lines above 3$\sigma$ were found, presumably due to the larger linewidth ($\sim$3~km~s$^{-1}$), and thus the more severe blending, compared to IRAS 16293B.

For C$_3$H$_6$, 21 unblended lines were found in the different frequency ranges of the observations. The upper-level energies of the detected transitions lie between 84 and 264~K. 34 lines above 3$\sigma$ are present in the spectrum, but they are significantly blended with other species, thus they were discarded from the fit. The transition at 349.153~GHz, with an upper-level energy of 355~K, was used in the fit to better assess the excitation temperature, even though the intensity of the line is lower than 3$\sigma$. All these transitions are listed in Table \ref{app-linelist-C3H6}. The same lines were found to be completely blended towards IRAS 16293A. 

Figure \ref{fig:maps} shows the integrated line emission of C$_2$H$_3$CHO and C$_3$H$_6$. The spatial extent of their emission is marginally resolved and located in the hot corino region, consistently with the other O-bearing COMs detected in the same dataset \citep[e.g.][]{Jorgensen-2016, Lykke-2017, Calcutt-2018b}. 

HCCCHO, \emph{n}-C$_3$H$_7$OH and \emph{i}-C$_3$H$_7$OH, as well as 
C$_3$O and \emph{cis}-HC(O)CHO were searched for in the data, however, no line above 1$\sigma$ was found.

\subsection{Column densities}

\begin{table}[t]
\centering
\caption{\label{tab-N}Column densities of detected species and upper limits of non-detection towards the offset position from IRAS 16293B.}
\begin{tabular}{cccc}
\hline\hline
 Species & $T_\mathrm{ex}$ (K) & $N_\mathrm{tot}$ (cm$^{-2}$) & Ref. \\
\hline
 C$_2$H$_3$CHO & $125\pm25$ & $3.4\pm 0.7\times10^{14}$ & 1 \\
 C$_3$H$_6$ & $75\pm15$ & $4.2\pm0.8 \times10^{16}$ & 1 \\
 HCCCHO & 100\tablefootmark{a} & $<5.0\times10^{14}$ & 1\\
 \textit{n}-C$_3$H$_7$OH & 100\tablefootmark{a} & $<3.0\times10^{15}$ & 1, 2\\
 \textit{i}-C$_3$H$_7$OH & 100\tablefootmark{a} & $<3.0\times10^{15}$ & 1, 2\\
 C$_3$O & 100\tablefootmark{a} & $<2.0\times10^{13}$ & 1\\
 \emph{cis-}HC(O)CHO & 100\tablefootmark{a} & $<5.0\times10^{13}$ & 1 \\
 C$_3$H$_8$ & 100\tablefootmark{a} & $<8.0\times10^{16}$ & 1 \\
 CH$_3$CCH & $100\pm20$ & $1.1\pm0.2\times10^{16}$ & 3 \\ 
 C$_2$H$_5$CHO & $125\pm25$ & $2.2\pm1.1\times10^{15}$\tablefootmark{$\dagger$} & 4 \\
 CH$_3$CHO & $125\pm25$ & $7.0\pm3.5\times10^{16}$\tablefootmark{$\dagger$} & 4 \\
 CH$_3$COCH$_3$ & $125\pm25$ & $1.7\pm0.8\times10^{16}$\tablefootmark{$\dagger$} & 4 \\
C$_2$H$_5$OCH$_3$ & $100\pm20$ & $1.8\pm0.2\times10^{16}$\tablefootmark{$\dagger$} & 5 \\
\hline
\end{tabular}
\tablefoot{
\tablefoottext{$\dagger$}{The authors chose a conservative estimation of the relative uncertainty on the column density of 50\%.}
\tablefoottext{a}{The excitation temperature is fixed to the mean excitation temperature of C$_2$H$_3$CHO and C$_3$H$_6$ and consistent with the excitation temperature of CH$_3$CCH \citep{Calcutt-2019}.}
}
\tablebib{
(1) this work; (2) \citet{Qasim-2019}; (3) \citet{Calcutt-2019}; (4) \citet{Lykke-2017}; (5) \citet{Manigand-2020}.
}
\end{table}

The identification of C$_2$H$_3$CHO and C$_3$H$_6$ has been confirmed by modelling the molecular emission under the LTE assumption and comparing the spectrum extracted at an offset position from the continuum peak position of IRAS 16293B. The model has been used in \cite{Manigand-2020}, where more substantial details can be found. This synthetic spectrum considers the gas as a homogeneous slab under LTE conditions and uses the optically thin molecular lines to fit the observed spectrum and to estimate the column density, the rotational temperature, the peak velocity shift, the FWHM and the source size, which is fixed to 0{\farcs}5, based on the previous PILS studies.

We run a grid of models in column densities from 10$^{13}$ to 10$^{17}$ cm$^{-2}$ and for excitation temperatures ranging from 50 to 300~K in steps of 25~K. The peak velocity shift and FWHM were set to 2.7 and 0.8~km~s$^{-1}$, respectively, for C$_2$H$_3$CHO and 2.5 and 0.8~km~s$^{-1}$, respectively, for C$_3$H$_6$, after a visual inspection of the alignment of the observed lines and the model.  The best model was estimated through a $\chi^2$ minimisation, with a modified weighting factor as detailed in \cite{Manigand-2020}. The modified weighting factor favours the under-estimation of the intensity of the modelled spectrum, which reflects possible unexpected blending effects with other species.  A correction factor was applied to the derived column density to take into account the continuum contribution as detailed, for example, in \cite{Jorgensen-2018} and \cite{Calcutt-2018b}.  The relative uncertainty of the column density and the rotational temperature is 20\%.

The best agreement between the data and the model for C$_2$H$_3$CHO was found at an excitation temperature of $125\pm25$~K and a column density of $3.4\pm 0.7\times10^{14}$ cm$^{-2}$. All the lines used in the fit are shown in Figure~\ref{fig-fit}. The lower intensity of the lines in Band 7 compared to those in Band 6 emphasises the relatively low rotational temperature derived from the model.

\begin{figure*}[t]\centering
\includegraphics[width=\textwidth]{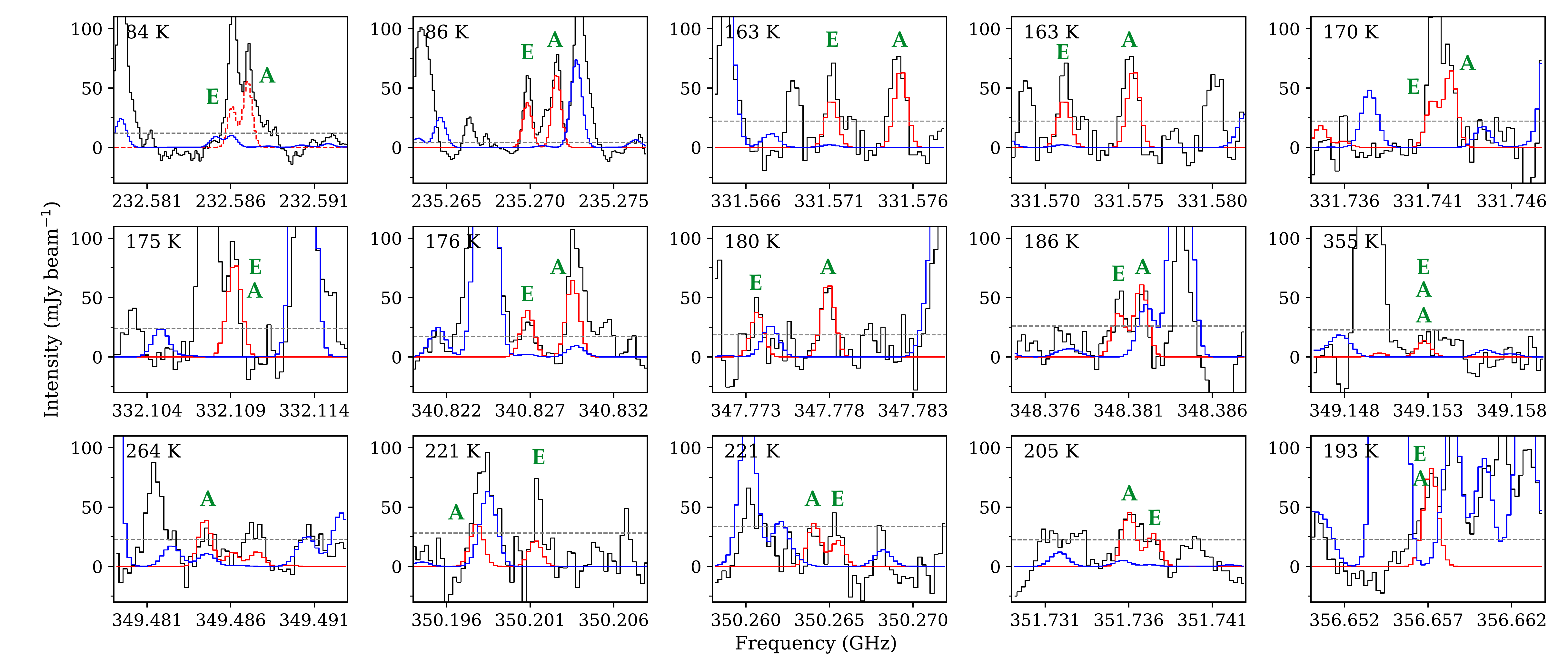}
\caption{\label{fig-fit-propene-split}Identified lines of C$_3$H$_6$ synthetic spectrum, in red, including the \emph{E}/\emph{A} spin weight ratio of $0.6\pm0.1$ are plotted along with the offset position spectrum towards IRAS 16293B. The dashed lines correspond to optically thick lines. The straight dashed black line represents the detection limit at the intensity of 3$\sigma$. The upper-level energy of the transition is noted in the upper left corner of each panel. The torsional state `E' or `A' of each transition is noted in green on the spectra. The reference spectrum in blue takes into account the previous species reported in PILS (see Figure \ref{fig-fit} and reference therein).}
\end{figure*}

The rotational temperature and column density of C$_3$H$_6$ was found to be $75\pm15$~K and $4.2\pm0.8\times10^{16}$~cm$^{-2}$, respectively. All the brightest lines in Band 6 are marginally optically thick for the best fit physical conditions. Nevertheless, the optically thick line at 235.272~GHz shows very good agreement with the observed line, which consolidates the unusually low value of 75~K found for the rotational temperature towards the IRAS 16293B hot corino region. The C$_3$H$_6$ lines used to assess the column density and the rotational temperature are shown in Figure \ref{fig-fit-propene}.

The majority of C$_3$H$_6$ transitions are split into \emph{A} and \emph{E} torsional sub-states due to the CH$_3$-- group in the molecule. This kind of split is common to most of the species having the CH$_3$ group, such as CH$_3$OH or CH$_3$OCHO. 
For C$_3$H$_6$, the intensities of the lines at 340.827, 347.774 and 351.738~GHz, corresponding to E-transitions, are overestimated by the LTE model. The A-transitions that have the same quantum numbers, in particular the same upper-level energies and Einstein's coefficients, are in good agreement with the model. This overestimation of the E-transitions alone may suggest that the spin weight ratio between the transitions arising from \emph{A} and \emph{E} substates is not equal to unity. 

To test this hypothesis, we added the spin weight of the \emph{E}-transitions with respect to the \emph{A}-transitions in the LTE model and assessed this ratio, given the best fit found for the excitation temperature, velocity shift and FWHM. This spin weight factor $g_\text{E/A}$ is multiplied to the upper state degeneracy of the \emph{E}-transitions. The best fit was found at the same column density, i.e. $4.2\pm 0.8\times10^{16}$ cm$^{-2}$, and a spin weight ratio of $g_\text{E/A} = 0.6\pm0.1$. Figure \ref{fig-fit-propene-split} shows the LTE model, including the discrepancy of the \emph{E}-transitions with respect to the \emph{A}-transitions. 
This difference between the A- and E-transitions could reflect an abundance asymmetry between the different conformers, due to low temperature at the moment of their formation, for example. Similar E/A ratios were observed towards dark clouds \citep{Friberg-1988}, cold prestellar clumps \citep{Menten-1988} and more recently towards young stellar objects \citep{Wirstrom-2011}. Nevertheless, in this study, there are still too few isolated transitions to be able to rule out any bias from the spectroscopic data that could explain this asymmetric E/A spin ratio.

For the sake of the chemical modelling comparison, the upper limits on the column density were derived for HCCCHO, C$_3$H$_7$OH, C$_3$O, C$_3$H$_8$, and \emph{cis}-HC(O)CHO. These upper limits correspond to the column density reached in the model when the most intense lines reached 3$\sigma$ at $\sim$100~K, which is the average excitation temperature of C$_2$H$_3$CHO and C$_3$H$_6$. The upper limit of cis-HC(O)CHO leads to an upper limit of the total amount of HC(O)CHO of $<$ 2.6 $\times~10^{23}$~cm$^{-2}$ in the gas phase, which does not provide any real constraint on this molecule.
Table \ref{tab-N} summarises the column densities and rotational temperatures of the newly detected species towards source B. Previous detections of chemically related species, namely acetaldehyde (CH$_3$CHO), acetone (CH$_3$COCH$_3$), ethyl methyl ether (C$_2$H$_5$OCH$_3$), C$_2$H$_5$CHO and CH$_3$CCH, reported from the same dataset by \cite{Lykke-2017} and \cite{Calcutt-2019}, are also indicated in the table and have been used for a comparison with the chemical model. \cite{Qasim-2019} already reported a column density upper limit for Ga-n-C$_3$H$_7$OH of $<1.2\times10^{15}$~cm$^{-2}$ at 300~K and $<7.4\times10^{14}$~cm$^{-2}$ at 125~K, using the Band 6 dataset of \cite{Taquet-2018}. Given their same dataset and the correction factor to include the contribution of the other conformers, this estimate is consistent with the results presented here. In the following, we consider the more conservative column density upper limit of $<3.0\times10^{15}$~cm$^{-2}$.

\section{\label{sec-NAUTILUS}The chemical model}

In this section, we compare the abundance, and the upper limits, derived from the observations towards IRAS 16293B with a physico-chemical model of a typical low-mass forming star. The core of the simulation uses the Nautilus code \citep{Ruaud-2016}, a 3-phase (gas, dust ice surfaces and dust ice mantle) time-dependent chemical model. The chemical study presented here is very similar, in terms of realisation, chemical code and astronomical source, to the studies of \cite{Andron-2018} and \cite{Coutens-2019}. The majority of the differences resides in the extended chemical network used here. 

The following subsections present the physical model that describes the evolution of physical conditions of the dust and gas during the simulation, the chemical network that computes the abundance variation caused by the reactions included in the network and the results obtained from the simulations.  

\subsection{Physical model}

Two successive evolutionary stages of a low-mass protostar are simulated: a uniform and constant stage, corresponding to the pre-stellar phase, or the cold-core phase, followed by an evolving stage during which the density and the temperature at the centre of the cloud increases while the envelope is collapsing towards the centre until breaking the hydrostatic equilibrium to form the protostar, i.e. the collapse phase.  

The cold-core phase consists of a homogeneous and static cloud at a temperature of 10~K for both gas and dust, a gas density of $2\times 10^{4}$~cm$^{-3}$, a visual extinction of 4.5 mag, a cosmic-ray ionisation rate of $1.3\times10^{-17}$~s$^{-1}$, and a standard external UV field of 1~G$_0$. The collapse phase is the same as used by \cite{Aikawa-2008, Aikawa-2012}, \cite{Wakelam-2014}, \cite{Andron-2018}, and \cite{Coutens-2019}. The physical structure used in their study was derived from a 1D radiative hydrodynamical model \citep{Masunaga-2000}, starting from the infall of the cold core, until the collapse stage of the first hydrostatic core, triggered by the dissociation of H$_2$ by collisions. While the physical structure evolves with time, several parcels of gas are tracked in time and position and their physical properties are traced throughout the collapse. Each parcel ends up at a different radius at the end of the simulation, which allows the evolution of the distribution of the molecular compounds included in the chemical network to be followed. In this study, the chosen parcel of gas starts the collapse at a radius of > $10^{4}$~au, where it stays for the major part of the collapse and ends-up at a radius of 15~au \citep[see Figure 5 in][]{Aikawa-2008}, where the gas reaches a temperature of $\sim265$~K and a density of $4\times10^{8}$~cm$^{-3}$. Figure~\ref{fig-phys} shows the evolution of the physical conditions of the targeted parcel of dust and gas. The parcel spends only the last couple of hundred years at a temperature higher than 100~K at a radius <$130$~au.

\subsection{Chemical network}

The chemical network includes the kida.uva.2014 gas-grain-phases network as a base  (\cite{Wakelam-2012, Wakelam-2015} with updates from \cite{Ruaud-2015, Hincelin-2015, Loison-2016, Hickson-2015, Hickson-2016a, Wakelam-2017, Loison-2017, Vidal-2017}). The initial abundances used in the simulations are listed in Table \ref{tab-init}.

\begin{table}[t]
\centering
\caption{\small \label{tab-init}Initial abundances used in the chemical model. }
\begin{tabular}{cc@{\qquad\quad}cc}
\hline\hline
Species & n~/~n(H) & Species & n~/~n(H)\\
\hline
H$_2$ & $5.0\times10^{-1}$ & S$^+$ & $1.5\times10^{-6}$ \\
He & $9.0\times10^{-2}$ & Fe$^+$ & $1.0\times10^{-8}$ \\
O & $2.4\times10^{-4}$ & F & $6.68\times10^{-9}$ \\
C$^+$ & $1.7\times10^{-4}$ & Cl$^+$ & $1.0\times10^{-9}$ \\
N & $6.2\times10^{-5}$ & & \\
\hline
\end{tabular}
\end{table}

Despite the inclusion of dozens of new species, hundred of new reactions, and the inclusion of hundreds of theoretical calculations to determine reaction barriers, there are still large uncertainties on branching ratios due to the lack of experimental studies and the difficulty of determining these branching ratios theoretically. Besides, the diffusion energies of radicals on surfaces are not well-known, which again leads to great uncertainties in determining which reactions becomes the most important when the temperature rises. More details on the upgrades made to the chemical network can be found in Appendix \ref{app-sec-Network}.

The chemical model distinguishes the ice surface and the ice mantle during the computation of the chemical reactions and the interaction processes between the different phases. For the comparison with the observations, the ice surface and ice mantle are considered as a single phase. We consequently summed the ice surface and ice mantle abundances to get the total ice abundances. The species in the solid phase are annotated with `s-' in front of their names.

To discuss the formation and destruction paths of the different species, the abundances relative to H$_2$ have been used in the comparison. This representation is convenient for discussing the evolution of the abundances with the modification of the density and the temperature along with the collapse of the protostellar envelope. 
The H$_2$ column density can be derived from the dust continuum radiation, assuming a dust size and an opacity distributions. However, the high continuum absorption seen towards IRAS 16293B in the PILS observations suggests that the dust emission is optically thick. \cite{Jorgensen-2016} estimated the lower limit of $>$1.2 $\times$ 10$^{25}$~cm$^{-2}$ for the H$_2$ column density.
Therefore, we chose to use the abundances relative to CH$_3$OH when comparing the final state of the simulation to the observational results.
The abundances derived from the observations are taken from \cite{Jorgensen-2016, Jorgensen-2018, Lykke-2017, Manigand-2020} and this study.

\subsection{Simulation runs}

Different reactions have been considered to reproduce the observations similarly to what has been done in other studies \citep[e.g.][]{Coutens-2018b}.
Among the important reactions investigated in this paper, the gas-phase formation reaction involving OH is crucial in the formation of C$_3$O from the consumption of C$_3$ and has been studied by \cite{Loison-2017}:
\begin{equation}\label{eq:C3-OH}
\mathrm{OH + C_3} \longrightarrow \mathrm{C_3O + H}
\end{equation}
The surface production pathways are marginally contributing to the formation of C$_3$O radicals present on the ice surfaces, during the pre-stellar phase.
The major destructive path of C$_3$ occurs through the gas-phase reaction: 
\begin{equation}\label{eq-C3O}
\mathrm{O + C_3} \longrightarrow \mathrm{C_2 + CO}
\end{equation}
This reaction is thought to play an important role in the abundance of C$_3$H$_x$, C$_3$O, and C$_2$H$_x$CHO species \citep{Hickson-2015}. 
The radical-radical addition reactions on grain surfaces are marginally contributing to the final abundances of these complex species.  

To investigate the chemical network involved in the formation of C$_3$-species, we run several simulations, depending on the tested chemical branch:
\begin{itemize}
\item[$\bullet$] A1: This simulation is the fiducial model, with which the other models can be compared. The common COMs, such as CH$_3$OCH$_3$ and CH$_3$OCHO, are mainly formed on grain surfaces but gas-phase formation pathways are also included, except for CH$_2$(OH)CHO, (CH$_2$OH)$_2$ and CH$_3$OCH$_2$OH. The hydroxyl radical OH contributes only to the formation of C$_3$O and HCCCHO in the gas phase. A reaction rate of $1.0\times10^{-12}$~cm$^{3}$\,molecule$^{-1}$\,s$^{-1}$ is considered for the gas-phase reaction \eqref{eq-C3O}.
\item[$\bullet$] B1: The hydrogenation of HC(O)CHO, based on \cite{Chuang-2016}, has been added to the grain surface reaction network:
\begin{eqnarray}\label{eq:Glyoxal-Hydrog}
\mathrm{\text{s-}HC(O)CHO + 2\ \text{s-}H} &\longrightarrow & \mathrm{\text{s-}CH_2(OH)CHO}\\
\mathrm{\text{s-}CH_2(OH)CHO + 2\ \text{s-}H} &\longrightarrow & \mathrm{\text{s-}(CH_2OH)_2}
\end{eqnarray}
In addition, we have included the hydrogenation of CH$_3$OCHO, forming CH$_3$OCH$_2$OH, to the network:
\begin{equation}\label{eq:MF-Hydrog}
\mathrm{\text{s-}CH_3OCHO + 2\ \text{s-}H} \quad \longrightarrow \quad \mathrm{\text{s-}CH_3OCH_2OH}
\end{equation}
based on the hydrogenation of CH$_2$(OH)CHO (reaction \ref{eq:Glyoxal-Hydrog}).
\item[$\bullet$] C1: This run uses the same reaction network as the fiducial run A1. In this case, the pre-stellar phase lasts longer than the other runs, i.e. $3\times10^6$~yr instead of $1\times10^6$~yr, which gives the cold chemistry more time to proceed. 
\item[$\bullet$] D1: In order to evaluate the contribution rate of the radical-radical addition on the grain surface to the abundance of HCCCHO, C$_2$H$_3$CHO and C$_2$H$_5$CHO, the following reactions have been removed from the network:\\
\begin{eqnarray}\label{eq:HCO-propenal}
\mathrm{\text{s-}HCO} + \mathrm{\text{s-}C_2H_3} &\longrightarrow & \mathrm{\text{s-}C_2H_3CHO} \\
 &\longrightarrow & \mathrm{\text{s-}C_2H_4 + \text{s-}CO}
\end{eqnarray}
\begin{eqnarray}\label{eq:HCO-propanal}
\mathrm{\text{s-}HCO} + \mathrm{\text{s-}C_2H_5} &\longrightarrow & \mathrm{\text{s-}C_2H_5CHO} \\
 &\longrightarrow & \mathrm{\text{s-}C_2H_6 + \text{s-}CO}
\end{eqnarray}
The addition reaction of HCO and C$_2$H on grains is not contributing to the formation of HCCCHO in our chemical network because the amount of available C$_2$H on the ice surfaces is limited by the very high activation barrier of the hydrogen abstraction of C$_2$H$_2$ \citep[>56$\,$000 K,][]{Zhou-2008}.
\item[$\bullet$] E1: This run takes into account the hydrogenation of HC(O)CHO and CH$_3$OCHO, as it is for run B1, and the activation energies for hydrogenation and abstraction reactions of H$_2$CO, CH$_3$OH, C$_2$H$_2$, C$_2$H$_4$, C$_2$H$_6$, C$_2$H$_3$CHO, and C$_2$H$_5$CHO, taken from the recent laboratory study of \cite{Qasim-2019} about propanal and propanol formation on ice surfaces from CO, and the references therein \citep{Andersson-2011, Goumans-2011, Kobayashi-2017, Song-2017, Alvarez-Barcia-2018, Zaverkin-2018}. The activation energies taken from \cite{Qasim-2019} are, in general, lower than those used in the simulation runs A1 to D1.
\item[$\bullet$] A2 to E2: These runs are similar to the runs A1 to E1, but with a lower reaction rate of $10^{-16}$ cm$^{3}$~s$^{-1}$ for reaction \eqref{eq-C3O}.
\end{itemize}

\textcolor{black}{The list of the reactions added, extrapolated or removed in each run, with the associated branching ratios and reaction rates, is presented in Appendix \ref{app-sec-runs}. The formation enthalpies of the intermediate species that intervene in the new hydrogenation paths on grain surfaces are taken from \cite{Goldsmith-2012}. }

\subsection{Modelling results}

\begin{figure}[t]\centering
\includegraphics[width=0.48\textwidth]{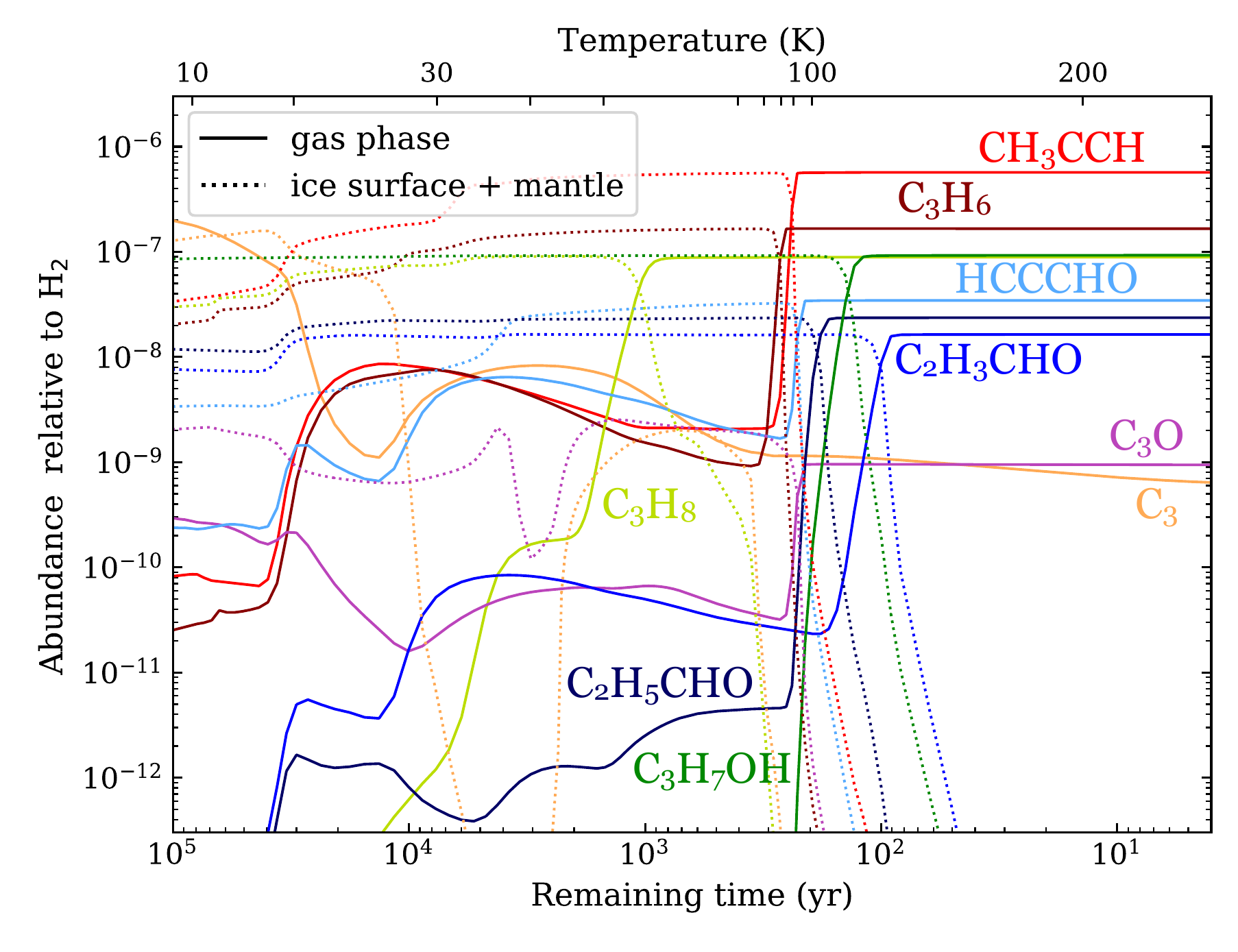}\\
\includegraphics[width=0.48\textwidth]{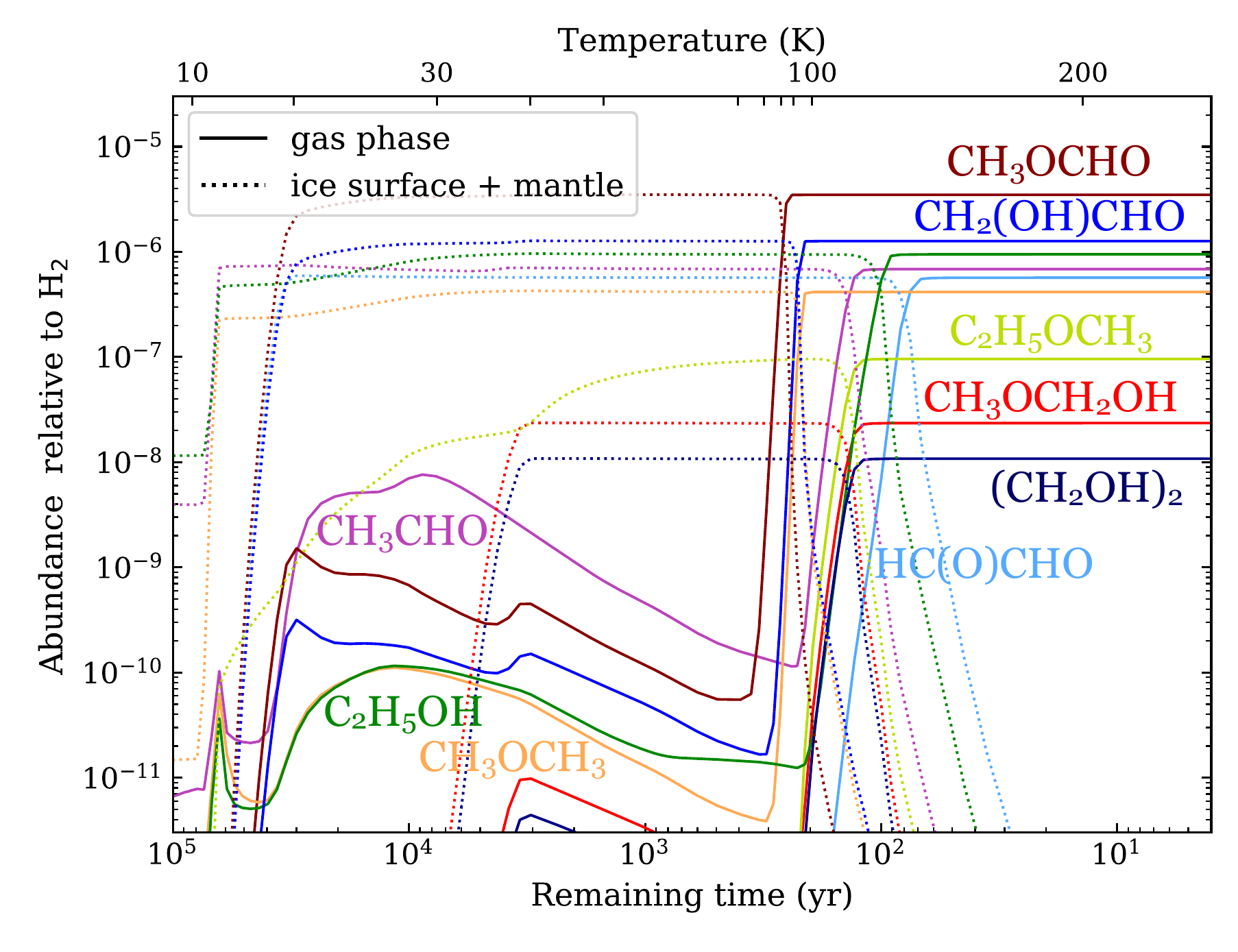}
\caption{\label{fig-simu-newchem}\small Time evolution of abundances in the gas phase (solid lines) and on the grain surface (dotted) of several species during the collapse phase of the A1 run. The time axis is reversed to better visualise the abundances evolution. Both panels show the results of the same simulation but with a different selection of species, with their name annotated following the colour code of the curves.}
\end{figure}

\begin{figure*}[h]\centering
\adjustbox{trim= 0 {0.03\height} 0 0, clip=true}{\includegraphics[width=0.24\textwidth]{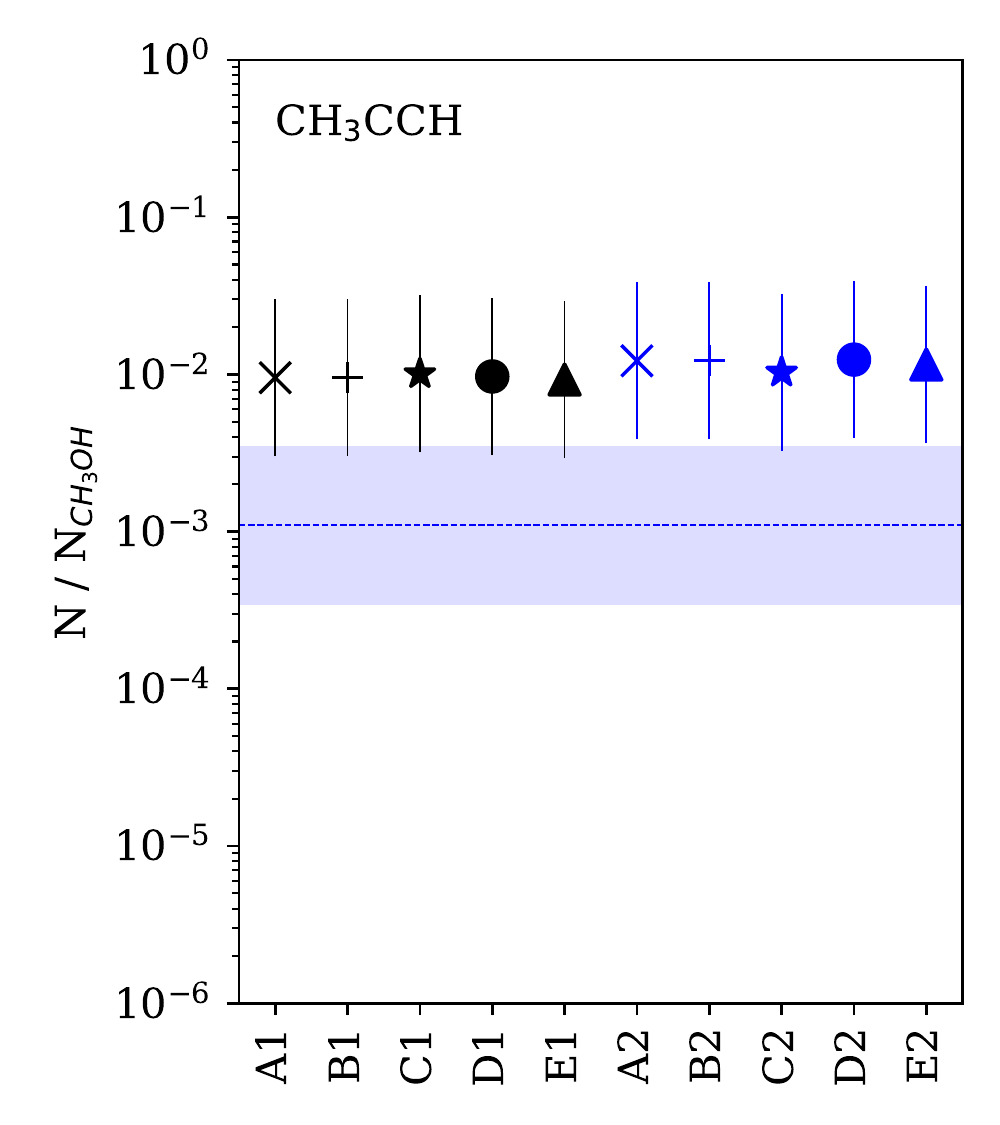}}
\adjustbox{trim= {0.12\width} {0.03\height} 0 0, clip=true}{\includegraphics[width=0.24\textwidth]{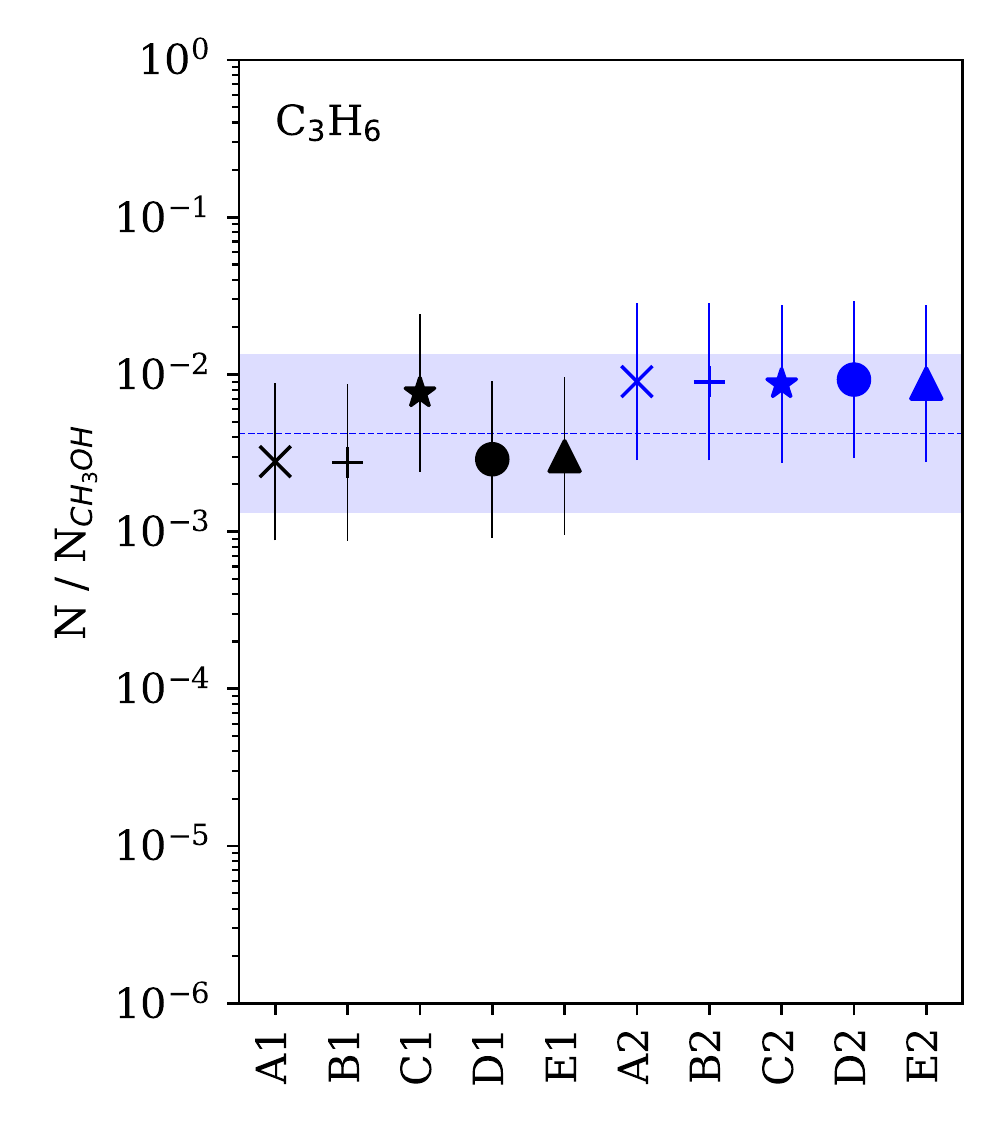}}
\adjustbox{trim= {0.12\width} {0.03\height} 0 0, clip=true}{\includegraphics[width=0.24\textwidth]{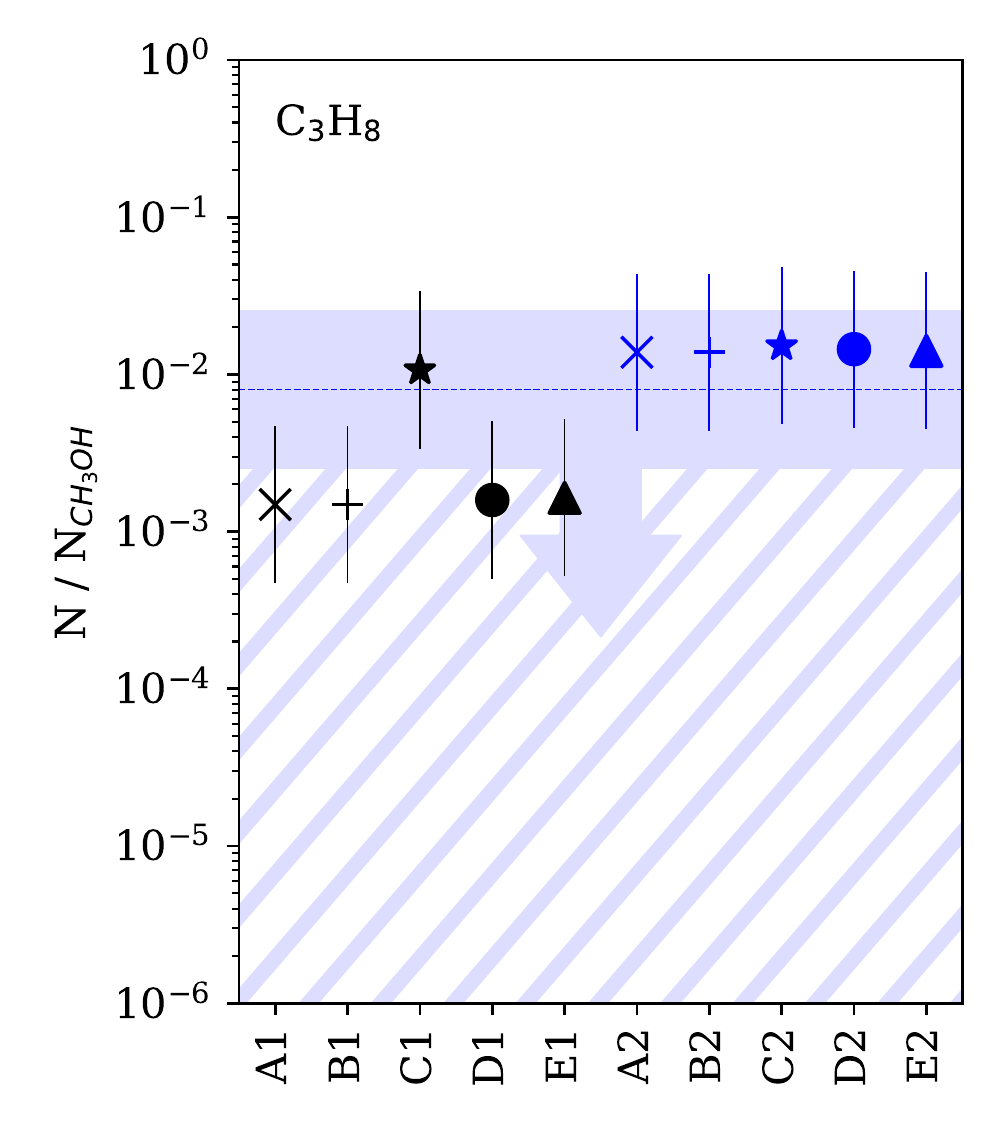}}
\adjustbox{trim= {0.12\width} {0.03\height} 0 0, clip=true}{\includegraphics[width=0.24\textwidth]{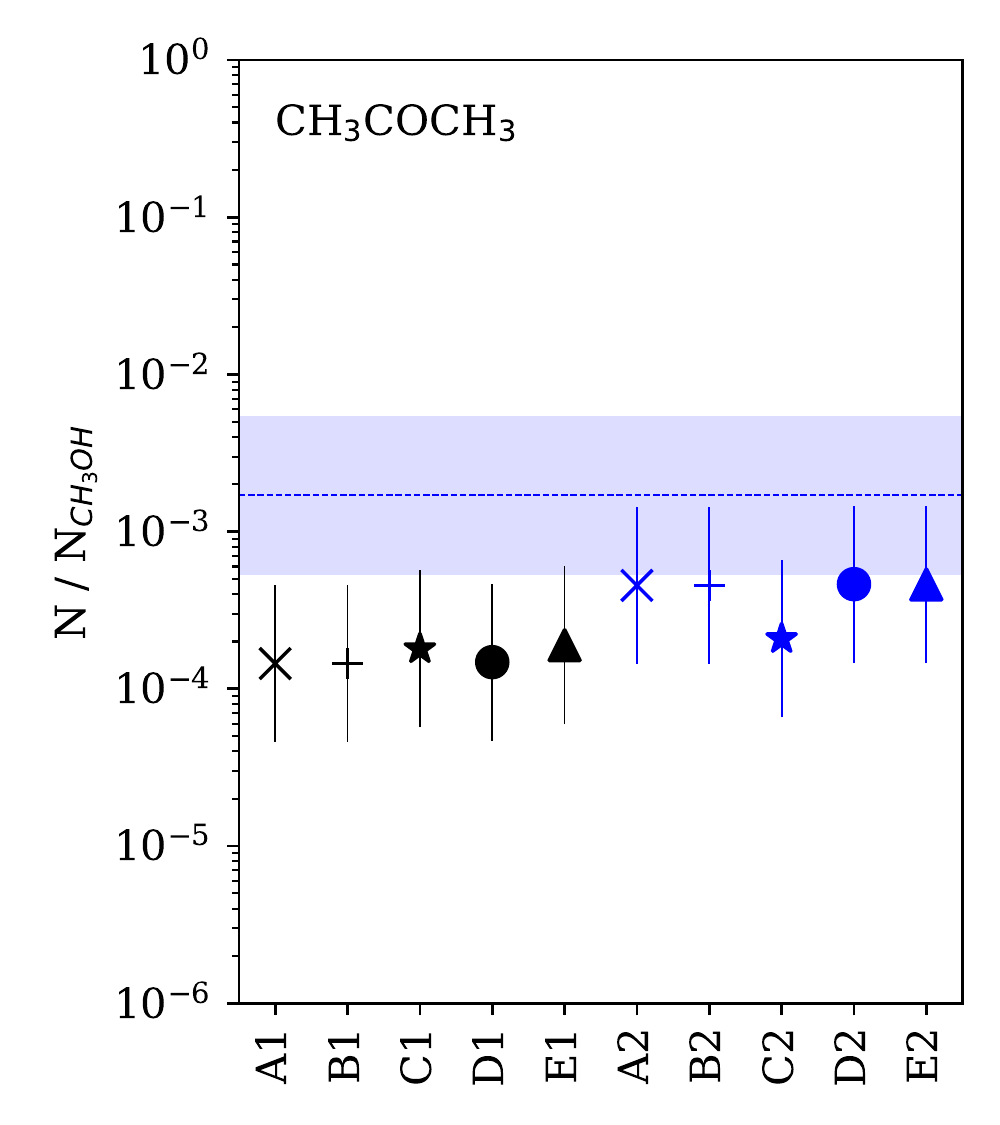}}\\
\adjustbox{trim= 0 {0.03\height} 0 0, clip=true}{\includegraphics[width=0.24\textwidth]{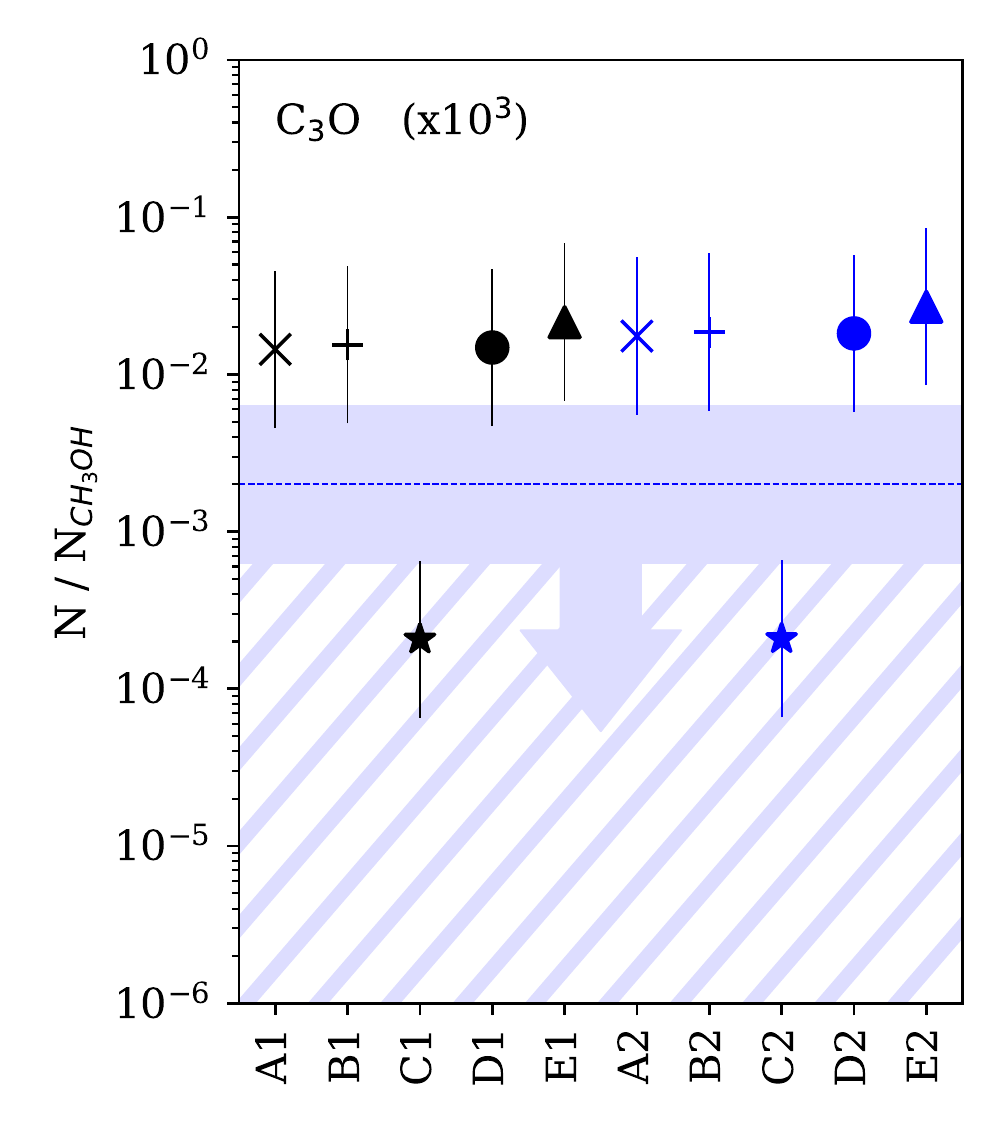}}
\adjustbox{trim= {0.12\width} {0.03\height} 0 0, clip=true}{\includegraphics[width=0.24\textwidth]{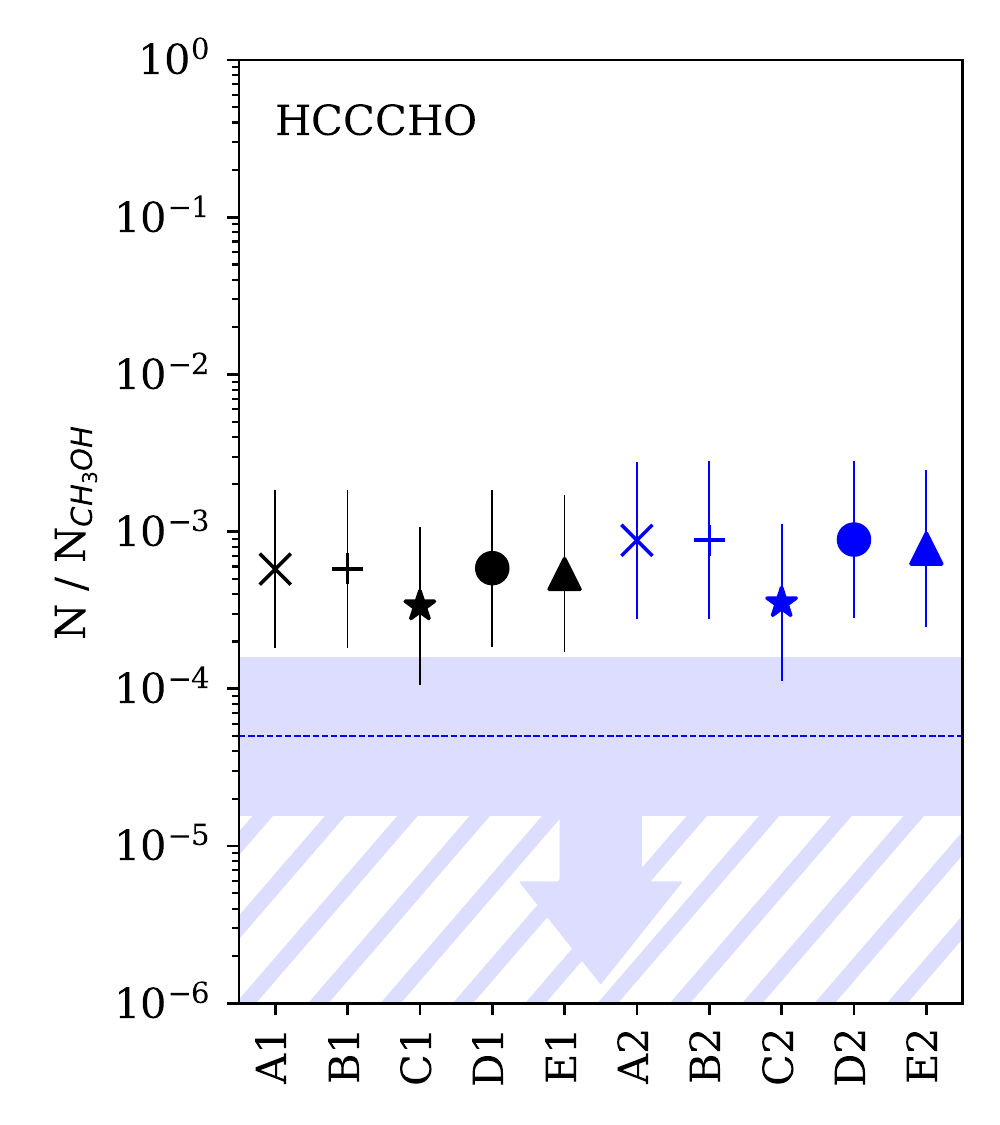}}
\adjustbox{trim= {0.12\width} {0.03\height} 0 0, clip=true}{\includegraphics[width=0.24\textwidth]{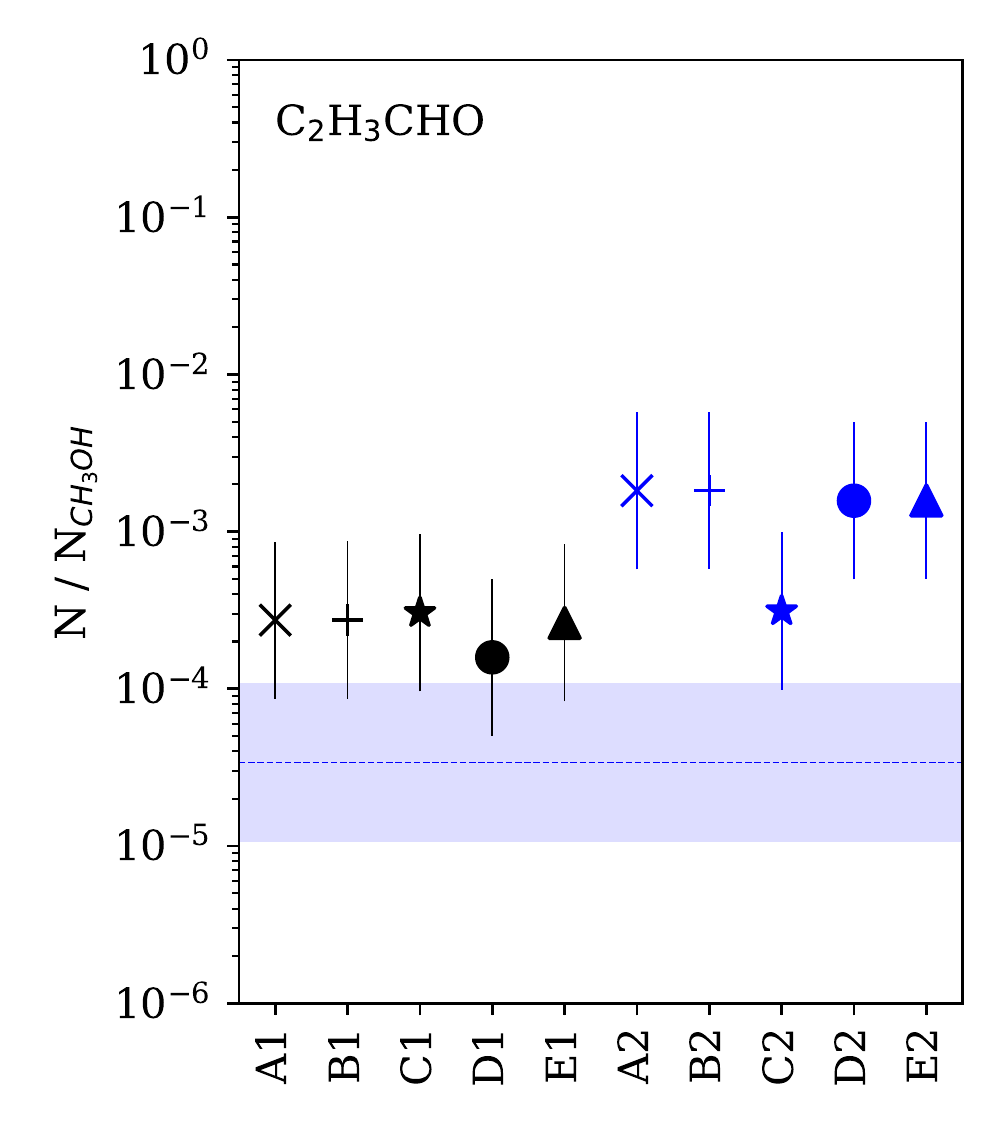}}
\adjustbox{trim= {0.12\width} {0.03\height} 0 0, clip=true}{\includegraphics[width=0.24\textwidth]{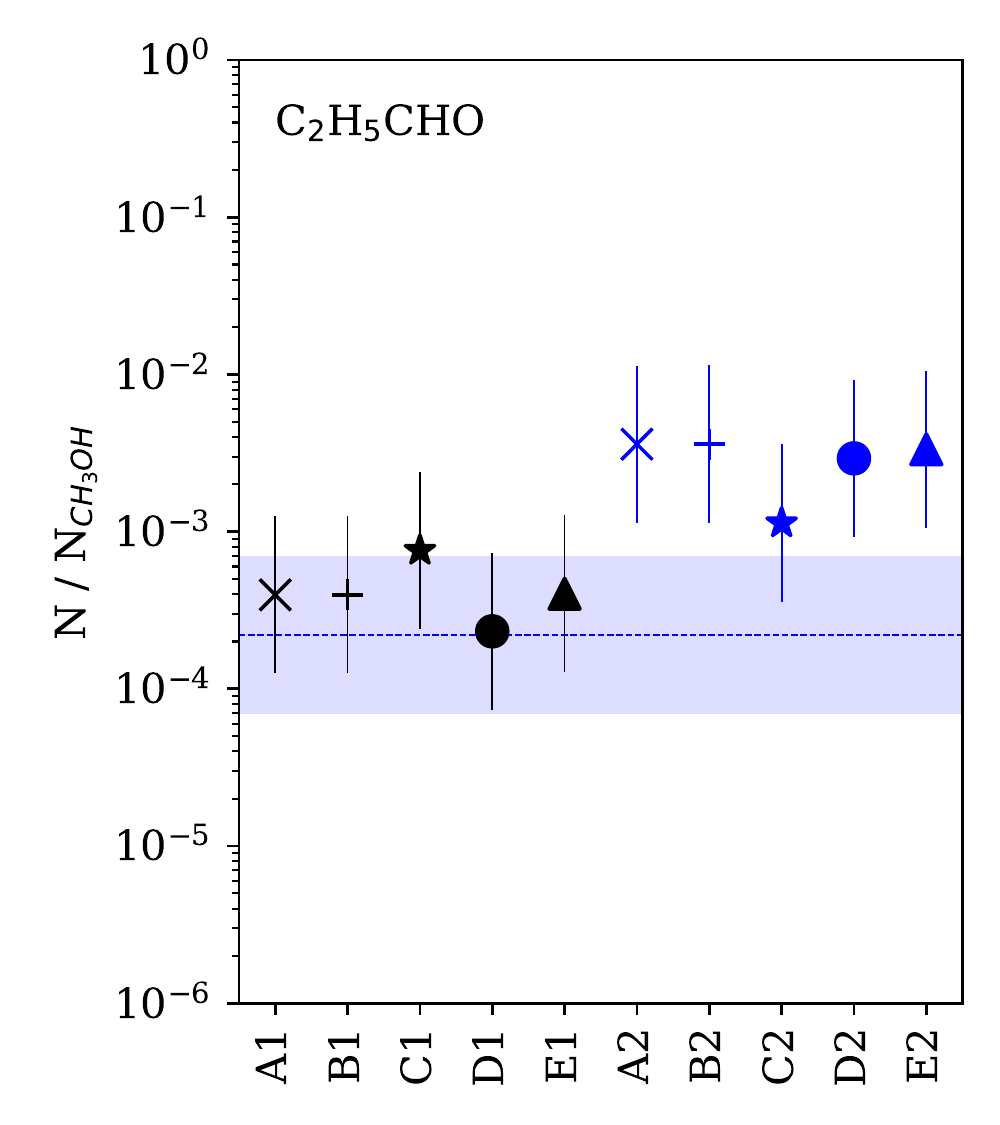}}\\
\adjustbox{trim= 0 {0.03\height} 0 0, clip=true}{\includegraphics[width=0.24\textwidth]{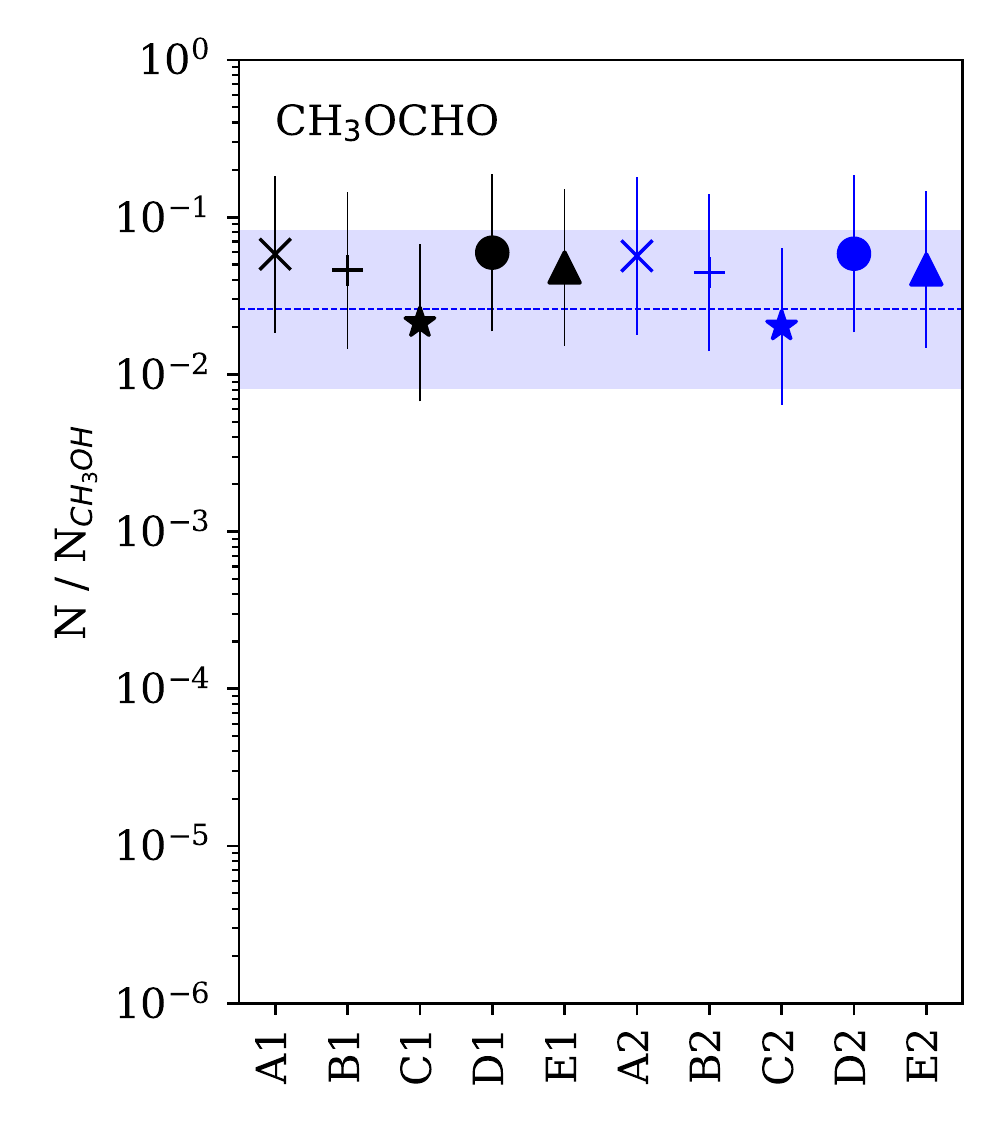}}
\adjustbox{trim= {0.12\width} {0.03\height} 0 0, clip=true}{\includegraphics[width=0.24\textwidth]{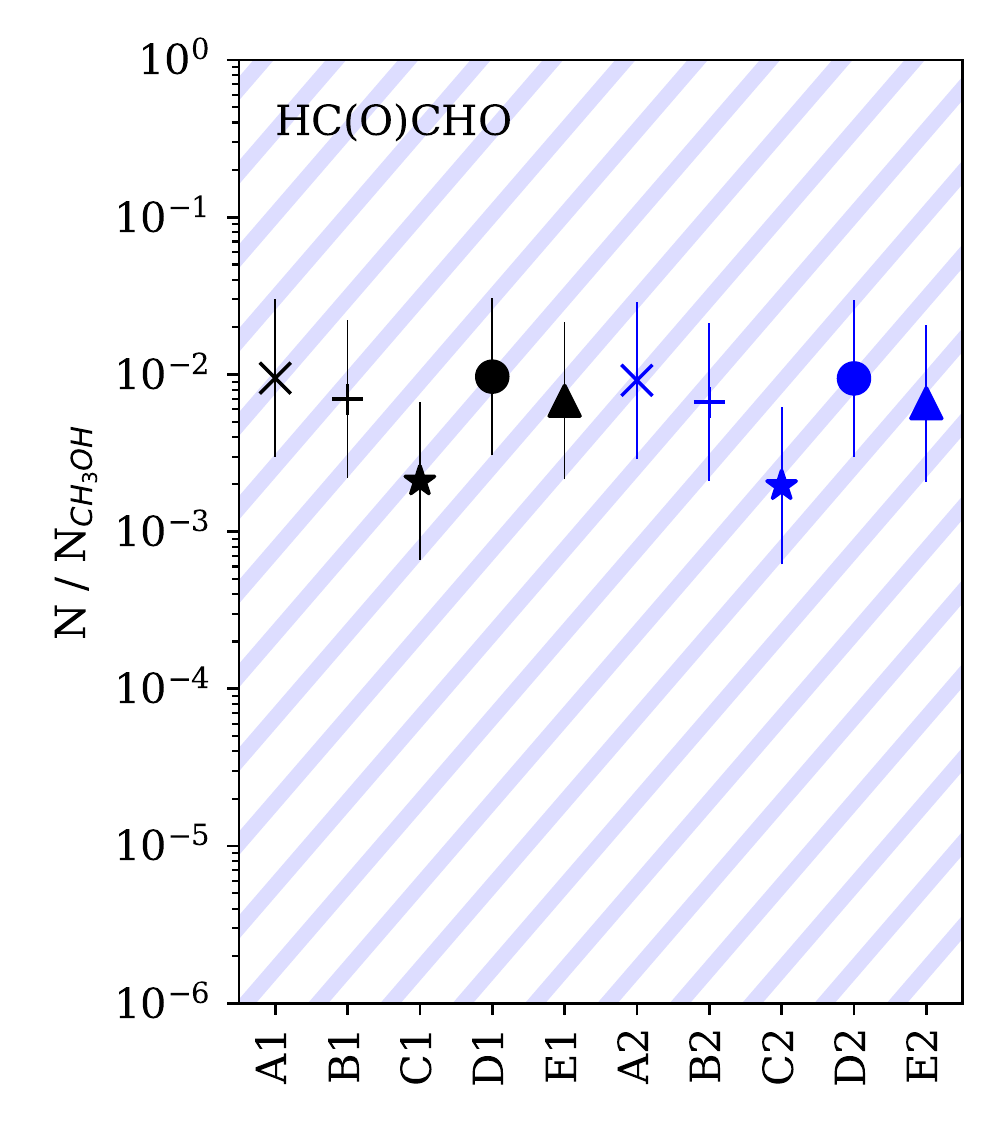}}
\adjustbox{trim= {0.12\width} {0.03\height} 0 0, clip=true}{\includegraphics[width=0.24\textwidth]{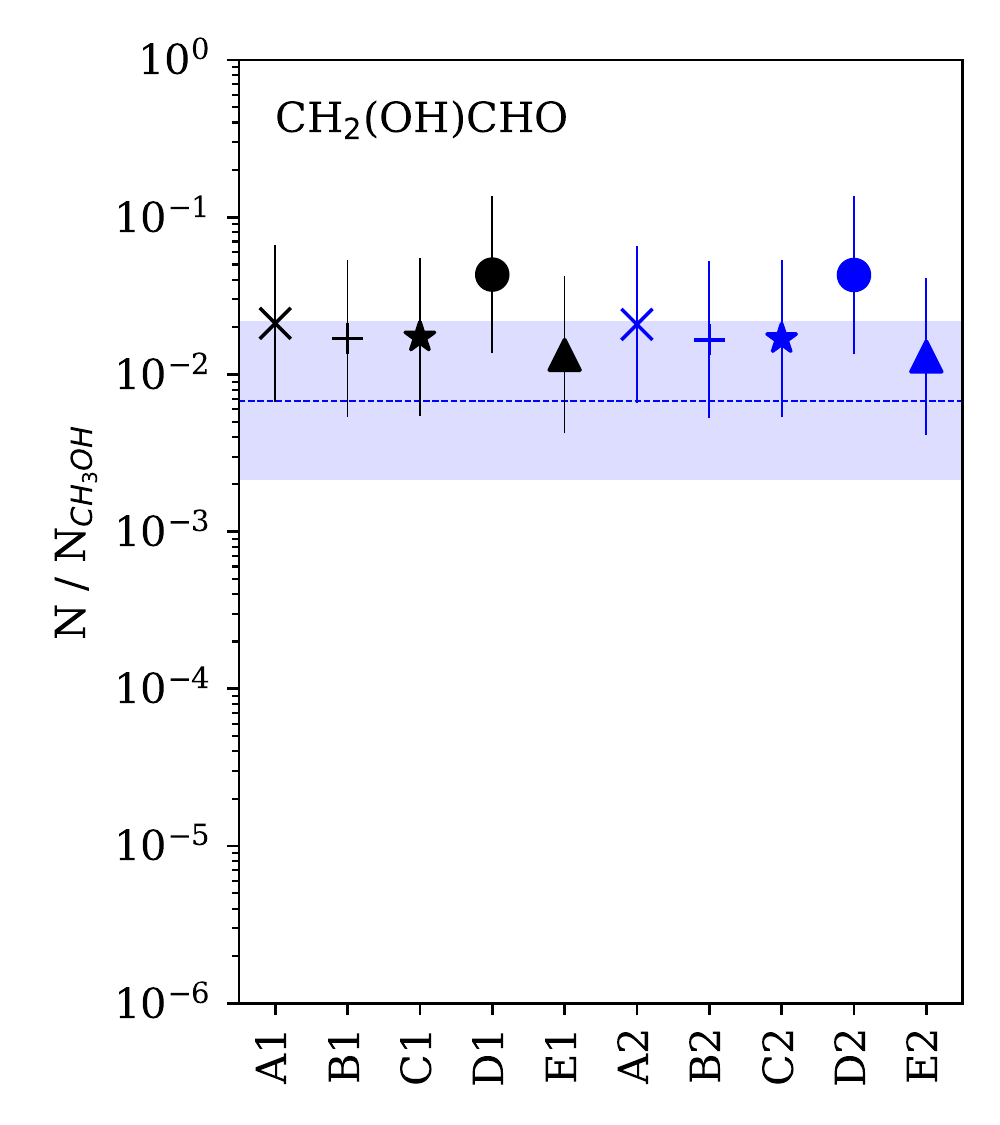}}
\adjustbox{trim= {0.12\width} {0.03\height} 0 0, clip=true}{\includegraphics[width=0.24\textwidth]{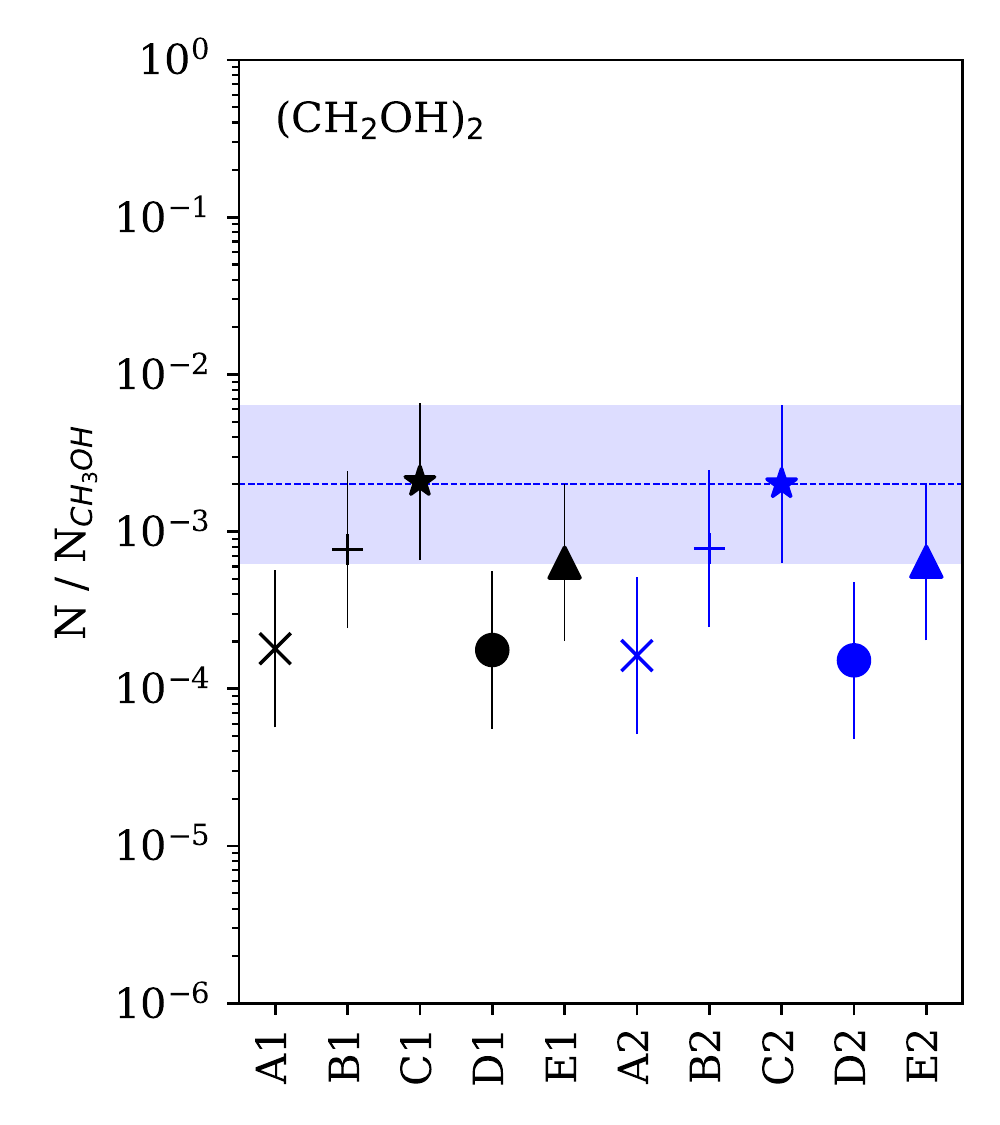}}\\
\adjustbox{trim= 0 {0.00\height} 0 0, clip=true}{\includegraphics[width=0.24\textwidth]{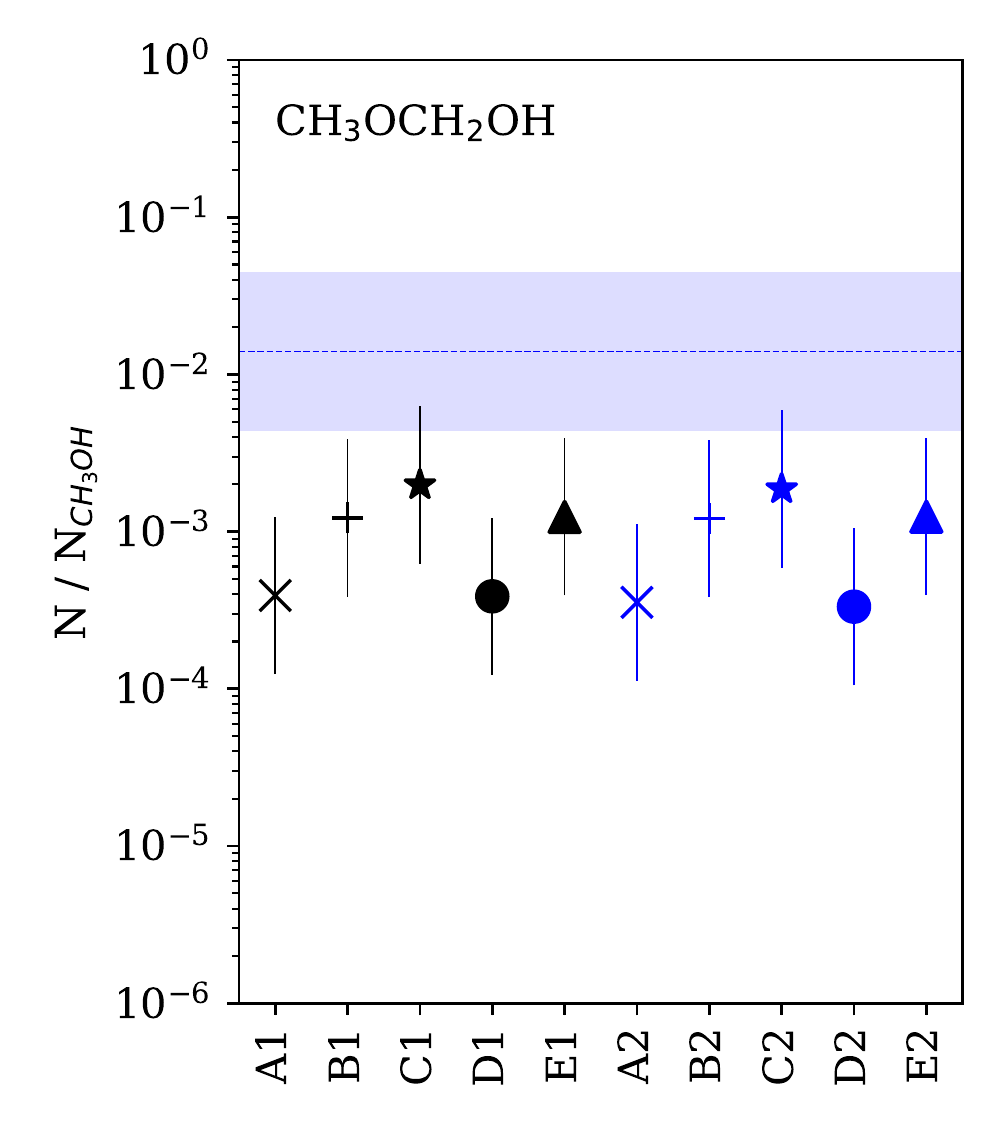}}
\adjustbox{trim= {0.12\width} {0.00\height} 0 0, clip=true}{\includegraphics[width=0.24\textwidth]{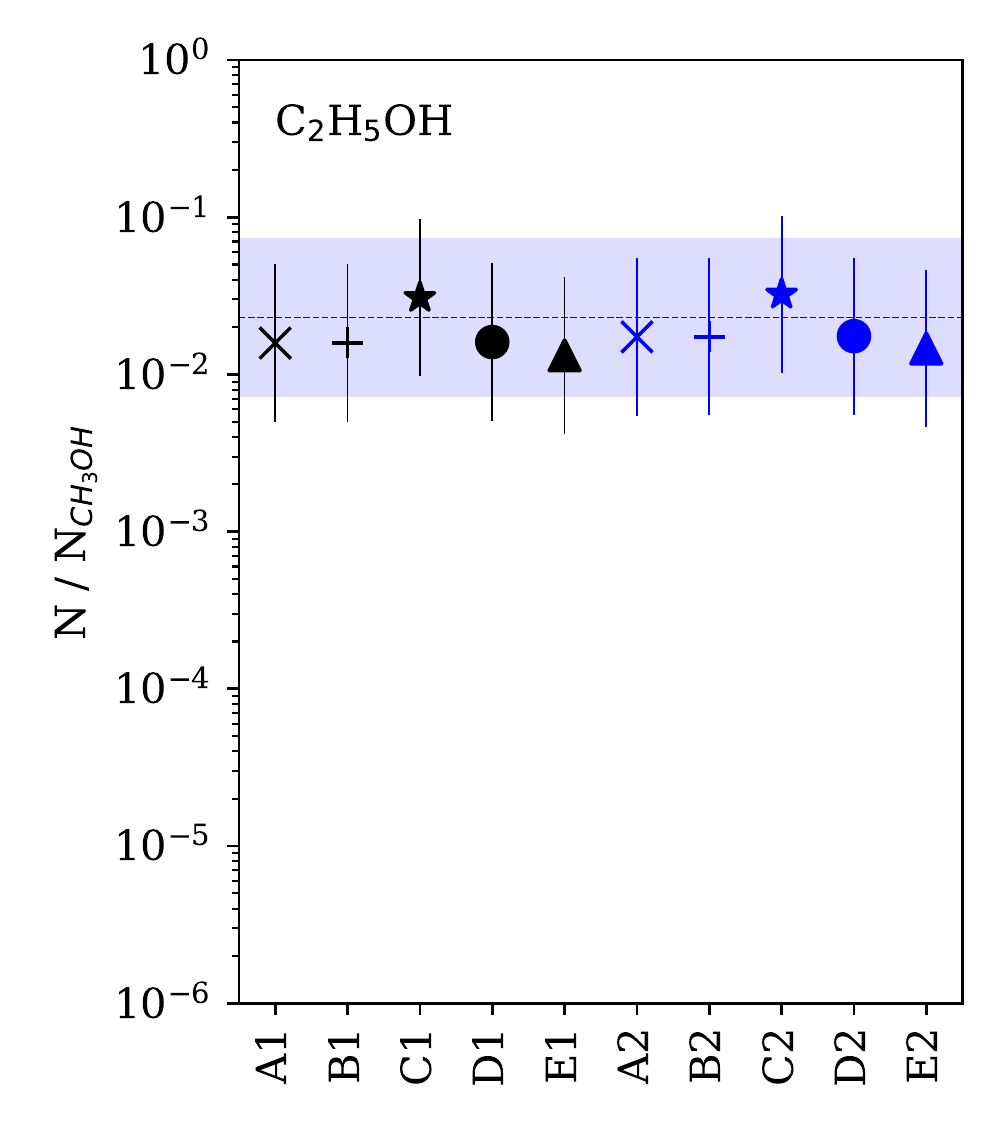}}
\adjustbox{trim= {0.12\width} {0.00\height} 0 0, clip=true}{\includegraphics[width=0.24\textwidth]{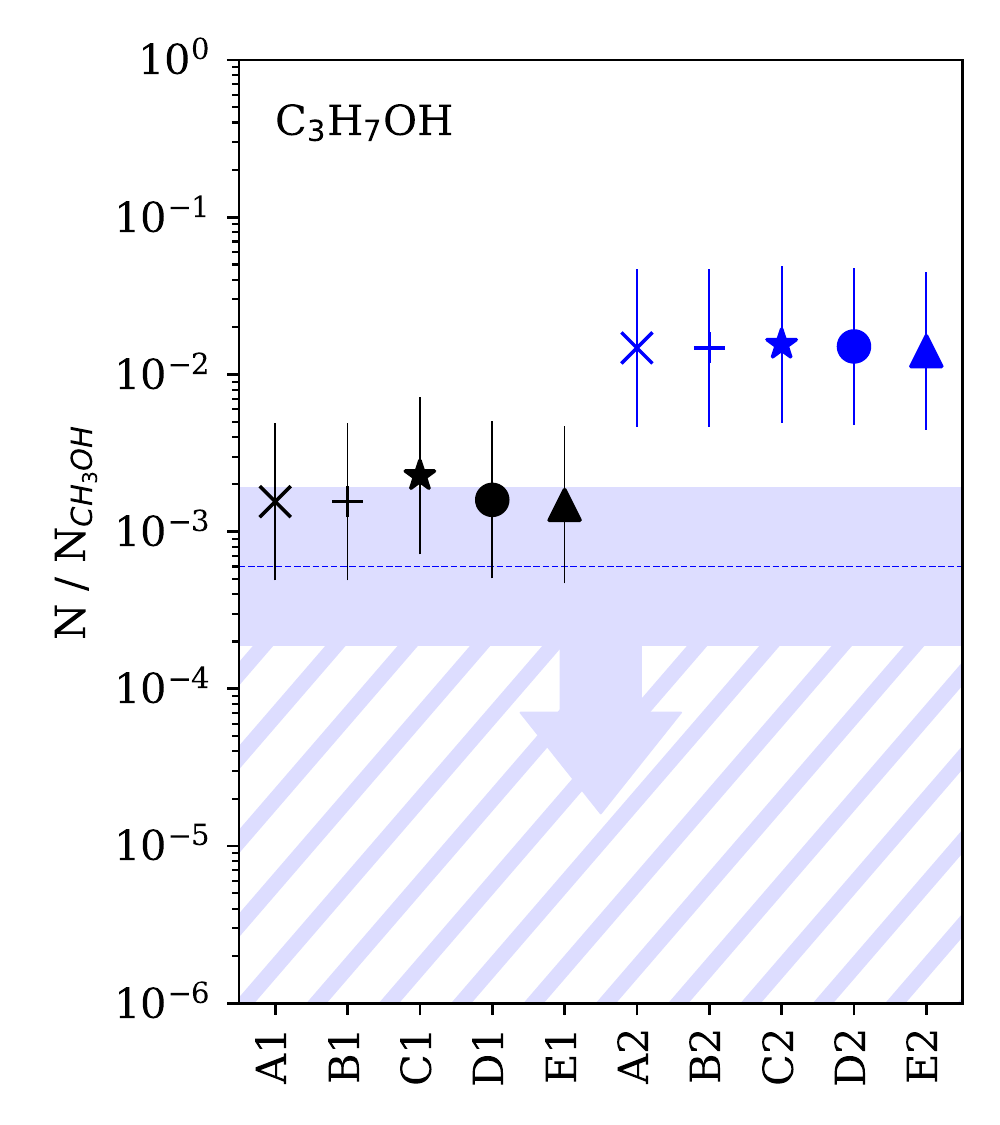}}
\adjustbox{trim= {0.12\width} {0.00\height} 0 0, clip=true}{\includegraphics[width=0.24\textwidth]{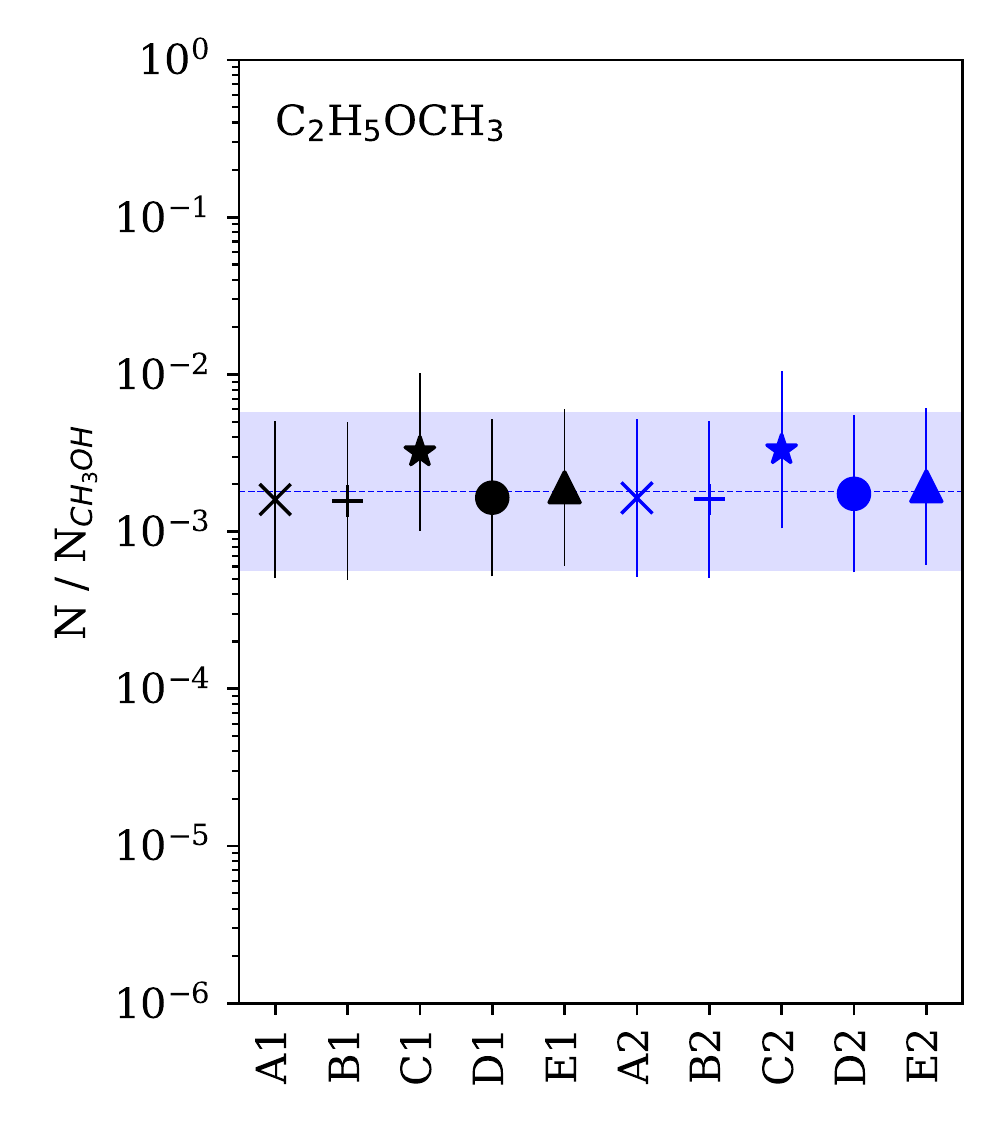}}
\caption{\small \label{fig-Runs}Final abundances reached at the end of each simulation run for the targeted species. The dashed blue line shows the observed abundance of the species towards IRAS 16293B. The error bars and the blue area correspond to one half order of magnitude confidence limit, thus a 1$\sigma$-difference between observations and simulations corresponds to one order of magnitude. The dashed blue region represents the abundance values that are consistent with the corresponding upper limit. The abundance scale of C$_3$O is lower than the other plots. The abundance upper limit of HC(O)CHO is $<$ 2.6 $\times10^4$ relative to CH$_3$OH, therefore, there is no observational constraint on the abundance of HC(O)CHO. The abundance upper limit of C$_3$H$_7$OH is the sum of n-C$_3$H$_7$OH and i-C$_3$H$_7$OH abundance upper limits.}
\end{figure*}

The evolution of the abundances during the collapse phase of the run A1 is shown in Figure \ref{fig-simu-newchem}. For the COMs formed in a hot corino region, most of the molecules targeted are produced on grain surfaces and are released in the gas-phase when the temperature reaches the desorption temperature, between 80 to 130~K depending on the species. 

This production scheme leads to a jump of several orders of magnitude in gas-phase abundances of the species around such temperatures. Concerning the C$_3$-species, the jump in gas-phase abundances is smaller which indicates significant gas-phase production route contributions. These gas-phase contributions seem higher for the most unsaturated species, such as HCCCHO, C$_3$ and C$_3$O. At the end of the simulation, all the formed molecules are in the gas phase.

Figure \ref{fig-Runs} summarises the gas-phase abundances relative to CH$_3$OH of the C$_3$-species that are under investigation in this study and a few other COMs, such as CH$_3$OCHO, CH$_2$(OH)CHO, (CH$_2$OH)$_2$, CH$_3$OCH$_2$OH, C$_2$H$_5$OH, CH$_3$COCH$_3$, and C$_2$H$_5$OCH$_3$, at the end of the collapse phase for every simulation. 
Because of the assumptions from the physical model compared to the actual physical structure of IRAS 16293B, which is still not perfectly known, an abundance difference of less than one order of magnitude is considered as an agreement between the simulation and the observations.  

\subsubsection{C$_3$-species}

The abundances of C$_3$--species, that are CH$_3$CCH, C$_3$H$_6$, C$_3$H$_8$, HCCCHO, C$_2$H$_3$CHO, and C$_2$H$_5$CHO, are correlated to those of their precursors, C$_3$ and C$_3$O. As C$_3$ is efficiently produced and is not a very reactive species in the gas phase, s-CH$_3$CCH and s-C$_3$H$_6$ are quite abundant in the solid phase.  
The comparison between the runs A1--E1 and A2--E2 shows that most of the C$_3$-species are sensitive to a decrease in the reaction rate of the C$_3$ destruction reaction \eqref{eq-C3O}. 
As the reaction rate decreases, there is more C$_3$ available to form C$_3$-species, thus their abundances increase. The increase in the abundances is more pronounced for the more saturated species.  

The abundances of the C$_3$-species at the end of the simulation run C1 are significantly different from those at the end of the fiducial run. The longer time spent at very low temperatures during the prestellar phase pushes the hydrogenation reactions to produce more saturated C$_3$-species by consuming the unsaturated precursors. This is particularly shown by the lower abundances of C$_3$O and HCCCHO contrasting the higher abundances of C$_2$H$_3$CHO and C$_2$H$_5$CHO.

The ice-surface formation pathways of C$_2$H$_3$CHO and C$_2$H$_5$CHO, through reactions \eqref{eq:HCO-propenal} and \eqref{eq:HCO-propanal}, are deactivated in the simulation run D1. The final abundances show a decrease compared to the simulation where the molecules are produced through ice-surface reactions, among others. The updated activation energies, taken from the study of \cite{Qasim-2019} and tested in the simulation E1, do not change the final abundances of the C$_3$-species significantly.

The C$_3$H$_7$OH species is a particular case as there is an alternative production pathway apart from successive hydrogenation of C$_3$O, mostly the O + C$_3$H$_8$ reaction. 
The slight overestimate for C$_3$H$_7$OH may be due to an overestimate of the s-OH + s-C$_3$H$_7$ recombination branching ratio versus the disproportionation, that is the exchange reaction of one H atom from one radical to the other:
\begin{equation}
\text{s-OH} + \text{s-}\mathrm{C_3H_7} \longrightarrow \text{s-}\mathrm{H_2O} + \text{s-}\mathrm{C_3H_6}
\end{equation}
Its isomer C$_2$H$_5$OCH$_3$, however, is well reproduced by the models, despite the fact that there is only the formation pathway through radical-radical addition that are implemented in the chemical network (see Table \ref{tab-BR} in Appendix). The variation of the final abundance of C$_2$H$_5$OCH$_3$ is very similar to that of C$_2$H$_5$OH which suggests that they share the same type of formation reactions on ice surfaces, assuming that the major formation pathways of C$_2$H$_5$OH effectively take place on ice surfaces as well.

\subsubsection{COMs hydrogenation}

Concerning the species formed through radical-radical addition, the slight to large overestimation or underestimation of the calculated abundances when considering the fiducial run A1 likely reflects the uncertainties of the association channel branching ratios compared to the disproportionation channel.
 
The underestimation of (CH$_2$OH)$_2$ is unexpected considering the large amount of s-CH$_2$OH as well as the large branching ratio used for the reaction  s-CH$_2$OH + s-CH$_2$OH $\rightarrow$ s-(CH$_2$OH)$_2$, that is 50\%, in the fiducial chemical network.  
Despite the approximate agreement between the simulations and the observations for CH$_3$OCHO and CH$_2$(OH)CHO, these two species are over-produced by the radical-radical additions whereas (CH$_2$OH)$_2$ and CH$_3$OCH$_2$OH, which are only produced by radical-radical additions on grain surfaces, are both under-produced.

The run B1 includes the hydrogenation formation pathways on grain surfaces of HC(O)CHO and CH$_3$OCHO. These reactions lead to the consumption of the unsaturated species and the production of saturated ones. The agreement of this run with observations is better than for the fiducial chemical network in terms of abundances of CH$_2$(OH)CHO, (CH$_2$OH)$_2$, CH$_3$OCHO, and CH$_3$OCH$_2$OH, suggesting that hydrogenation reactions play an important role in formation of such COMs in hot corino regions. 
The hydrogenation of CH$_3$OCHO has been recently studied in laboratory by \cite{Krim-2018}. They reported a calculated energy barrier of 32.7~kJ~mol$^{-1}$ for the ice-surface reaction s-H + s-CH$_3$OCHO and concluded that this pathway did not seem to contribute to the formation of CH$_3$OCH$_2$OH. The experimental result of Krim et al. mainly shows that this reaction is slower than the s-H + s-H$_2$CO $\rightarrow$ s-CH$_3$O reaction but does not completely exclude its role in interstellar ice conditions. In our simulation, we include this reaction with a barrier equal to 3000~K (25~kJ~mol$^{-1}$) as average between the value calculated in this work, the value calculated by Krim et al., and the value calculated by \cite{Alvarez-Barcia-2018}, i.e. 4960~K (41.2 kJ~mol$^{-1}$). This reaction, like all H reactions with barriers, is partly allowed by the tunnel effect.

The reaction s-HCO + s-HCO has been suggested to produce s-HC(O)CHO \citep{Fedoseev-2015, Chuang-2016} to explain the formation of CH$_2$(OH)CHO and (CH$_2$OH)$_2$ in experimental surface hydrogenation of CO molecules. HC(O)CHO production through s-HCO + s-HCO reaction is also achieved by \cite{Simons-2020} using density functional theory \citep[DFT,][]{Scuseria-2005} calculations in a model in which the surface molecules were not explicitly taken into account. 
Additionally, the recent experimental work of \cite{Butscher-2017} indicates that the reaction of two HCO radicals on ice surfaces does not lead to HC(O)CHO but rather to H$_2$CO and CO. Considering the conflicting results from different studies, it is difficult to estimate the proportion of HC(O)CHO produced by the s-HCO + s-HCO reaction. In this study, we consider a branching ratio of 5\% for HC(O)CHO formation through the s-HCO + s-HCO reaction. This leads to a significant production of HC(O)CHO due to the great abundance of HCO on the grains, which is around 1\% of the CH$_3$OH (see Figure \ref{fig-Runs}). The comparison with the observations is delicate because the most stable isomer of HC(O)CHO, the trans- one, has no dipole moment and the population fraction of cis-HC(O)CHO is equal to $1.9\times10^{-10}$ at 100~K, considering that the thermal equilibrium is reached, and is not detected in the observations.

The updated activation energies from \cite{Qasim-2019} of abstraction reactions of CH$_3$OH and H$_2$CO affect the final abundances of the COMs that are predominantly formed through radical-radical addition reactions. In particular, the lower final abundance of (CH$_2$OH)$_2$ in run E1 compared to B1 is affected by the newly implemented abstraction reaction of CH$_3$OH which produces CH$_3$O radicals.

\section{Discussion}

In most of the models, C$_3$ reaches a high abundance in the gas phase, due to its low reactivity. This leads to a high abundance of s-C$_3$ in solid phase, which results in high abundances of all the C$_3$-species in the gas phase, especially CH$_3$CCH, C$_3$O, and HCCCHO, which are much more abundant than observed. A better agreement to the observations can be found when the rate constant for the $\mathrm{O + C_3} \longrightarrow \mathrm{C_2 + CO}$  gas-phase reaction is set to $1\times10^{-12}$~cm$^3$~s$^{-1}$, assuming a lower activation barrier than in the theoretical work of \cite{Woon-1996}.

The agreement for CH$_3$CCH, C$_3$H$_6$, C$_2$H$_3$CHO, and C$_2$H$_5$CHO is rather good between the fiducial run A1 and the observations for IRAS 16293. However, this leads to an underestimation of CH$_3$CCH and C$_3$H$_6$ abundances in the case of dense molecular clouds, the agreement being notably less good with observations of TMC-1 \citep{Marcelino-2007, Markwick-2002} as shown by \cite{Hickson-2016b}. Alternatively, the abundances of C$_3$O and HCCCHO are in good agreement with the observations for dense clouds \citep{Herbst-1984, Irvine-1988, Ohishi-1998, Loison-2016}, but they are overproduced for IRAS 16293B. The overproduction of C$_3$O and HCCCHO for the case of IRAS 16293B may be due to under-represented consumption pathways of these species through ice-surface reactions and demonstrates that laboratory studies in this area are required. 

The duration of the prestellar phase has a significant impact on the COM abundances at the end of the simulation. 
A longer time spent in the prestellar phase enhances the impact of efficient reactions at cold temperature (10~K) on the final abundance. This effect is presumably due to the ongoing hydrogenation at cold temperature, and the addition of nearby radicals, with a non-efficient thermal diffusion on the ice surfaces at a such temperature. The agreement with the observations for C$_3$H$_6$ and C$_3$H$_8$ is particularly better with a longer prestellar phase in the simulation. The impact on the abundance of C$_2$H$_3$CHO and C$_2$H$_5$CHO is less important in comparison with the observations.

Concerning the other COMs, a longer prestellar phase leads to a better match of the simulated and observed abundances, in particular for (CH$_2$OH)$_2$. 
The absence of strong outflows from IRAS 16293 B has been suggested as due to this protostar being the less evolved source compared to its companion \citep{vanderWiel-2019}, which shows at least two major outflows \citep{Yeh-2008, Kristensen-2013, Girart-2014}. The abundance ratio of vinyl cyanide (C$_2$H$_3$CN) and ethyl cyanide (C$_2$H$_5$CN) observed by \cite{Calcutt-2018b} suggests that the warm-up timescale was shorter for IRAS 16293A, the more luminous binary component, or it has a higher accretion rate compared to IRAS 16293B. Another scenario could be that IRAS 16293A started to collapse earlier than IRAS 16293B. This delay of the beginning of the collapse of each protostar of a binary system could be the result of the fragmentation of the parent cloud when the first protostar started to form. This scenario is in agreement with the dynamical modelling study of \cite{Kuffmeier-2019}, in which the CO-rich bridge structure between the two protostars has been interpreted as a product of the multiple system dynamics and could be a remnant of the parent cloud fragmentation. 

The agreement with the observations of the abundance ratios of unsaturated species over saturated ones between the three C$_2$H$_x$CHO species is associated with an overall overproduction of these species during the prestellar phase. This overproduction suggests that the common precursors C$_3$, and then C$_3$O, are also overproduced.
The over-abundance of C$_3$ and C$_3$O could be a result of an underestimation of the consumption of these species by a lack of reactions in the chemical network.
The largest carbon-chain species taken into account in the chemical network are C$_5$H$_x$ in linear forms, and a few benzene-like molecules, such as C$_6$H$_5$CN, associated to their recent detection in the ISM reported in the study of \cite{McGuire-2018b}. The larger poly-aromatic hydrocarbons (PAH), which are expected to be abundant in space \citep{Puget-1989}, are not included in the calculation. This limitation in the size of carbon-chain species might be the origin of the over-production, or underconsumption, of the small carbon-chain species, which are the building blocks of these PAHs at cold temperatures as suggested by the recent studies of \cite{Joblin-2018} and \cite{Kaiser-2015}.

\section{Conclusion}

This study presents the observations and the chemical modelling of C$_3$-species around the Class 0 protostar IRAS 16293B. Using ALMA observations from the PILS survey, the molecular emission is analysed by comparing the spectrum to LTE models to determine the excitation temperatures and column densities of these species. Then, the abundances relative to CH$_3$OH are compared to the three-phase chemical model Nautilus to investigate the different formation pathways of such molecules under the typical physical conditions of the hot corino region. Different models were tested, each one focusing on a single aspect of the chemical network. The key findings are summarised in the following: 

\begin{enumerate}
\item We report the first detection of C$_2$H$_3$CHO and C$_3$H$_6$  towards the protostar IRAS 16293B. Their emission has an excitation temperature of  125 and 75~K, respectively, and their column densities are found to be $3.4\pm 0.7\times10^{14}$ and $4.2\pm 0.8\times10^{16}$~cm$^{-2}$, respectively.
These molecules were not detected towards the other member of the binary system. 
For C$_3$H$_6$, the E-transitions were found to exhibit a lower intensity with respect to the A-transitions, leading to a ratio of the E-transitions over A-transitions emitting molecules, being $0.6\pm 0.1$ in terms of column densities. The column density upper limits of several chemically related species, that are HCCCHO, C$_3$H$_7$OH, C$_3$O, HC(O)CHO, and C$_3$H$_8$, are also reported to provide constraints on the chemical simulations.
\item Most of the simulations reproduce the abundances observed towards IRAS 16293B within an order of magnitude.
The final abundances of the different simulations are sensitive to the duration of the prestellar phase, especially when the successive hydrogenation reactions play an important role in the formation of more saturated species. This seems to be the case for CH$_3$CCH, HC(O)CHO and CH$_3$OCHO which form C$_3$H$_6$, C$_3$H$_8$, CH$_2$(OH)CHO, (CH$_2$OH)$_2$ and CH$_3$OCH$_2$OH. The longer the cloud spends at low temperatures, the more abundant the saturated species, and the more depleted the precursors. 
\item On ice surfaces, the successive hydrogenation reactions of C$_3$O, forming HCCCHO, C$_2$H$_3$CHO, and C$_2$H$_5$CHO, and the radical-radical additions of HCO and C$_2$H$_3$ or C$_2$H$_5$ equally contribute to the amount of C$_3$-species observed in the hot corino of IRAS 16293B. 
\item A high gas-phase reaction rate of $1\times10^{-12}$~cm$^{3}$~s$^{-1}$ for the gas-phase reaction $\mathrm{C_3+O} \rightarrow \mathrm{C_2 + CO}$ is necessary to fit the final abundances of C$_3$-species detected in the observations. 
The very efficient production of C$_3$ and C$_3$O, along with the overall overproduction of C$_3$-species, suggests that consumption pathways of C$_3$ species are missing in the chemical network. These chemical routes could be related to the production and growth of PAHs at cold temperatures, which are not included in the present chemical model. 
\end{enumerate}

Most of the models reproduce the abundances of the C$_3$-species towards IRAS 16293B, which emphasises the contribution of the grain-surfaces production pathways under the hot corino physical conditions. 
Besides, the formation of these species on grain surfaces suggests that they can be incorporated into the protostellar disk as the protostar evolves into a Class I protostar and contribute to the formation of prebiotic molecules. Deeper observations of Class I protostellar disks and the detection of more complex species towards Class I objects are necessary to clearly state on the presence of these molecular species in such objects. Finally, the overproduction of unsaturated C$_3$-species, in general, suggests the lack of longer carbon-chain molecules formation pathways on grain surfaces and in gas phase and indicate the need of laboratory and theoretical studies of such reactions.

\begin{acknowledgement}
The authors thank the referee for useful comments that improved the quality of this manuscript. This paper makes use of the following ALMA data: ADS/JAO.ALMA{\#}2012.1.00712.S, ADS/JAO.ALMA{\#}2013.1.00278.S, and ADS/JAO.ALMA{\#}2016.1.01150.S. ALMA is a partnership of ESO (representing its member states), NSF (USA) and NINS (Japan), together with NRC (Canada), NSC and ASIAA (Taiwan), and KASI (Republic of Korea), in cooperation with the Republic of Chile. The Joint ALMA Observatory is operated by ESO, AUI/NRAO and NAOJ.
The group of JKJ acknowledges support from the H2020 European Research Council (ERC) (grant agreement No 646908) through ERC Consolidator Grant ``S4F''. Research at Centre for Star and Planet Formation is funded by the Danish National Research Foundation. 
AC acknowledges financial support from the Agence Nationale de la Recherche (grant ANR-19-ERC7-0001-01).
SFW. acknowledges financial support by the Swiss National Science Foundation (SNSF) Eccellenza Professorial Fellowship PCEFP2\_181150.
MND and BMK acknowledge the Swiss National Science Foundation (SNSF) Ambizione grant 180079.
MND also acknowledges the Center for Space and Habitability (CSH) Fellowship, and the IAU Gruber Foundation Fellowship.
This research has made use of NASA's Astrophysics Data System and VizierR catalogue access tool, CDS, Strasbourg, France \citep{Vizier}, as well as community-developed core Python packages for astronomy and scientific computing including 
Astropy \citep{Astropy}, 
Scipy \citep{Scipy}, 
Numpy \citep{Numpy} and 
Matplotlib \citep{Matplotlib}. 
\end{acknowledgement}

\bibliographystyle{aa} 
\bibliography{biblio} 

\begin{thebibliography}{134}
\expandafter\ifx\csname natexlab\endcsname\relax\def\natexlab#1{#1}\fi

\bibitem[{{Aikawa} {et~al.}(2008){Aikawa}, {Wakelam}, {Garrod}, \&
  {Herbst}}]{Aikawa-2008}
{Aikawa}, Y., {Wakelam}, V., {Garrod}, R.~T., \& {Herbst}, E. 2008, \apj, 674,
  984

\bibitem[{{Aikawa} {et~al.}(2012){Aikawa}, {Wakelam}, {Hersant}, {Garrod}, \&
  {Herbst}}]{Aikawa-2012}
{Aikawa}, Y., {Wakelam}, V., {Hersant}, F., {Garrod}, R.~T., \& {Herbst}, E.
  2012, \apj, 760, 40

\bibitem[{{{\'A}lvarez-Barcia} {et~al.}(2018){{\'A}lvarez-Barcia}, {Russ},
  {K{\"a}stner}, \& {Lamberts}}]{Alvarez-Barcia-2018}
{{\'A}lvarez-Barcia}, S., {Russ}, P., {K{\"a}stner}, J., \& {Lamberts}, T.
  2018, \mnras, 479, 2007

\bibitem[{{Andersson} {et~al.}(2011){Andersson}, {Goumans}, \&
  {Arnaldsson}}]{Andersson-2011}
{Andersson}, S., {Goumans}, T.~P.~M., \& {Arnaldsson}, A. 2011, Chemical
  Physics Letters, 513, 31

\bibitem[{{Andron} {et~al.}(2018){Andron}, {Gratier}, {Majumdar}, {Vidal},
  {Coutens}, {Loison}, \& {Wakelam}}]{Andron-2018}
{Andron}, I., {Gratier}, P., {Majumdar}, L., {et~al.} 2018, \mnras, 481, 5651

\bibitem[{{\'A}sgeirsson {et~al.}(2017){\'A}sgeirsson, J{\'o}nsson, \&
  Wikfeldt}]{Asgeirsson-2017}
{\'A}sgeirsson, V., J{\'o}nsson, H., \& Wikfeldt, K.~T. 2017, The Journal of
  Physical Chemistry C, 121, 1648

\bibitem[{Baggott {et~al.}(1987)Baggott, Frey, Lightfoot, \&
  Walsh}]{Baggott-1987}
Baggott, J.~E., Frey, H.~M., Lightfoot, P.~D., \& Walsh, R. 1987, 91, 3386

\bibitem[{Bermúdez {et~al.}(2013)Bermúdez, Peña, Cabezas, Daly, \&
  Alonso}]{Bermudez-2013}
Bermúdez, C., Peña, I., Cabezas, C., Daly, A.~M., \& Alonso, J.~L. 2013,
  ChemPhysChem, 14, 893

\bibitem[{{Bestmann} {et~al.}(1985{\natexlab{a}}){Bestmann}, {Dreizler},
  {Vacherand}, {Boucher}, {Eijck}, \& {Demaison}}]{Bestmann-1985b}
{Bestmann}, G., {Dreizler}, H., {Vacherand}, J.~M., {et~al.}
  1985{\natexlab{a}}, Zeitschrift Naturforschung Teil A, 40, 508

\bibitem[{{Bestmann} {et~al.}(1985{\natexlab{b}}){Bestmann}, {Lalowski}, \&
  {Dreizler}}]{Bestmann-1985a}
{Bestmann}, G., {Lalowski}, W., \& {Dreizler}, H. 1985{\natexlab{b}},
  Zeitschrift Naturforschung Teil A, 40, 271

\bibitem[{{Bizzocchi} {et~al.}(2008){Bizzocchi}, {Degli Esposti}, \&
  {Dore}}]{Bizzocchi-2008}
{Bizzocchi}, L., {Degli Esposti}, C., \& {Dore}, L. 2008, \aap, 492, 875

\bibitem[{{Blom} {et~al.}(1984){Blom}, {Grassi}, \& {Bauder}}]{Blom-1984}
{Blom}, C.~E., {Grassi}, G., \& {Bauder}, A. 1984, J. Am. Chem. Soc., 106, 7427

\bibitem[{Brown {et~al.}(1983)Brown, Eastwood, Elmes, \& Godfrey}]{Brown-1983}
Brown, R.~D., Eastwood, F.~W., Elmes, P.~S., \& Godfrey, P.~D. 1983, Journal of
  the American Chemical Society, 105, 6496

\bibitem[{{Brown} \& {Godfrey}(1984)}]{Brown-1984}
{Brown}, R.~D. \& {Godfrey}, P.~D. 1984, Australian Journal of Chemistry, 37,
  1951

\bibitem[{{Butscher} {et~al.}(2017){Butscher}, {Duvernay}, {Rimola},
  {Segado-Centellas}, \& {Chiavassa}}]{Butscher-2017}
{Butscher}, T., {Duvernay}, F., {Rimola}, A., {Segado-Centellas}, M., \&
  {Chiavassa}, T. 2017, Physical Chemistry Chemical Physics (Incorporating
  Faraday Transactions), 19, 2857

\bibitem[{{Calcutt} {et~al.}(2018{\natexlab{a}}){Calcutt}, {Fiechter},
  {Willis}, {M{\"u}ller}, {Garrod}, {J{\o}rgensen}, {Wampfler}, {Bourke},
  {Coutens}, {Drozdovskaya}, {Ligterink}, \& {Kristensen}}]{Calcutt-2018a}
{Calcutt}, H., {Fiechter}, M.~R., {Willis}, E.~R., {et~al.} 2018{\natexlab{a}},
  \aap, 617, A95

\bibitem[{{Calcutt} {et~al.}(2018{\natexlab{b}}){Calcutt}, {J{\o}rgensen},
  {M{\"u}ller}, {Kristensen}, {Coutens}, {Bourke}, {Garrod}, {Persson}, {van
  der Wiel}, {van Dishoeck}, \& {Wampfler}}]{Calcutt-2018b}
{Calcutt}, H., {J{\o}rgensen}, J.~K., {M{\"u}ller}, H.~S.~P., {et~al.}
  2018{\natexlab{b}}, \aap, 616, A90

\bibitem[{{Calcutt} {et~al.}(2019){Calcutt}, {Willis}, {J{\o}rgensen},
  {Bjerkeli}, {Ligterink}, {Coutens}, {M{\"u}ller}, {Garrod}, {Wampfler}, \&
  {Drozdovskaya}}]{Calcutt-2019}
{Calcutt}, H., {Willis}, E.~R., {J{\o}rgensen}, J.~K., {et~al.} 2019, \aap,
  631, A137

\bibitem[{{Caux} {et~al.}(2011){Caux}, {Kahane}, {Castets}, {Coutens},
  {Ceccarelli}, {Bacmann}, {Bisschop}, {Bottinelli}, {Comito}, {Helmich},
  {Lefloch}, {Parise}, {Schilke}, {Tielens}, {van Dishoeck}, {Vastel},
  {Wakelam}, \& {Walters}}]{Caux-2011}
{Caux}, E., {Kahane}, C., {Castets}, A., {et~al.} 2011, \aap, 532, A23

\bibitem[{{Cazaux} {et~al.}(2003){Cazaux}, {Tielens}, {Ceccarelli}, {Castets},
  {Wakelam}, {Caux}, {Parise}, \& {Teyssier}}]{Cazaux-2003}
{Cazaux}, S., {Tielens}, A.~G.~G.~M., {Ceccarelli}, C., {et~al.} 2003, \apjl,
  593, L51

\bibitem[{{Cherniak} \& {Costain}(1966)}]{Cherniak-1966}
{Cherniak}, E.~A. \& {Costain}, C.~C. 1966, \jcp, 45, 104

\bibitem[{{Christen} \& {M{\"u}ller}(2003)}]{Christen-2003}
{Christen}, D. \& {M{\"u}ller}, H.~S.~P. 2003, Physical Chemistry Chemical
  Physics (Incorporating Faraday Transactions), 5

\bibitem[{{Chuang} {et~al.}(2016){Chuang}, {Fedoseev}, {Ioppolo}, {van
  Dishoeck}, \& {Linnartz}}]{Chuang-2016}
{Chuang}, K.~J., {Fedoseev}, G., {Ioppolo}, S., {van Dishoeck}, E.~F., \&
  {Linnartz}, H. 2016, \mnras, 455, 1702

\bibitem[{{Colberg} \& {Friedrichs}(2006)}]{Colberg-2006}
{Colberg}, M. \& {Friedrichs}, G. 2006, Journal of Physical Chemistry A, 110,
  160

\bibitem[{{Costain} \& {Morton}(1959)}]{Costain-1959}
{Costain}, C.~C. \& {Morton}, J.~R. 1959, \jcp, 31, 389

\bibitem[{{Coutens} {et~al.}(2016){Coutens}, {J{\o}rgensen}, {van der Wiel},
  {M{\"}ller}, {Lykke}, {Bjerkeli}, {Bourke}, {Calcutt}, {Drozdovskaya},
  {Favre}, {Fayolle}, {Garrod}, {Jacobsen}, {Ligterink}, {{\"}Oberg},
  {Persson}, {van Dishoeck}, \& {Wampfler}}]{Coutens-2016}
{Coutens}, A., {J{\o}rgensen}, J.~K., {van der Wiel}, M. H.~D., {et~al.} 2016,
  \aap, 590, 1

\bibitem[{{Coutens} {et~al.}(2019){Coutens}, {Ligterink}, {Loison}, {Wakelam},
  {Calcutt}, {Drozdovskaya}, {J{\o}rgensen}, {M{\"u}ller}, {van Dishoeck}, \&
  {Wampfler}}]{Coutens-2019}
{Coutens}, A., {Ligterink}, N.~F.~W., {Loison}, J.~C., {et~al.} 2019, \aap,
  623, L13

\bibitem[{{Coutens} {et~al.}(2018{\natexlab{a}}){Coutens}, {Viti}, {Rawlings},
  {Beltr{\'a}n}, {Holdship}, {Jim{\'e}nez-Serra}, {Qu{\'e}nard}, \&
  {Rivilla}}]{Coutens-2018b}
{Coutens}, A., {Viti}, S., {Rawlings}, J.~M.~C., {et~al.} 2018{\natexlab{a}},
  \mnras, 475, 2016

\bibitem[{{Coutens} {et~al.}(2018{\natexlab{b}}){Coutens}, {Willis}, {Garrod},
  {M{\"u}ller}, {Bourke}, {Calcutt}, {Drozdovskaya}, {J{\o}rgensen},
  {Ligterink}, {Persson}, {St{\'e}phan}, {van der Wiel}, {van Dishoeck}, \&
  {Wampfler}}]{Coutens-2018}
{Coutens}, A., {Willis}, E.~R., {Garrod}, R.~T., {et~al.} 2018{\natexlab{b}},
  \aap, 612, A107

\bibitem[{{Craig} {et~al.}(2016){Craig}, {Groner}, {Conrad}, {Gurusinghe}, \&
  {Tubergen}}]{Craig-2016}
{Craig}, N.~C., {Groner}, P., {Conrad}, A.~R., {Gurusinghe}, R., \& {Tubergen},
  M.~J. 2016, Journal of Molecular Spectroscopy, 328, 1

\bibitem[{{Daly} {et~al.}(2015){Daly}, {Berm{\'u}dez}, {Kolesnikov{\'a}}, \&
  {Alonso}}]{Daly-2015}
{Daly}, A.~M., {Berm{\'u}dez}, C., {Kolesnikov{\'a}}, L., \& {Alonso}, J.~L.
  2015, \apjs, 218, 30

\bibitem[{{Drouin} {et~al.}(2006){Drouin}, {Pearson}, {Walters}, \&
  {Lattanzi}}]{Drouin-2006}
{Drouin}, B.~J., {Pearson}, J.~C., {Walters}, A., \& {Lattanzi}, V. 2006,
  Journal of Molecular Spectroscopy, 240, 227

\bibitem[{{Drozdovskaya} {et~al.}(2018){Drozdovskaya}, {van Dishoeck},
  {J{\o}rgensen}, {Calmonte}, {van der Wiel}, {Coutens}, {Calcutt},
  {M{\"u}ller}, {Bjerkeli}, {Persson}, {Wampfler}, \&
  {Altwegg}}]{Drozdovskaya-2018}
{Drozdovskaya}, M.~N., {van Dishoeck}, E.~F., {J{\o}rgensen}, J.~K., {et~al.}
  2018, \mnras, 476, 4949

\bibitem[{{Enrique-Romero} {et~al.}(2020){Enrique-Romero},
  {{\'A}lvarez-Barcia}, {Kolb}, {Rimola}, {Ceccarelli}, {Balucani}, {Meisner},
  {Ugliengo}, {Lamberts}, \& {K{\"a}stner}}]{EnriqueRomero-2020}
{Enrique-Romero}, J., {{\'A}lvarez-Barcia}, S., {Kolb}, F.~J., {et~al.} 2020,
  \mnras, 493, 2523

\bibitem[{{Enrique-Romero} {et~al.}(2016){Enrique-Romero}, {Rimola},
  {Ceccarelli}, \& {Balucani}}]{EnriqueRomero-2016}
{Enrique-Romero}, J., {Rimola}, A., {Ceccarelli}, C., \& {Balucani}, N. 2016,
  \mnras, 459, L6

\bibitem[{{Fayolle} {et~al.}(2017){Fayolle}, {{\"O}berg}, {J{\o}rgensen},
  {Altwegg}, {Calcutt}, {M{\"u}ller}, {Rubin}, {van der Wiel}, {Bjerkeli},
  {Bourke}, {Coutens}, {van Dishoeck}, {Drozdovskaya}, {Garrod}, {Ligterink},
  {Persson}, {Wampfler}, \& {Rosina Team}}]{Fayolle-2017}
{Fayolle}, E.~C., {{\"O}berg}, K.~I., {J{\o}rgensen}, J.~K., {et~al.} 2017,
  Nature Astronomy, 1, 703

\bibitem[{{Fedoseev} {et~al.}(2015){Fedoseev}, {Cuppen}, {Ioppolo}, {Lamberts},
  \& {Linnartz}}]{Fedoseev-2015}
{Fedoseev}, G., {Cuppen}, H.~M., {Ioppolo}, S., {Lamberts}, T., \& {Linnartz},
  H. 2015, \mnras, 448, 1288

\bibitem[{{Friberg} {et~al.}(1988){Friberg}, {Madden}, {Hjalmarson}, \&
  {Irvine}}]{Friberg-1988}
{Friberg}, P., {Madden}, S.~C., {Hjalmarson}, A., \& {Irvine}, W.~M. 1988,
  \aap, 195, 281

\bibitem[{{Gargaud} {et~al.}(2007){Gargaud}, {Amils}, {Cernicharo}, {Cleaves
  II}, {Irvine}, {Pinti}, \& {Viso}}]{astrobiology}
{Gargaud}, M., {Amils}, R., {Cernicharo}, J., {et~al.} 2007, Encyclopedia of
  astrobiology (Springer-Verlag Berlin Heidelberg)

\bibitem[{{Garrod}(2013)}]{Garrod-2013}
{Garrod}, R.~T. 2013, \apj, 778, 158

\bibitem[{{Girart} {et~al.}(2014){Girart}, {Estalella}, {Palau}, {Torrelles},
  \& {Rao}}]{Girart-2014}
{Girart}, J.~M., {Estalella}, R., {Palau}, A., {Torrelles}, J.~M., \& {Rao}, R.
  2014, \apjl, 780, L11

\bibitem[{{Goldsmith} {et~al.}(2012){Goldsmith}, {Magoon}, \&
  {Green}}]{Goldsmith-2012}
{Goldsmith}, C.~F., {Magoon}, G.~R., \& {Green}, W.~H. 2012, Journal of
  Physical Chemistry A, 116, 9033

\bibitem[{{Goumans} \& {K{\"a}stner}(2011)}]{Goumans-2011}
{Goumans}, T.~P.~M. \& {K{\"a}stner}, J. 2011, Journal of Physical Chemistry A,
  115, 10767

\bibitem[{{Grefenstette}(2017)}]{Grefenstette-2017}
{Grefenstette}, N.~M. 2017, PhD thesis, University College London

\bibitem[{{Herbst} {et~al.}(1984){Herbst}, {Smith}, \& {Adams}}]{Herbst-1984}
{Herbst}, E., {Smith}, D., \& {Adams}, N.~G. 1984, \aap, 138, L13

\bibitem[{{Herbst} \& {van Dishoeck}(2009)}]{Herbst-2009}
{Herbst}, E. \& {van Dishoeck}, E.~F. 2009, \araa, 47, 427

\bibitem[{{Hickson} {et~al.}(2015){Hickson}, {Loison}, {Bourgalais}, {Capron},
  {Le Picard}, {Goulay}, \& {Wakelam}}]{Hickson-2015}
{Hickson}, K.~M., {Loison}, J.-C., {Bourgalais}, J., {et~al.} 2015, \apj, 812,
  107

\bibitem[{{Hickson} {et~al.}(2016{\natexlab{a}}){Hickson}, {Loison},
  {Nu{\~n}ez-Reyes}, \& {M{\'e}reau}}]{Hickson-2016a}
{Hickson}, K.~M., {Loison}, J.-C., {Nu{\~n}ez-Reyes}, D., \& {M{\'e}reau}, R.
  2016{\natexlab{a}}, The Journal of Physical Chemistry Letters, 7, 3641

\bibitem[{{Hickson} {et~al.}(2016{\natexlab{b}}){Hickson}, {Wakelam}, \&
  {Loison}}]{Hickson-2016b}
{Hickson}, K.~M., {Wakelam}, V., \& {Loison}, J.-C. 2016{\natexlab{b}},
  Molecular Astrophysics, 3, 1

\bibitem[{{Hincelin} {et~al.}(2015){Hincelin}, {Chang}, \&
  {Herbst}}]{Hincelin-2015}
{Hincelin}, U., {Chang}, Q., \& {Herbst}, E. 2015, \aap, 574, A24

\bibitem[{{Hirota}(1966)}]{Hirota-1966}
{Hirota}, E. 1966, \jcp, 45, 1984

\bibitem[{{Hirota}(1979)}]{Hirota-1979}
{Hirota}, E. 1979, \jcp, 83, 1457

\bibitem[{{Hollis} {et~al.}(2004){Hollis}, {Jewell}, {Lovas}, {Remijan}, \&
  {M{\o}llendal}}]{Hollis-2004}
{Hollis}, J.~M., {Jewell}, P.~R., {Lovas}, F.~J., {Remijan}, A., \&
  {M{\o}llendal}, H. 2004, \apjl, 610, L21

\bibitem[{{H{\"u}bner} {et~al.}(1997){H{\"u}bner}, {Leeser}, {Burkert},
  {Ramsay}, \& {H{\"u}ttner}}]{Hubner-1997}
{H{\"u}bner}, H., {Leeser}, A., {Burkert}, A., {Ramsay}, D.~A., \&
  {H{\"u}ttner}, W. 1997, Journal of Molecular Spectroscopy, 184, 221

\bibitem[{{Hudson} \& {Moore}(1999)}]{Hudson-1999}
{Hudson}, R.~L. \& {Moore}, M.~H. 1999, \icarus, 140, 451

\bibitem[{{Hunter}(2007)}]{Matplotlib}
{Hunter}, J.~D. 2007, Comput. Sci. Eng., 9, 90

\bibitem[{{Irvine} {et~al.}(1988){Irvine}, {Brown}, {Cragg}, {Friberg},
  {Godfrey}, {Kaifu}, {Matthews}, {Ohishi}, {Suzuki}, \& {Takeo}}]{Irvine-1988}
{Irvine}, W.~M., {Brown}, R.~D., {Cragg}, D.~M., {et~al.} 1988, \apjl, 335, L89

\bibitem[{{Joblin} \& {Cernicharo}(2018)}]{Joblin-2018}
{Joblin}, C. \& {Cernicharo}, J. 2018, Science, 359, 156

\bibitem[{Jones {et~al.}(2001)Jones, Oliphant, Peterson, {et~al.}}]{Scipy}
Jones, E., Oliphant, T., Peterson, P., {et~al.} 2001, {SciPy}: Open source
  scientific tools for {Python}, [Online; accessed \today]

\bibitem[{{J{\o}rgensen} {et~al.}(2018){J{\o}rgensen}, {M{\"u}ller}, {Calcutt},
  {Coutens}, {Drozdovskaya}, {{\"O}berg}, {Persson}, {Taquet}, {van Dishoeck},
  \& {Wampfler}}]{Jorgensen-2018}
{J{\o}rgensen}, J.~K., {M{\"u}ller}, H.~S.~P., {Calcutt}, H., {et~al.} 2018,
  \aap, 620, A170

\bibitem[{{J{\o}rgensen} {et~al.}(2016){J{\o}rgensen}, {van der Wiel},
  {Coutens}, {Lykke}, {M{\"u}ller}, {van Dishoeck}, {Calcutt}, {Bjerkeli},
  {Bourke}, {Drozdovskaya}, {Favre}, {Fayolle}, {Garrod}, {Jacobsen},
  {{\"O}berg}, {Persson}, \& {Wampfler}}]{Jorgensen-2016}
{J{\o}rgensen}, J.~K., {van der Wiel}, M.~H.~D., {Coutens}, A., {et~al.} 2016,
  \aap, 595, A117

\bibitem[{Kahn \& Bruice(2005)}]{Kahn-2005}
Kahn, K. \& Bruice, T.~C. 2005, ChemPhysChem, 6, 487

\bibitem[{{Kaiser} {et~al.}(2015){Kaiser}, {Parker}, \& {Mebel}}]{Kaiser-2015}
{Kaiser}, R.~I., {Parker}, D. S.~N., \& {Mebel}, A.~M. 2015, Annual Review of
  Physical Chemistry, 66, 43

\bibitem[{{Kisiel} {et~al.}(2010){Kisiel}, {Dorosh}, {Maeda}, {Medvedev}, {De
  Lucia}, {Herbst}, {Drouin}, {Pearson}, \& {Shipman}}]{Kisiel-2010}
{Kisiel}, Z., {Dorosh}, O., {Maeda}, A., {et~al.} 2010, Physical Chemistry
  Chemical Physics (Incorporating Faraday Transactions), 12, 8329

\bibitem[{{Klebsch} {et~al.}(1985){Klebsch}, {Bester}, {Yamada}, {Winnewisser},
  \& {Joentgen}}]{Klebsch-1985}
{Klebsch}, W., {Bester}, M., {Yamada}, K.~M.~T., {Winnewisser}, G., \&
  {Joentgen}, W. 1985, \aap, 152, L12

\bibitem[{{Kobayashi} {et~al.}(2017){Kobayashi}, {Hidaka}, {Lamberts}, {Hama},
  {Kawakita}, {K{\"a}stner}, \& {Watanabe}}]{Kobayashi-2017}
{Kobayashi}, H., {Hidaka}, H., {Lamberts}, T., {et~al.} 2017, \apj, 837, 155

\bibitem[{{Krim} {et~al.}(2019){Krim}, {Guillemin}, \& {Woon}}]{Krim-2019}
{Krim}, L., {Guillemin}, J.-C., \& {Woon}, D.~E. 2019, \mnras, 485, 5210

\bibitem[{{Krim} {et~al.}(2018){Krim}, {Jonusas}, {Guillemin}, {Y{\'a}{\~n}ez},
  \& {Lamsabhi}}]{Krim-2018}
{Krim}, L., {Jonusas}, M., {Guillemin}, J.-C., {Y{\'a}{\~n}ez}, M., \&
  {Lamsabhi}, A.~M. 2018, Physical Chemistry Chemical Physics (Incorporating
  Faraday Transactions), 20, 19971

\bibitem[{{Kristensen} {et~al.}(2013){Kristensen}, {Klaassen}, {Mottram},
  {Schmalzl}, \& {Hogerheijde}}]{Kristensen-2013}
{Kristensen}, L.~E., {Klaassen}, P.~D., {Mottram}, J.~C., {Schmalzl}, M., \&
  {Hogerheijde}, M.~R. 2013, \aap, 549, L6

\bibitem[{{Kuffmeier} {et~al.}(2019){Kuffmeier}, {Calcutt}, \&
  {Kristensen}}]{Kuffmeier-2019}
{Kuffmeier}, M., {Calcutt}, H., \& {Kristensen}, L.~E. 2019, \aap, 628, A112

\bibitem[{{Lide}(1960)}]{Lide-1960}
{Lide}, David~R., J. 1960, \jcp, 33, 1514

\bibitem[{{Lide} \& {Mann}(1957)}]{Lide-1957}
{Lide}, David~R., J. \& {Mann}, D.~E. 1957, \jcp, 27, 868

\bibitem[{{Ligterink} {et~al.}(2017){Ligterink}, {Coutens}, {Kofman},
  {M{\"u}ller}, {Garrod}, {Calcutt}, {Wampfler}, {J{\o}rgensen}, {Linnartz}, \&
  {van Dishoeck}}]{Ligterink-2017}
{Ligterink}, N.~F.~W., {Coutens}, A., {Kofman}, V., {et~al.} 2017, \mnras, 469,
  2219

\bibitem[{{Loison} {et~al.}(2016){Loison}, {Ag{\'u}ndez}, {Marcelino},
  {Wakelam}, {Hickson}, {Cernicharo}, {Gerin}, {Roueff}, \&
  {Gu{\'e}lin}}]{Loison-2016}
{Loison}, J.-C., {Ag{\'u}ndez}, M., {Marcelino}, N., {et~al.} 2016, \mnras,
  456, 4101

\bibitem[{{Loison} {et~al.}(2017){Loison}, {Ag{\'u}ndez}, {Wakelam}, {Roueff},
  {Gratier}, {Marcelino}, {Reyes}, {Cernicharo}, \& {Gerin}}]{Loison-2017}
{Loison}, J.-C., {Ag{\'u}ndez}, M., {Wakelam}, V., {et~al.} 2017, \mnras, 470,
  4075

\bibitem[{{Loison} {et~al.}(2014){Loison}, {Wakelam}, {Hickson}, {Bergeat}, \&
  {Mereau}}]{Loison-2014}
{Loison}, J.-C., {Wakelam}, V., {Hickson}, K.~M., {Bergeat}, A., \& {Mereau},
  R. 2014, \mnras, 437, 930

\bibitem[{{Lykke} {et~al.}(2017){Lykke}, {Coutens}, {J{\o}rgensen}, {van der
  Wiel}, {Garrod}, {M{\"u}ller}, {Bjerkeli}, {Bourke}, {Calcutt},
  {Drozdovskaya}, {Favre}, {Fayolle}, {Jacobsen}, {{\"O}berg}, {Persson}, {van
  Dishoeck}, \& {Wampfler}}]{Lykke-2017}
{Lykke}, J.~M., {Coutens}, A., {J{\o}rgensen}, J.~K., {et~al.} 2017, \aap, 597,
  A53

\bibitem[{{Maeda} {et~al.}(2006{\natexlab{a}}){Maeda}, {De Lucia}, {Herbst},
  {Pearson}, {Riccobono}, {Trosell}, \& {Bohn}}]{Maeda-2006-n}
{Maeda}, A., {De Lucia}, F.~C., {Herbst}, E., {et~al.} 2006{\natexlab{a}},
  \apjs, 162, 428

\bibitem[{{Maeda} {et~al.}(2006{\natexlab{b}}){Maeda}, {Medvedev}, {De Lucia},
  \& {Herbst}}]{Maeda-2006-i}
{Maeda}, A., {Medvedev}, I.~R., {De Lucia}, F.~C., \& {Herbst}, E.
  2006{\natexlab{b}}, \apjs, 166, 650

\bibitem[{{Manigand} {et~al.}(2019){Manigand}, {Calcutt}, {J{\o}rgensen},
  {Taquet}, {M{\"u}ller}, {Coutens}, {Wampfler}, {Ligterink}, {Drozdovskaya},
  {Kristensen}, {van der Wiel}, \& {Bourke}}]{Manigand-2019}
{Manigand}, S., {Calcutt}, H., {J{\o}rgensen}, J.~K., {et~al.} 2019, \aap, 623,
  A69

\bibitem[{{Manigand} {et~al.}(2020){Manigand}, {J{\o}rgensen}, {Calcutt},
  {M{\"u}ller}, {Ligterink}, {Coutens}, {Drozdovskaya}, {van Dishoeck}, \&
  {Wampfler}}]{Manigand-2020}
{Manigand}, S., {J{\o}rgensen}, J.~K., {Calcutt}, H., {et~al.} 2020, \aap, 635,
  A48

\bibitem[{{Marcelino} {et~al.}(2007){Marcelino}, {Cernicharo}, {Ag{\'u}ndez},
  {Roueff}, {Gerin}, {Mart{\'\i}n-Pintado}, {Mauersberger}, \&
  {Thum}}]{Marcelino-2007}
{Marcelino}, N., {Cernicharo}, J., {Ag{\'u}ndez}, M., {et~al.} 2007, \apjl,
  665, L127

\bibitem[{{Markwick} {et~al.}(2002){Markwick}, {Millar}, \&
  {Charnley}}]{Markwick-2002}
{Markwick}, A.~J., {Millar}, T.~J., \& {Charnley}, S.~B. 2002, \aap, 381, 560

\bibitem[{{Masunaga} \& {Inutsuka}(2000)}]{Masunaga-2000}
{Masunaga}, H. \& {Inutsuka}, S.-i. 2000, \apj, 536, 406

\bibitem[{{McGuire} {et~al.}(2018){McGuire}, {Burkhardt}, {Kalenskii},
  {Shingledecker}, {Remijan}, {Herbst}, \& {McCarthy}}]{McGuire-2018b}
{McGuire}, B.~A., {Burkhardt}, A.~M., {Kalenskii}, S., {et~al.} 2018, Science,
  359, 202

\bibitem[{{McKellar} {et~al.}(2008){McKellar}, {Watson}, {Chu}, \&
  {Lee}}]{McKellar-2008}
{McKellar}, A.~R.~W., {Watson}, J.~K.~G., {Chu}, L.-K., \& {Lee}, Y.-P. 2008,
  Journal of Molecular Spectroscopy, 252, 230

\bibitem[{{Menten} {et~al.}(1988){Menten}, {Walmsley}, {Henkel}, \&
  {Wilson}}]{Menten-1988}
{Menten}, K.~M., {Walmsley}, C.~M., {Henkel}, C., \& {Wilson}, T.~L. 1988,
  \aap, 198, 253

\bibitem[{{Moldoveanu}(2010)}]{Moldoveanu-2010}
{Moldoveanu}, S. 2010, Pyrolysis of Organic Molecules: Applications to Health
  and Environmental Issues, Vol.~28 (Amsterdam: Elsevier)

\bibitem[{{M{\"u}ller} {et~al.}(2005){M{\"u}ller}, {Schl{\"o}der}, {Stutzki},
  \& {Winnewisser}}]{cdms-2}
{M{\"u}ller}, H.~S.~P., {Schl{\"o}der}, F., {Stutzki}, J., \& {Winnewisser}, G.
  2005, J. Mol. Struct., 742, 215

\bibitem[{{M{\"u}ller} {et~al.}(2001){M{\"u}ller}, {Thorwirth}, {Roth}, \&
  {Winnewisser}}]{cdms-1}
{M{\"u}ller}, H.~S.~P., {Thorwirth}, S., {Roth}, D.~A., \& {Winnewisser}, G.
  2001, \aap, 370, L49

\bibitem[{{Neish} {et~al.}(2010){Neish}, {Somogyi}, \& {Smith}}]{Neish-2010}
{Neish}, C.~D., {Somogyi}, {\'A}., \& {Smith}, M.~A. 2010, Astrobiology, 10,
  337

\bibitem[{{Nelsestuen}(1980)}]{Nelsestuen-1980}
{Nelsestuen}, G.~L. 1980, Journal of Molecular Evolution, 15, 59

\bibitem[{{Ochsenbein} {et~al.}(2000){Ochsenbein}, {Bauer}, \&
  {Marcout}}]{Vizier}
{Ochsenbein}, F., {Bauer}, P., \& {Marcout}, J. 2000, \aaps, 143, 23

\bibitem[{{Ohishi} \& {Kaifu}(1998)}]{Ohishi-1998}
{Ohishi}, M. \& {Kaifu}, N. 1998, Faraday Discussions, 109, 205

\bibitem[{{Pearson} {et~al.}(1994){Pearson}, {Sastry}, {Herbst}, \&
  {Delucia}}]{Pearson-1994}
{Pearson}, J.~C., {Sastry}, K.~V.~L.~N., {Herbst}, E., \& {Delucia}, F.~C.
  1994, Journal of Molecular Spectroscopy, 166, 120

\bibitem[{{Persson} {et~al.}(2018){Persson}, {J{\o}rgensen}, {M{\"u}ller},
  {Coutens}, {van Dishoeck}, {Taquet}, {Calcutt}, {van der Wiel}, {Bourke}, \&
  {Wampfler}}]{Persson-2018}
{Persson}, M.~V., {J{\o}rgensen}, J.~K., {M{\"u}ller}, H.~S.~P., {et~al.} 2018,
  \aap, 610, A54

\bibitem[{{Pickett} {et~al.}(1998){Pickett}, {Poynter}, {Cohen}, {Delitsky},
  {Pearson}, \& {M{\"u}ller}}]{jpl_0}
{Pickett}, H.~M., {Poynter}, R.~L., {Cohen}, E.~A., {et~al.} 1998, \jqsrt, 60,
  883

\bibitem[{{Puget} \& {Leger}(1989)}]{Puget-1989}
{Puget}, J.~L. \& {Leger}, A. 1989, \araa, 27, 161

\bibitem[{{Qasim} {et~al.}(2019){Qasim}, {Fedoseev}, {Chuang}, {Taquet},
  {Lamberts}, {He}, {Ioppolo}, {van Dishoeck}, \& {Linnartz}}]{Qasim-2019}
{Qasim}, D., {Fedoseev}, G., {Chuang}, K.~J., {et~al.} 2019, \aap, 627, A1

\bibitem[{{Requena-Torres} {et~al.}(2008){Requena-Torres},
  {Mart{\'\i}n-Pintado}, {Mart{\'\i}n}, \& {Morris}}]{Requena-Torres-2008}
{Requena-Torres}, M.~A., {Mart{\'\i}n-Pintado}, J., {Mart{\'\i}n}, S., \&
  {Morris}, M.~R. 2008, \apj, 672, 352

\bibitem[{{Robitaille} {et~al.}(2013){Robitaille}, {Tollerud}, {Greenfield},
  {Droettboom}, {Bray}, {Aldcroft}, {Davis}, {Ginsburg}, {Price-Whelan},
  {Kerzendorf}, {Conley}, {Crighton}, {Barbary}, {Muna}, {Ferguson},
  {Grollier}, {Parikh}, {Nair}, {Unther}, {Deil}, {Woillez}, {Conseil},
  {Kramer}, {Turner}, {Singer}, {Fox}, {Weaver}, {Zabalza}, {Edwards}, {Azalee
  Bostroem}, {Burke}, {Casey}, {Crawford}, {Dencheva}, {Ely}, {Jenness},
  {Labrie}, {Lim}, {Pierfederici}, {Pontzen}, {Ptak}, {Refsdal}, {Servillat},
  \& {Streicher}}]{Astropy}
{Robitaille}, T.~P., {Tollerud}, E.~J., {Greenfield}, P., {et~al.} 2013, \aap,
  558, A33

\bibitem[{{Ruaud} {et~al.}(2015){Ruaud}, {Loison}, {Hickson}, {Gratier},
  {Hersant}, \& {Wakelam}}]{Ruaud-2015}
{Ruaud}, M., {Loison}, J.~C., {Hickson}, K.~M., {et~al.} 2015, \mnras, 447,
  4004

\bibitem[{{Ruaud} {et~al.}(2016){Ruaud}, {Wakelam}, \& {Hersant}}]{Ruaud-2016}
{Ruaud}, M., {Wakelam}, V., \& {Hersant}, F. 2016, \mnras, 459, 3756

\bibitem[{{Sakai} {et~al.}(2008){Sakai}, {Sakai}, {Hirota}, \&
  {Yamamoto}}]{Sakai-2008}
{Sakai}, N., {Sakai}, T., {Hirota}, T., \& {Yamamoto}, S. 2008, \apj, 672, 371

\bibitem[{{Scuseria} \& {Staroverov}(2005)}]{Scuseria-2005}
{Scuseria}, G.~E. \& {Staroverov}, V.~N. 2005, in Theory and Applications of
  Computational Chemistry, ed. C.~E. Dykstra, G.~Frenking, K.~S. Kim, \& G.~E.
  Scuseria (Amsterdam: Elsevier), 669 -- 724

\bibitem[{{Shibasaki} {et~al.}(2008){Shibasaki}, {Kanai}, \&
  {Mita}}]{Shibasaki-2008}
{Shibasaki}, M., {Kanai}, M., \& {Mita}, T. 2008, The Catalytic Asymmetric
  Strecker Reaction (American Cancer Society), 1--119

\bibitem[{{Simons} {et~al.}(2020){Simons}, {Lamberts}, \&
  {Cuppen}}]{Simons-2020}
{Simons}, M.~A.~J., {Lamberts}, T., \& {Cuppen}, H.~M. 2020, \aap, 634, A52

\bibitem[{{Song} \& {K{\"a}stner}(2017)}]{Song-2017}
{Song}, L. \& {K{\"a}stner}, J. 2017, \apj, 850, 118

\bibitem[{{Strecker}(1850)}]{Strecker-1850}
{Strecker}, A. 1850, Justus Liebigs Annalen der Chemie, 75, 27

\bibitem[{{Strecker}(1854)}]{Strecker-1854}
{Strecker}, A. 1854, Justus Liebigs Annalen der Chemie, 91, 349

\bibitem[{{Tang} {et~al.}(1985){Tang}, {Inokuchi}, {Saito}, {Yamada}, \&
  {Hirota}}]{Tang-1985}
{Tang}, T.~B., {Inokuchi}, H., {Saito}, S., {Yamada}, C., \& {Hirota}, E. 1985,
  Chemical Physics Letters, 116, 83

\bibitem[{{Taquet} {et~al.}(2018){Taquet}, {van Dishoeck}, {Swayne}, {Harsono},
  {J{\o}rgensen}, {Maud}, {Ligterink}, {M{\"u}ller}, {Codella}, {Altwegg},
  {Bieler}, {Coutens}, {Drozdovskaya}, {Furuya}, {Persson}, {van't Hoff},
  {Walsh}, \& {Wampfler}}]{Taquet-2018}
{Taquet}, V., {van Dishoeck}, E.~F., {Swayne}, M., {et~al.} 2018, \aap, 618,
  A11

\bibitem[{{Turner}(1991)}]{Turner-1991}
{Turner}, B.~E. 1991, \apjs, 76, 617

\bibitem[{{Ulenikov} {et~al.}(1991){Ulenikov}, {Malikova}, {Qagar}, {Musaev},
  {Adilov}, \& {Mehtiev}}]{Ulenikov-1991}
{Ulenikov}, O.~N., {Malikova}, A.~B., {Qagar}, C.~O., {et~al.} 1991, Journal of
  Molecular Spectroscopy, 145, 262

\bibitem[{{van der Walt} {et~al.}(2011){van der Walt}, {Colbert}, \&
  {Varoquaux}}]{Numpy}
{van der Walt}, S., {Colbert}, S.~C., \& {Varoquaux}, G. 2011, Computing in
  Science Engineering, 13, 22

\bibitem[{{van der Wiel} {et~al.}(2019){van der Wiel}, {Jacobsen},
  {J{\o}rgensen}, {Bourke}, {Kristensen}, {Bjerkeli}, {Murillo}, {Calcutt},
  {M{\"u}ller}, {Coutens}, {Drozdovskaya}, {Favre}, \&
  {Wampfler}}]{vanderWiel-2019}
{van der Wiel}, M.~H.~D., {Jacobsen}, S.~K., {J{\o}rgensen}, J.~K., {et~al.}
  2019, \aap, 626, A93

\bibitem[{{van Dishoeck} {et~al.}(1995){van Dishoeck}, {Blake}, {Jansen}, \&
  {Groesbeck}}]{VanDishoeck-1995}
{van Dishoeck}, E.~F., {Blake}, G.~A., {Jansen}, D.~J., \& {Groesbeck}, T.~D.
  1995, \apj, 447, 760

\bibitem[{{van Trump} \& {Miller}(1972)}]{vanTrump-1972}
{van Trump}, J.~E. \& {Miller}, S.~L. 1972, Science, 178, 859

\bibitem[{{Vidal} {et~al.}(2017){Vidal}, {Loison}, {Jaziri}, {Ruaud},
  {Gratier}, \& {Wakelam}}]{Vidal-2017}
{Vidal}, T. H.~G., {Loison}, J.-C., {Jaziri}, A.~Y., {et~al.} 2017, \mnras,
  469, 435

\bibitem[{{Wakelam} {et~al.}(2012){Wakelam}, {Herbst}, {Loison}, {Smith},
  {Chandrasekaran}, {Pavone}, {Adams}, {Bacchus-Montabonel}, {Bergeat},
  {B{\'e}roff}, {Bierbaum}, {Chabot}, {Dalgarno}, {van Dishoeck}, {Faure},
  {Geppert}, {Gerlich}, {Galli}, {H{\'e}brard}, {Hersant}, {Hickson},
  {Honvault}, {Klippenstein}, {Le Picard}, {Nyman}, {Pernot}, {Schlemmer},
  {Selsis}, {Sims}, {Talbi}, {Tennyson}, {Troe}, {Wester}, \&
  {Wiesenfeld}}]{Wakelam-2012}
{Wakelam}, V., {Herbst}, E., {Loison}, J.~C., {et~al.} 2012, \apjs, 199, 21

\bibitem[{{Wakelam} {et~al.}(2015){Wakelam}, {Loison}, {Herbst}, {Pavone},
  {Bergeat}, {B{\'e}roff}, {Chabot}, {Faure}, {Galli}, \&
  {Geppert}}]{Wakelam-2015}
{Wakelam}, V., {Loison}, J.~C., {Herbst}, E., {et~al.} 2015, \apjs, 217, 20

\bibitem[{{Wakelam} {et~al.}(2017){Wakelam}, {Loison}, {Mereau}, \&
  {Ruaud}}]{Wakelam-2017}
{Wakelam}, V., {Loison}, J.~C., {Mereau}, R., \& {Ruaud}, M. 2017, Molecular
  Astrophysics, 6, 22

\bibitem[{{Wakelam} {et~al.}(2014){Wakelam}, {Vastel}, {Aikawa}, {Coutens},
  {Bottinelli}, \& {Caux}}]{Wakelam-2014}
{Wakelam}, V., {Vastel}, C., {Aikawa}, Y., {et~al.} 2014, \mnras, 445, 2854

\bibitem[{{White}(1975)}]{White-1975}
{White}, W.~F. 1975, {Microwave spectra of some volatile organic compounds},
  Tech. rep.

\bibitem[{{Winnewisser}(1973)}]{Winnewisser-1973}
{Winnewisser}, G. 1973, Journal of Molecular Spectroscopy, 46, 16

\bibitem[{{Winnewisser} {et~al.}(1975){Winnewisser}, {Winnewisser}, {Honda}, \&
  {Hirota}}]{Winnewisser-1975}
{Winnewisser}, M., {Winnewisser}, G., {Honda}, T., \& {Hirota}, E. 1975,
  Zeitschrift Naturforschung Teil A, 30, 1001

\bibitem[{{Wirstr{\"o}m} {et~al.}(2011){Wirstr{\"o}m}, {Geppert}, {Hjalmarson},
  {Persson}, {Black}, {Bergman}, {Millar}, {Hamberg}, \&
  {Vigren}}]{Wirstrom-2011}
{Wirstr{\"o}m}, E.~S., {Geppert}, W.~D., {Hjalmarson}, {\r{A}}., {et~al.} 2011,
  \aap, 533, A24

\bibitem[{{Wlodarczak} {et~al.}(1994){Wlodarczak}, {Demaison}, {Heineking}, \&
  {Csaszar}}]{Wlodarczak-1994}
{Wlodarczak}, G., {Demaison}, J., {Heineking}, N., \& {Csaszar}, A.~G. 1994,
  Journal of Molecular Spectroscopy, 167, 239

\bibitem[{{Woon} \& {Herbst}(1996)}]{Woon-1996}
{Woon}, D.~E. \& {Herbst}, E. 1996, \apj, 465, 795

\bibitem[{{Yeh} {et~al.}(2008){Yeh}, {Hirano}, {Bourke}, {Ho}, {Lee}, {Ohashi},
  \& {Takakuwa}}]{Yeh-2008}
{Yeh}, S. C.~C., {Hirano}, N., {Bourke}, T.~L., {et~al.} 2008, \apj, 675, 454

\bibitem[{{Zaverkin} {et~al.}(2018){Zaverkin}, {Lamberts}, {Markmeyer}, \&
  {K{\"a}stner}}]{Zaverkin-2018}
{Zaverkin}, V., {Lamberts}, T., {Markmeyer}, M.~N., \& {K{\"a}stner}, J. 2018,
  \aap, 617, A25

\bibitem[{{Zhao} \& {Truhlar}(2008)}]{Zhao-2008}
{Zhao}, Y. \& {Truhlar}, D.~G. 2008, Theoretical Chemistry Accounts, 120, 215

\bibitem[{{Zhou} {et~al.}(2008){Zhou}, {Kaiser}, {Gao}, {Chang}, {Liang}, \&
  {Yung}}]{Zhou-2008}
{Zhou}, L., {Kaiser}, R.~I., {Gao}, L.~G., {et~al.} 2008, \apj, 686, 1493

\bibitem[{{Zucker} {et~al.}(2019){Zucker}, {Speagle}, {Schlafly}, {Green},
  {Finkbeiner}, {Goodman}, \& {Alves}}]{Zucker-2019}
{Zucker}, C., {Speagle}, J.~S., {Schlafly}, E.~F., {et~al.} 2019, \apj, 879,
  125

\end{thebibliography}

\appendix

\section{\label{app-sec-phys}Physical model evolution}

\begin{figure}[t]
\includegraphics[width=0.49\textwidth]{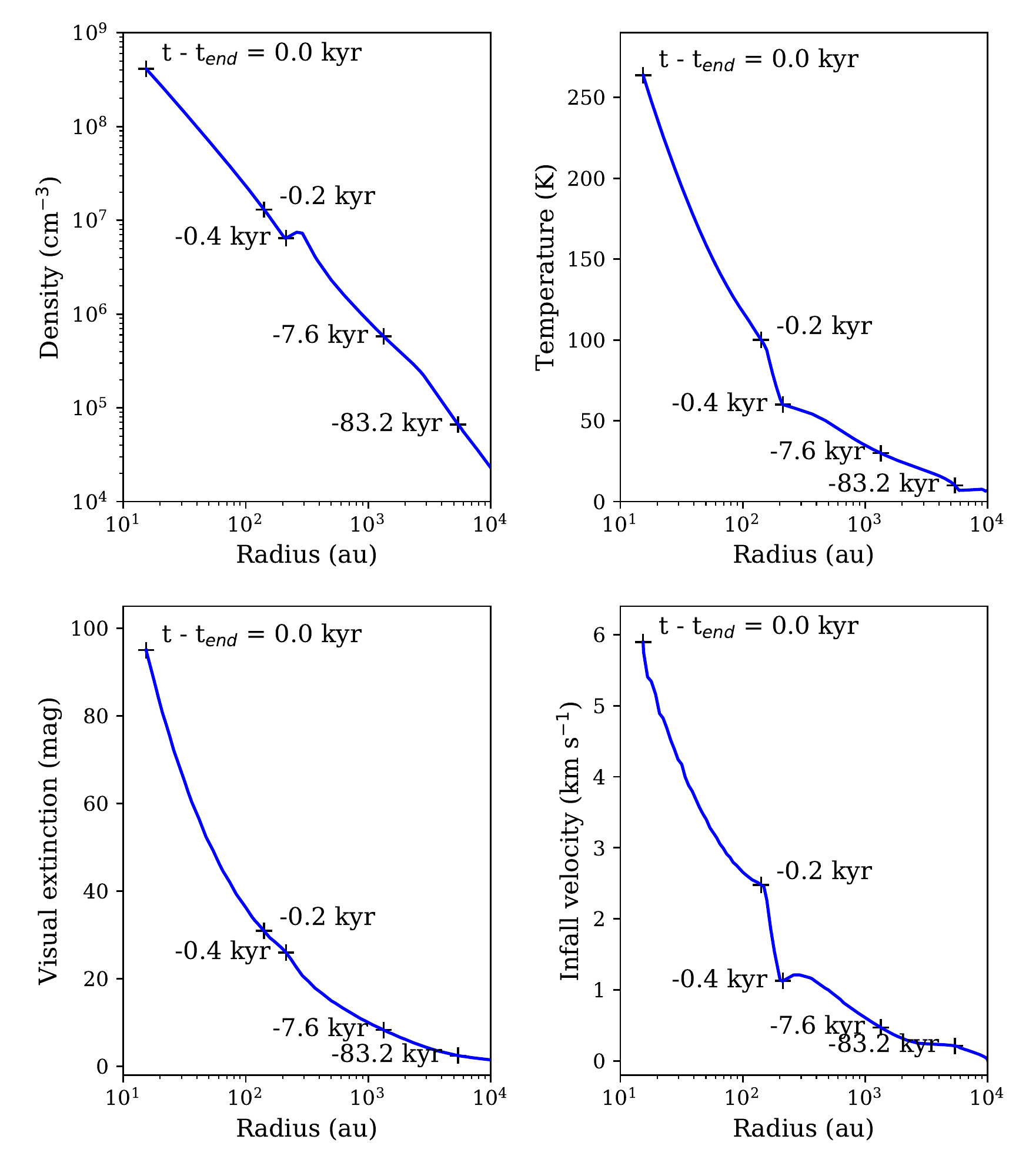}
\caption{\label{fig-phys}\small Evolution of the density, dust temperature, visual extinction and infall velocity across the envelope during the simulation. The time is indicated at several positions on the curves, with $t=0$~yr set to the end of the collapse phase in the simulation.}
\end{figure}

Figure \ref{fig-phys} shows the evolution of the density, the temperature, the extinction coefficient and the infall velocity of the gas during the collapse phase. 

\section{Chemical network}

\subsection{\label{app-sec-Network}Upgrades from the NAUTILUS public version}

The chemical network used here is limited to a skeleton up to five atoms of carbons for linear chains, including oxygen nitrogen and sulfur-bearing, neutral and ionic compounds. Besides, benzene, benzonitrile (C$_6$H$_5$CN) and ethylbenzene (C$_6$H$_5$C$_2$H) are also included with their ions. Special care has been taken with end-of-chain processing, which is an important source of error for finite models that want to account for infinite chemistry. In principle, 
there is no limit for the formation of very large molecules, that are then associated to a very large dilution and thus an abundance of large molecules in most cases very small by bottom-up chemistry. In other words, the limitations are imposed by the extent of the chemical network and the formation of molecules up to five carbons in their structure ensures a proper representation of the production of smaller molecules, such as the C$_3$-species.
Even though there are some notable changes in the gas-phase chemistry (new reactions, update of rate constants and branching ratios according to new measurements and theoretical calculations), the most important changes were made for the grain-surfaces chemistry. The chemistry will be detailed in future articles, thus we only present major changes that concern the species discussed in this work. 

We have started to homogenise the chemistry on grains completing the possible reactions for the most abundant species on grains. First, we systematically supplemented the hydrogenation reactions through s-H and s-H$_2$ reactions, which are critical because they are favoured by the completion between diffusion and reaction through tunnelling. To estimate the probabilities of the tunnelling effect, we systematically consider a barrier width of 1 Angstrom and we calculate the height of the reaction barriers in the gas phase using the density functional theory \citep[DFT,][]{Scuseria-2005} with the functional M06-2X and the AVTZ basis \citep{Zhao-2008}; we have calculated 240 barrier heights for s-H and s-H$_2$ reactions. 
The reactions with s-H$_2$ lead to quick consumption of s-CN, s-OH, s-NH$_2$, s-C, s-CH, s-CH$_2$, s-C$_2$, s-C$_2$H, s-C$_2$H$_3$ as long as there is a lot of s-H$_2$ on grains. Reactions of these species with s-H$_2$ are exothermic and are enhanced by tunnelling effect when an activation barrier exists.
For some low exothermic reactions, such as s-CH$_2$ + s-H$_2$, s-C$_2$H$_3$ + s-H$_2$, and s-NH$_2$ + s-H$_2$, a barrier of 1 Angstrom undoubtedly overestimates the tunnelling effect. However, considering the very high abundance of s-H$_2$, this should not be critical in the end, at least as long as there is a lot of s-H$_2$ on grains, although more accurate tunnelling treatment would be desirable in the future. We have also modified the Nautilus code to better take into account the tunnelling effect for reactions involving hydrogen atom transfer between heavy species. For example, the $\mathrm{O + C_3H_8 \rightarrow OH + C_3H_7 \rightarrow C_3H_7OH\ /\ C_3H_6 + H_2O }$ reaction is now an efficient source of propanol production through tunnelling of the first step followed by reaction between s-OH and s-C$_3$H$_7$ radicals which are supposed to stay close enough on the grains to react. We also consider diffusion by tunneling for all species, with a diffusion barrier width of 2.5 Angstrom according to \cite{Asgeirsson-2017} and a diffusion barrier height equal to 0.6 of the binding energy. The width of the diffusion barrier is not critical as long as it is greater than 2 Angstrom but limits the reactivity through tunnelling for reactions with a high barrier ($>$ 5000 K).

\begin{table}[t]
\caption{\small \label{tab-BR}Branching ratios of radical addition reactions.}
\begin{tabular}{llcc}
\hline\hline
Reacting species & \textcolor{white}{$\rightarrow\ $} Produced species & BR & E$_\text{a}$ \\
 & & (\%) & (K) \\
\hline
s-CH$_3$\tablefootmark{a} + s-CH$_3$ & $\rightarrow\ $  s-C$_2$H$_6$ & 100 & 0\vspace{5pt}\\
s-CH$_3$ + s-C$_2$H$_3$ & $\rightarrow\ $  s-C$_3$H$_6$ & 100 & 0 \\
 & $\rightarrow\ $  s-CH$_4$ + s-C$_2$H$_2$ & 0 & 0\vspace{5pt}\\
s-CH$_3$ + s-C$_2$H$_5$ & $\rightarrow\ $  s-C$_3$H$_8$ & 100 & 0 \\
 & $\rightarrow\ $  s-CH$_4$ + s-C$_2$H$_4$ & 0 & 0\vspace{5pt}\\
s-CH$_3$ + s-HCO & $\rightarrow\ $  s-CH$_3$CHO & 30 & 0 \\
 & $\rightarrow\ $  s-CH$_4$ + s-CO & 70 & 0\vspace{5pt}\\
s-CH$_3$ + s-CH$_3$O & $\rightarrow\ $   s-CH$_3$OCH$_3$ & 50 & 0 \\
 & $\rightarrow\ $  s-CH$_4$ + s-H$_2$CO & 50 & 0\vspace{5pt} \\
s-CH$_3$ + s-CH$_2$OH & $\rightarrow\ $  s-C$_2$H$_5$OH & 30 & 0 \\
 & $\rightarrow\ $  s-CH$_4$ + s-H$_2$CO & 70 & 0\vspace{5pt} \\
s-HCO + s-C$_2$H$_3$ & $\rightarrow\ $  s-C$_2$H$_3$CHO & 10 & 0 \\
 & $\rightarrow\ $  s-C$_2$H$_4$ + s-CO & 90 & 0 \\
 & $\rightarrow\ $  s-C$_2$H$_2$ + s-H$_2$CO & 0 & 0\vspace{5pt} \\
s-HCO + s-C$_2$H$_5$ & $\rightarrow\ $  s-C$_2$H$_5$CHO & 10 & 0 \\
 & $\rightarrow\ $  s-C$_2$H$_6$ + s-CO & 90 & 0 \\
 & $\rightarrow\ $  s-C$_2$H$_4$ + s-H$_2$CO & 0 & 0\vspace{5pt} \\
s-HCO + s-HCO & $\rightarrow\ $  s-HC(O)CHO & 5 & 0 \\
 & $\rightarrow\ $  s-CO + s-H$_2$CO & 95 & 0\vspace{5pt} \\
s-HCO + s-CH$_3$O & $\rightarrow\ $  s-CH$_3$OCHO & 50 & 0 \\
 & $\rightarrow\ $  s-CO + s-CH$_3$OH & 50 & 0 \\
 & $\rightarrow\ $  s-H$_2$CO + s-H$_2$CO & 0 & 0\vspace{5pt} \\
s-HCO + s-H$_2$CO & $\rightarrow\ $  s-CH$_2$(OH)CHO & 5 & 0 \\
 & $\rightarrow\ $  s-CO + s-CH$_3$OH & 95 & 0 \\
 & $\rightarrow\ $  s-H$_2$CO + s-H$_2$CO & 0 & 0\vspace{5pt} \\
s-CH$_3$O + s-CH$_3$O & $\rightarrow\ $  s-CH$_3$OOCH$_3$ & 50 & 0 \\
 & $\rightarrow\ $  \small s-CH$_3$OH + s-H$_2$CO & 50 & 0\vspace{5pt} \\
\small s-CH$_3$O + s-CH$_2$OH & $\rightarrow\ $  s-CH$_3$OCH$_2$OH & 50 & 0 \\
 & $\rightarrow\ $  \small s-CH$_3$OH + s-H$_2$CO & 50 & 0\vspace{5pt} \\
\small s-CH$_2$OH + s-CH$_2$OH & $\rightarrow\ $  s-(CH$_2$OH)$_2$ & 50 & 0 \\
 & $\rightarrow\ $  \small s-CH$_3$OH + s-H$_2$CO & 50 & 0\vspace{5pt} \\
s-O + s-C$_2$H$_6$ & $\rightarrow\ $  s-C$_2$H$_5$OH & 50 & 2400 \\
 & $\rightarrow\ $  s-H$_2$O + s-C$_2$H$_4$ & 50 & 2400\vspace{5pt} \\
s-O + s-C$_3$H$_8$ & $\rightarrow\ $  s-C$_3$H$_7$OH\tablefootmark{b} & 20 & 1000 \\
 & $\rightarrow\ $  s-H$_2$O + s-C$_3$H$_6$ & 80 & 1000\vspace{5pt}  \\
 s-CH$_3$O + s-C$_2$H$_5$ & $\rightarrow\ $  s-C$_2$H$_5$OCH$_3$ & 20 & 0 \\
 & $\rightarrow\ $  s-C$_3$OH + s-C$_2$H$_4$ & 20 & 0 \\
 & $\rightarrow\ $  s-H$_2$CO + s-C$_2$H$_6$ & 60 & 0\vspace{5pt}  \\
 s-CH$_3$ + s-CH$_3$OCH$_2$ & $\rightarrow\ $  s-C$_2$H$_5$OCH$_3$ & 100 & 0 \\
\hline
\end{tabular}
\tablefoot{
\tablefoottext{a}{The prefix `s-' means the species is in the solid phase, which corresponds to the ice surface and the ice mantle.}
\tablefoottext{b}{s-C$_3$H$_7$OH accounts for all the isomers.}
}
\end{table}

We have therefore introduced a large number of reactions involving s-HCO, s-CH$_3$, s-CH$_3$O, s-CH$_2$OH, s-NH, s-HS and to a lesser extent s-H$_2$NO, s-HNOH, s-H$_2$CN, s-HCNH, s-CH$_3$NH, s-CH$_2$NH$_2$, s-HOCO, s-H$_2$NS, s-HNSH, s-C$_2$H$_3$, s-C$_2$H$_5$, s-C$_3$H$_{x=0-3,5,7}$ and a few others. For s-CN, s-OH, s-C, s-CH, s-C$_2$, and s-C$_2$H we consider a limited number of reactions as these species react with s-H$_2$. For s-NH$_2$ and s-CH$_2$ we consider a relatively large number of reactions even if they can react with H$_2$ due to their potential role at high temperatures ($>$20~K) where s-H$_2$ is low and where their abundance is not negligible due to photodissociation of s-NH$_3$ and s-CH$_4$. Of all the possible reactions, those between radicals play an important role in the formation of COMs. As noted already by \cite{Garrod-2013}, the reaction between two radicals can either lead to a single molecule or to H atom exchange, called disproportionation, for example $\mathrm{\text{s-}HCO + \text{s-}HCO  \rightarrow \text{s-}HC(O)CHO}$ or $\mathrm{\text{s-}H_2CO + \text{s-}CO}$, respectively. The branching ratios are in general poorly known even in the gas phase. Most models favor, when they consider these reactions, formation of the single molecule or use statistical ratios \citep{Garrod-2013}. However, when branching ratio determinations exist in the gas phase, it leads to almost exclusively disproportionation except for reactions with CH$_3$ for which single molecule production is important \citep{Baggott-1987}. Moreover, the branching ratios are likely to be significantly different between the gas and grain processes. A recent experimental work \citep{Butscher-2017} suggested that the reaction of two s-HCO radicals on ice surfaces does not lead to HC(O)CHO but rather to s-H$_2$CO and s-CO. 
\textcolor{black}{Additionally, \cite{EnriqueRomero-2016} found that the interaction between s-HCO and s-CH$_3$ with water ice (through cluster water ice approximation) imposes an initial geometry leading to the formation of s-CH$_4$ and s-CO through Eley-Rideal reactive mechanism, hindering the formation of s-CH$_3$CHO at least at low temperatures ($\sim$10~K). However, a revision of the model \citep{EnriqueRomero-2020} suggests the possibility to still form s-CH$_3$CHO through reaction of s-CH$_3$ and s-HCO.} 
s-CH$_3$CHO molecules are likely produced at temperatures where s-HCO and s-CH$_3$ acquire some mobility. In our network, for reactions involving s-CH$_3$ we always consider a non-negligible branching ratio for the formation of a single molecule deduced from gas-phase measurements. For the radical reactions we largely favour the disproportionation channel except when it is the only effective way to produce some observed species such as CH$_3$OCHO, which is produced by s-CH$_3$O + s-HCO on ice surfaces. For the barrier-less addition reactions on unsaturated radicals, such as H + H$_2$CCCN, we consider addition on several sites due to non-localised radical character of the lonely electron ($\mathrm{ H_2C\text{=}C^\bullet\text{--}C{\scriptstyle\large\equiv }N \leftrightarrow  H_2C\text{=}C\text{=}C\text{=}N^\bullet}$) with branching ratios deduced from Mulliken atomic spin density given by theoretical calculations. This leads to nitrile (R--CN) and imine (R=NH) production \citep{Krim-2019}, but only low alcohol production in the case of s-H reaction through H atom addition on unsaturated aldehyde radicals \citep{Krim-2018}. 

The systematic inclusion of reactions between radicals leads to the formation of a very large number of species on grains. We have not systematically completed the chemistry of all the new species, this would lead to an almost infinite number of reactions and species. In a first approach we focused on the most abundant species. For some, but not all of them, we have included the corresponding desorption mechanisms and gas phase chemistry. They are namely C$_2$H$_5$CN, C$_3$H$_3$CN, HCOCN, C$_2$H$_3$CHO, C$_2$H$_5$CHO, HC(O)CHO, CH$_3$COOH, CH$_2$(OH)CHO and (CH$_2$OH)$_2$. At the same time, the gas chemistry of CH$_3$OCH$_3$, CH$_3$COCH$_3$, C$_2$H$_5$OH and C$_6$H$_6$ was fully incorporated in the network. Branching ratios used for radicals recombination in this work are summarised in Table \ref{tab-BR}. s-\textit{n}-C$_3$H$_7$OH is not included in the network by simplification as neither \text{n-}C$_3$H$_7$OH nor \textit{i-}C$_3$H$_7$OH have been firmly detected.

\subsection{\label{app-sec-runs}Discussed reactions}

\begin{table}[t]\centering
\caption{\small\label{app-tab-reactions-Chuang}Hydrogenation reactions of CH(O)CHO from \cite{Chuang-2016} and CH$_3$OCHO on grain surfaces.}
\begin{tabular}{lclc}
\hline\hline
Reacting species & & Produced species & BR  \\
 & & & (\%)  \\
\hline
s-H\tablefootmark{a} + s-HC(O)CHO & $\rightarrow$ & s-CH(OH)CHO & 50  \\
 & $\rightarrow$ & s-CH$_2$(O)CHO & 50  \\
s-H + s-CH(OH)CHO & $\rightarrow$ & s-CH$_2$(OH)CHO & 100  \\
s-H + s-CH$_2$(O)CHO & $\rightarrow$ & s-CH$_2$(OH)CHO & 100 \\
s-H + s-CH$_2$(OH)CHO & $\rightarrow$ & s-CH$_2$(OH)CH$_2$O & 50  \\
 & $\rightarrow$ & s-CH$_2$(OH)CHOH & 50  \\
s-H + s-CH$_2$(OH)CH$_2$O & $\rightarrow$ & s-(CH$_2$OH)$_2$ & 100  \\
s-H + s-CH$_2$(OH)CHOH & $\rightarrow$ & s-(CH$_2$OH)$_2$ & 100  \\
s-H + s-CH$_3$OCHO & $\rightarrow$ & s-CH$_3$OCH$_2$O & 50 \\
 & $\rightarrow$ & s-CH$_3$OCHOH & 50 \\
s-H + s-CH$_3$OCH$_2$O & $\rightarrow$ & s-CH$_3$OCH$_2$OH & 100 \\
s-H + s-CH$_3$OCHOH & $\rightarrow$ & s-CH$_3$OCH$_2$OH & 100 \\
\hline
\end{tabular}
\tablefoot{
\tablefoottext{a}{The prefix `s-' means the species is in the solid phase, which corresponds to the ice surface and the ice mantle.}
}
\end{table}
This section presents the list of involved reactions in each run of the chemical simulation. Table \ref{app-tab-reactions-Chuang} shows the hydrogenation reactions of HC(O)CHO \citep{Chuang-2016} and CH$_3$OCHO on grain surfaces. The branching ratios have been assumed to be 50\% for all the intermediate species. The high reactivity of these species should prevent them from having the time to leave the surface before reacting with another hydrogen atom. Thus, the chemical and physical desorption processes are included in the network only for the more stable species HC(O)CHO, CH$_2$(OH)CHO and (CH$_2$OH)$_2$. Formation enthalpies of the intermediate species are taken from \cite{Goldsmith-2012}. Energy barriers of hydrogenation reactions of HC(O)CHO, CH$_2$(OH)CHO and CH$_3$OCHO are based on studies of \cite{Colberg-2006}, \cite{Alvarez-Barcia-2018}, \cite{Krim-2018} and our calculations.
The reactions $\mathrm{\text{s-}HCO + \text{s-}C_2H_3}$ and $\mathrm{\text{s-}HCO + \text{s-}C_2H_5}$ that are removed in the runs D1 and D2 are shown in Table \ref{tab-BR}.

\onecolumn

\section{Main spectroscopic parameters}

The main spectroscopic parameters of the species analysed in this paper are shown in Table \ref{app-tab-spectro}. It includes the database, the tag number, the dipole moment $\left| \left| \vec{\mu}\right| \right|$, the principle axis rotational parameters $A$, $B$ and $C$, and the references.

\begin{table*}[h]\centering
\caption{\small\label{app-tab-spectro} Main spectroscopic parameters of the analysed species.}
\begin{tabular}{l cccccc l}
\hline\hline
Species & Database & TAG & $\left| \left| \vec{\mu}\right| \right|$ & $A$ & $B$ & $C$ & References \\
 & & & (D) & (MHz) & (MHz) & (MHz) & \\
\hline
 C$_2$H$_3$CHO & CDMS & 56519 & 3.116 & 47353.71 & 4659.49 & 4242.70 & 1, 2, 3, 4\\
 C$_3$H$_6$ & CDMS & 42516 & 0.363 & 46280.29 & 9305.24 & 8134.23 & 5, 6, 7, 8, 9\\
 HCCCHO & CDMS & 54510 & 2.778 & 68035.30 & 4826.22 & 4499.59 & 10, 11, 12, 13\\
 \textit{n}-C$_3$H$_7$OH & CDMS & 60505 & 1.415 & 14330.37 & 5119.28 & 4324.23 & 14, 15, 16, 17\\
 \textit{i}-C$_3$H$_7$OH & CDMS & 60518 & 1.564 & 8639.54 & 8063.22 & 4768.25 & 18, 19, 20, 21\\
 C$_3$O & CDMS & 52501 & 2.391 & 0 & 4810.89 & 0 & 22, 23, 24, 25 \\
\small \emph{cis-}HC(O)CHO & CDMS & 58513 & 3.40 & 26713.36 & 6190.77 & 5032.48 & 26 \\
 C$_3$H$_8$ & JPL & 44013 & 0.085 & 29207.47 & 8445.97 & 7459.00 & 27, 28, 29, 30\\
\hline
\end{tabular}
\tablebib{
(1) \citet{Daly-2015}; (2) \citet{Blom-1984}; (3) \citet{Winnewisser-1975}; (4) \citet{Cherniak-1966}; (5) \citet{Craig-2016}; (6) \citet{Wlodarczak-1994}; (7) \citet{Pearson-1994}; (8) \citet{Hirota-1966}; (9) \citet{Lide-1957}; (10) \citet{McKellar-2008}; (11) \citet{Brown-1984}; (12) \citet{Winnewisser-1973}; (13) \citet{Costain-1959}; (14) \citet{Kisiel-2010}; (15) \citet{Maeda-2006-n}; (16) \citet{Kahn-2005}; (17) \citet{White-1975}; (18) \citet{Maeda-2006-i}; (19) \citet{Christen-2003}; (20) \citet{Ulenikov-1991}; (21) \citet{Hirota-1979}; (22) \citet{Bizzocchi-2008}; (23) \citet{Klebsch-1985}; (24) \citet{Tang-1985}; (25) \citet{Brown-1983}; (26) \citet{Hubner-1997}; (27) \citet{Drouin-2006}; (28) \citet{Bestmann-1985b}; (29) \citet{Bestmann-1985a}; (30) \citet{Lide-1960}. 
}
\end{table*}

\section{Final abundances of the simulation runs}

\begin{figure}[h]\centering
\adjustbox{trim=0 {0.03\height} 0 {0.03\height}, clip=true}{\includegraphics[width=0.97\textwidth]{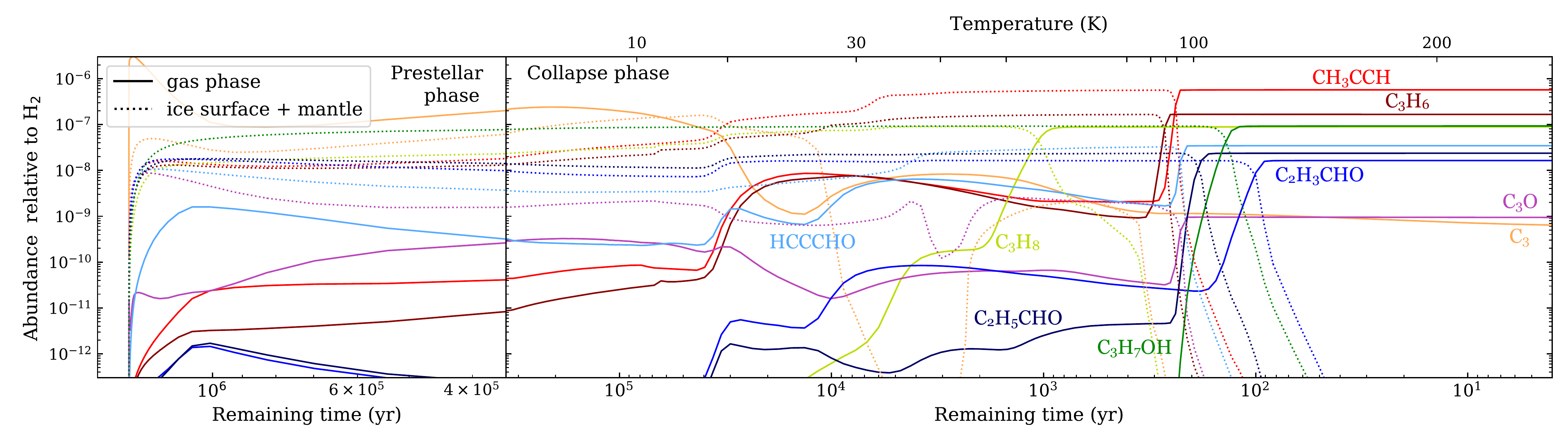}}\\
\adjustbox{trim=0 {0.03\height} 0 {0.03\height}, clip=true}{\includegraphics[width=0.97\textwidth]{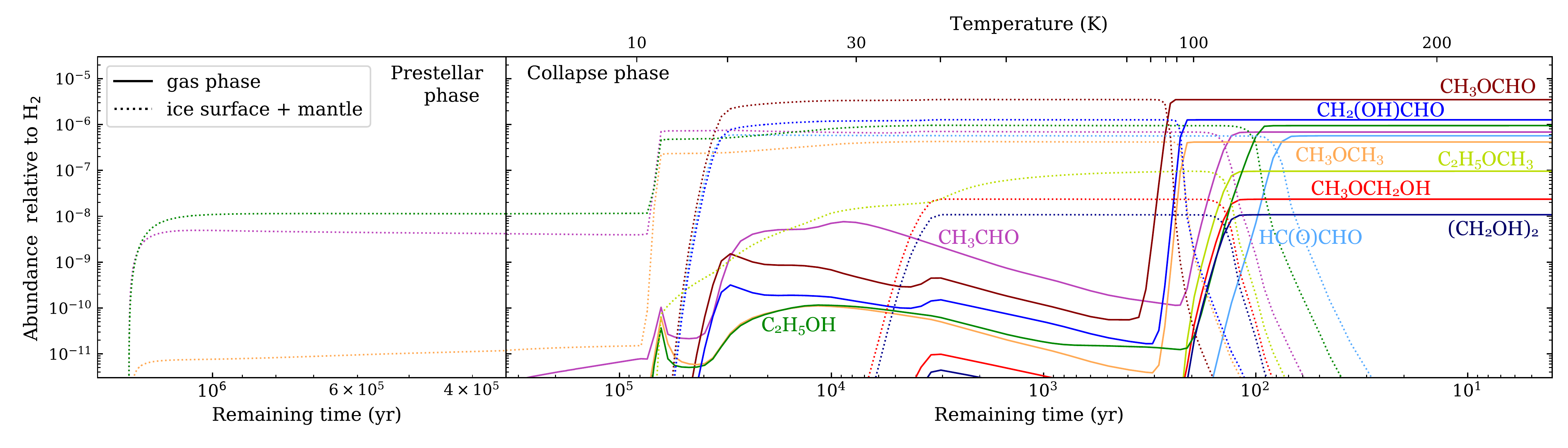}}
\caption{\label{fig:app:evol_C12_0}\small Evolution of the abundances of three-carbon species during the prestellar and the collapse phase of the simulation run A1. The time axis is reversed to better visualise the abundance evolution. Each color corresponds to a single species, with its abundance in the gas phase and on grain surfaces shown in solid and dotted lines, respectively.}
\end{figure}

\begin{figure}[h]\centering
\adjustbox{trim=0 {0.03\height} 0 {0.03\height}, clip=true}{\includegraphics[width=0.97\textwidth]{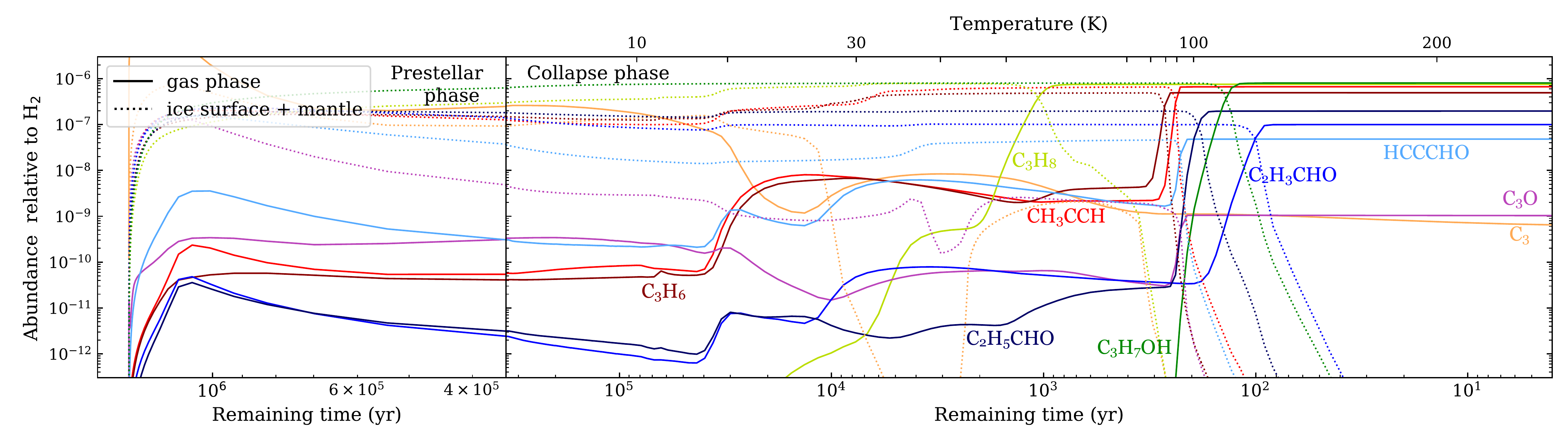}}\\
\adjustbox{trim=0 {0.03\height} 0 {0.03\height}, clip=true}{\includegraphics[width=0.97\textwidth]{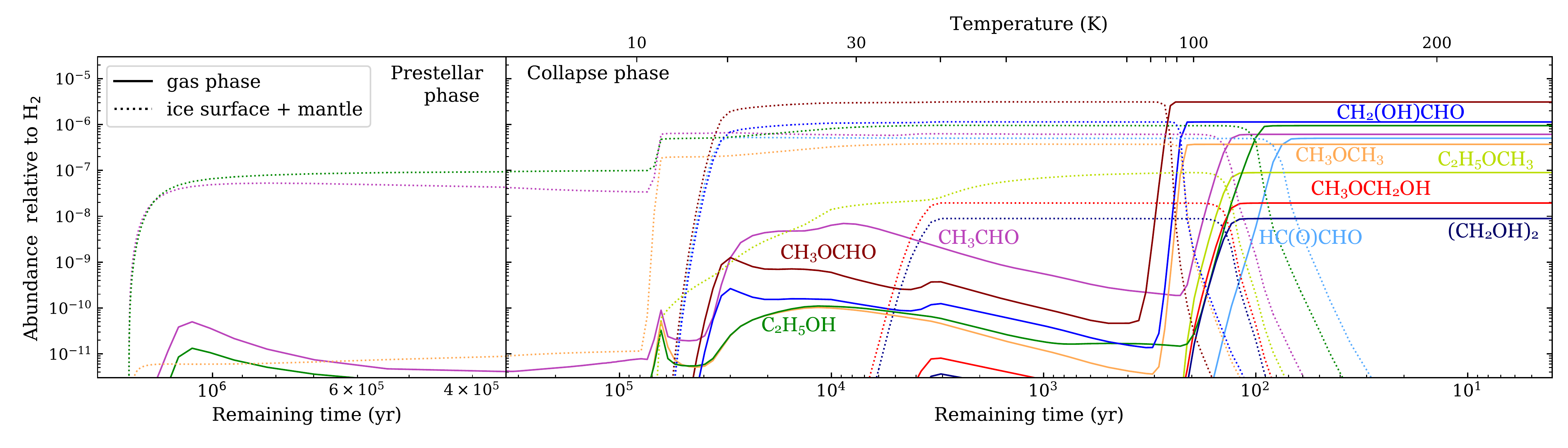}}
\caption{\label{fig:app:evol_C16_0}\small Evolution of the abundances of three-carbon species during the prestellar and the collapse phase of the simulation run A2. The time axis is reversed to better visualise the abundance evolution. Each color corresponds to a single species, with its abundance in the gas phase and on grain surfaces shown in solid and dotted lines, respectively.}
\end{figure}

\begin{figure}[h]\centering
\adjustbox{trim=0 {0.03\height} 0 {0.03\height}, clip=true}{\includegraphics[width=0.97\textwidth]{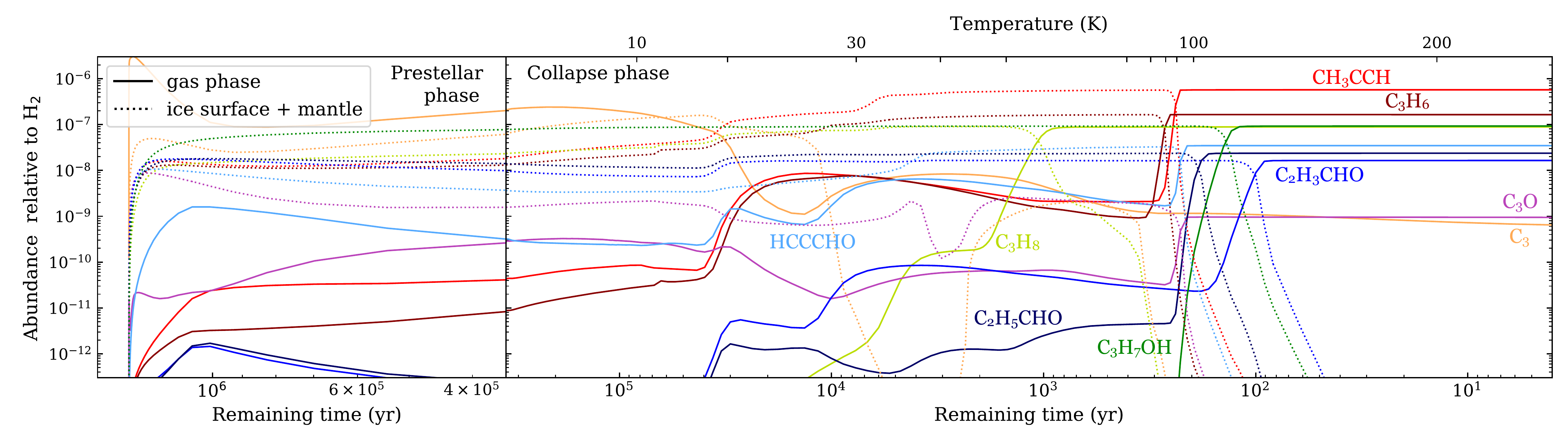}}\\
\adjustbox{trim=0 {0.03\height} 0 {0.03\height}, clip=true}{\includegraphics[width=0.97\textwidth]{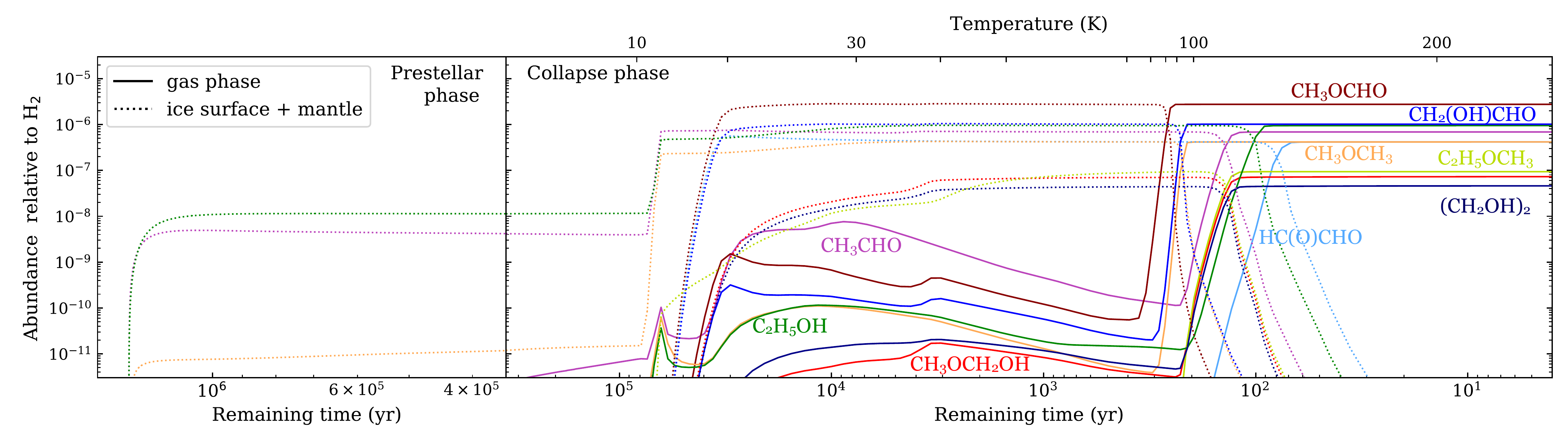}}
\caption{\label{fig:app:evol_C12_1}\small Evolution of the abundances of three-carbon species during the prestellar and the collapse phase of the simulation run B1. The time axis is reversed to better visualise the abundance evolution. Each color corresponds to a single species, with its abundance in the gas phase and on grain surfaces shown in solid and dotted lines, respectively.}
\end{figure}

\begin{figure}[h]\centering
\adjustbox{trim=0 {0.03\height} 0 {0.03\height}, clip=true}{\includegraphics[width=0.97\textwidth]{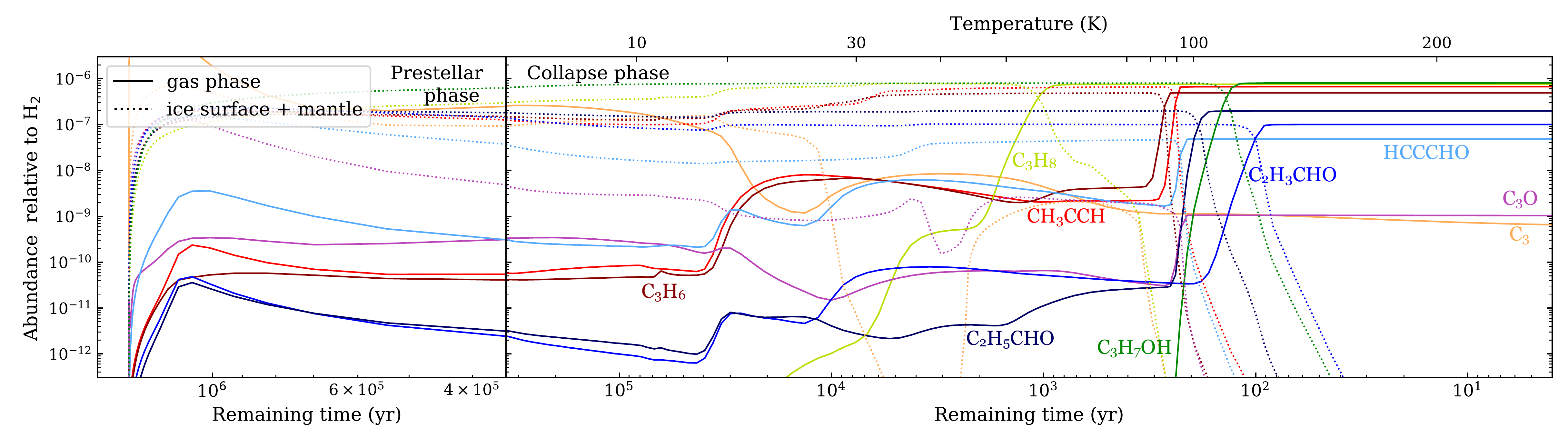}}\\
\adjustbox{trim=0 {0.03\height} 0 {0.03\height}, clip=true}{\includegraphics[width=0.97\textwidth]{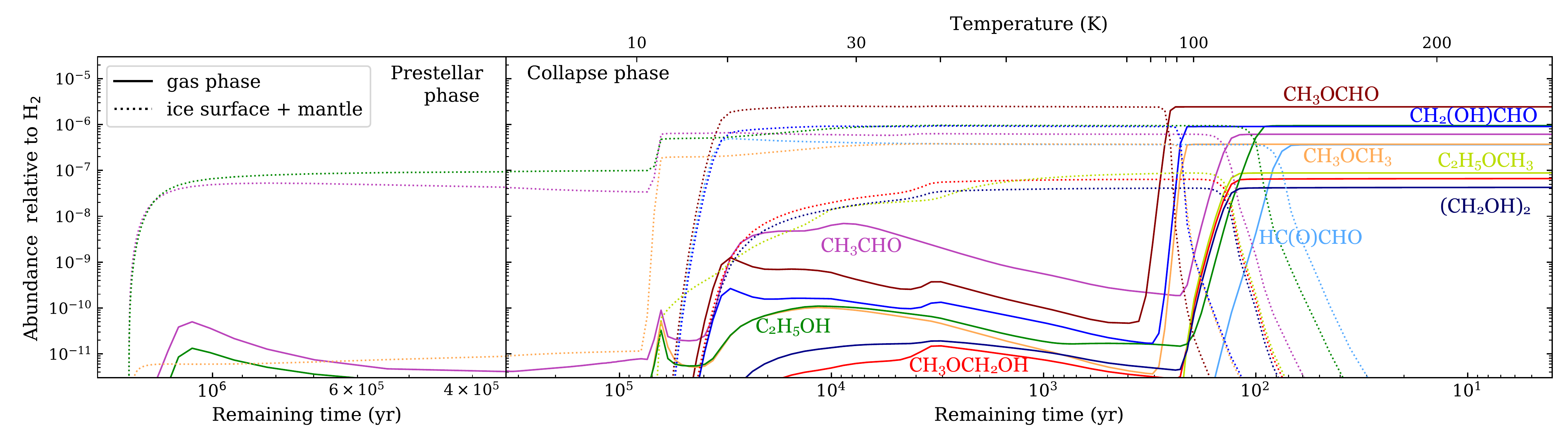}}
\caption{\label{fig:app:evol_C16_1}\small Evolution of the abundances of three-carbon species during the prestellar and the collapse phase of the simulation run B2. The time axis is reversed to better visualise the abundance evolution. Each color corresponds to a single species, with its abundance in the gas phase and on grain surfaces shown in solid and dotted lines, respectively.}
\end{figure}

\begin{figure}[h]\centering
\adjustbox{trim=0 {0.03\height} 0 {0.03\height}, clip=true}{\includegraphics[width=0.97\textwidth]{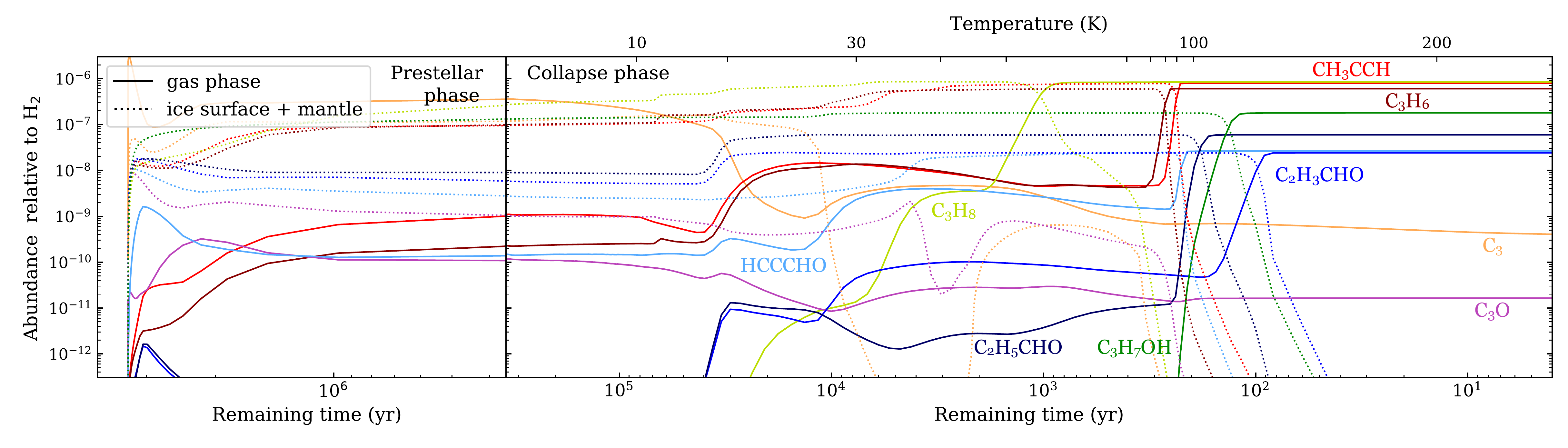}}\\
\adjustbox{trim=0 {0.03\height} 0 {0.03\height}, clip=true}{\includegraphics[width=0.97\textwidth]{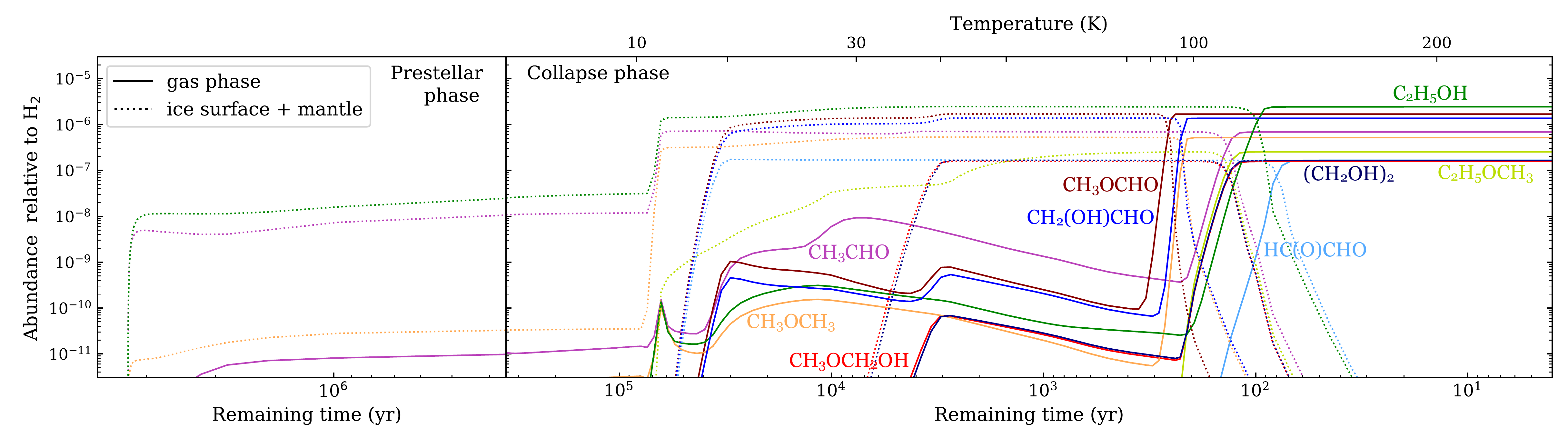}}
\caption{\label{fig:app:evol_C12_2}\small Evolution of the abundances of three-carbon species during the prestellar and the collapse phase of the simulation run C1. The time axis is reversed to better visualise the abundance evolution. Each color corresponds to a single species, with its abundance in the gas phase and on grain surfaces shown in solid and dotted lines, respectively.}
\end{figure}

\begin{figure}[h]\centering
\adjustbox{trim=0 {0.03\height} 0 {0.03\height}, clip=true}{\includegraphics[width=0.97\textwidth]{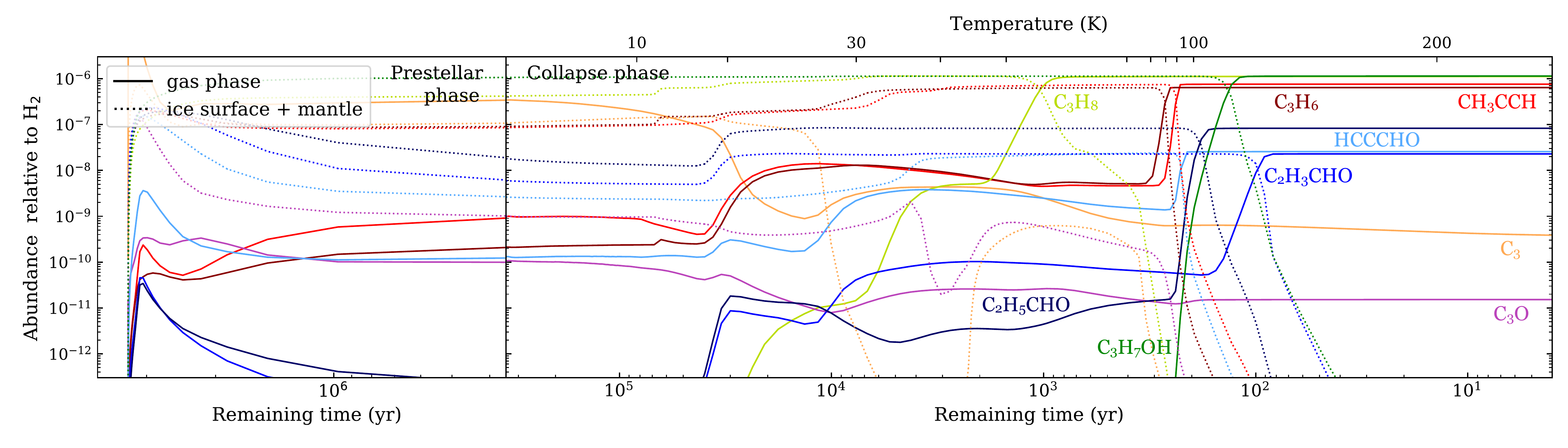}}\\
\adjustbox{trim=0 {0.03\height} 0 {0.03\height}, clip=true}{\includegraphics[width=0.97\textwidth]{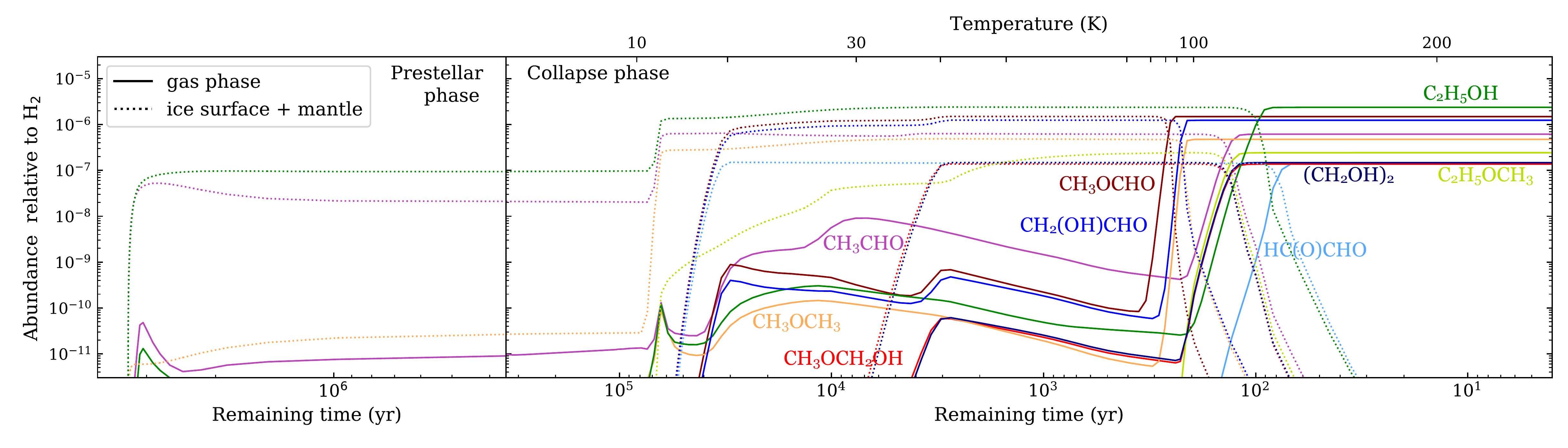}}
\caption{\label{fig:app:evol_C16_2}\small Evolution of the abundances of three-carbon species during the prestellar and the collapse phase of the simulation run C2. The time axis is reversed to better visualise the abundance evolution. Each color corresponds to a single species, with its abundance in the gas phase and on grain surfaces shown in solid and dotted lines, respectively.}
\end{figure}

\begin{figure}[h]\centering
\adjustbox{trim=0 {0.03\height} 0 {0.03\height}, clip=true}{\includegraphics[width=0.97\textwidth]{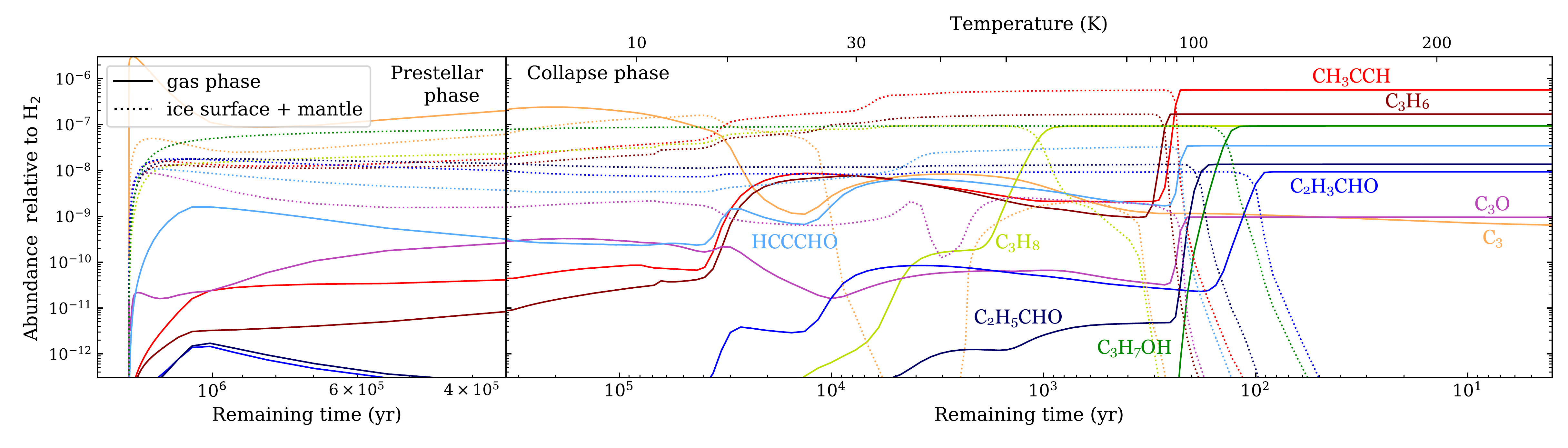}}\\
\adjustbox{trim=0 {0.03\height} 0 {0.03\height}, clip=true}{\includegraphics[width=0.97\textwidth]{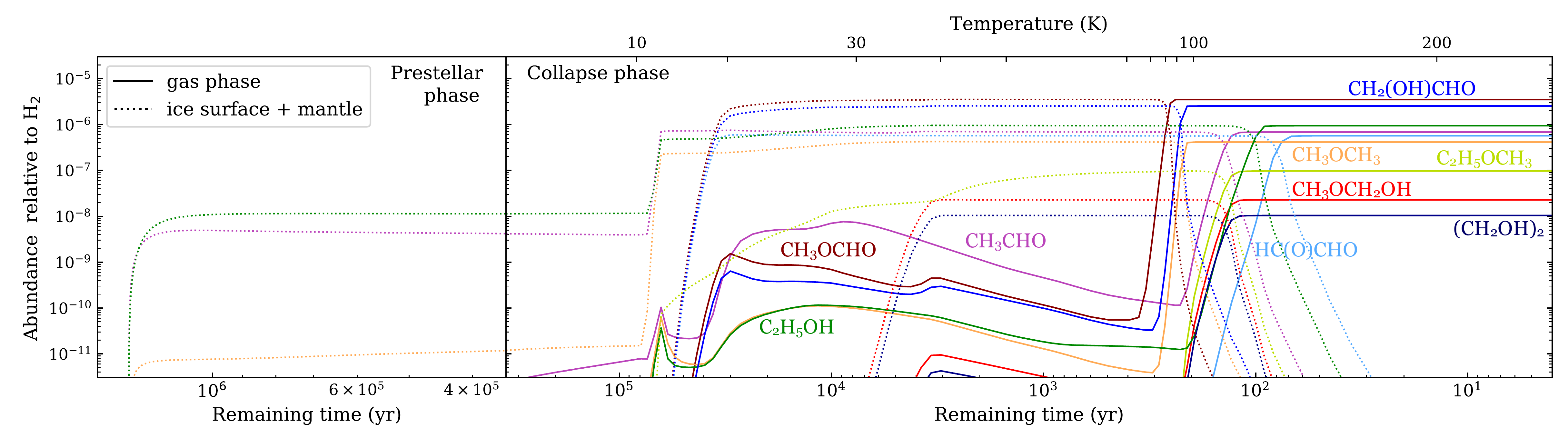}}
\caption{\label{fig:app:evol_C12_3}\small Evolution of the abundances of three-carbon species during the prestellar and the collapse phase of the simulation run D1. The time axis is reversed to better visualise the abundance evolution. Each color corresponds to a single species, with its abundance in the gas phase and on grain surfaces shown in solid and dotted lines, respectively.}
\end{figure}

\begin{figure}[h]\centering
\adjustbox{trim=0 {0.03\height} 0 {0.03\height}, clip=true}{\includegraphics[width=0.97\textwidth]{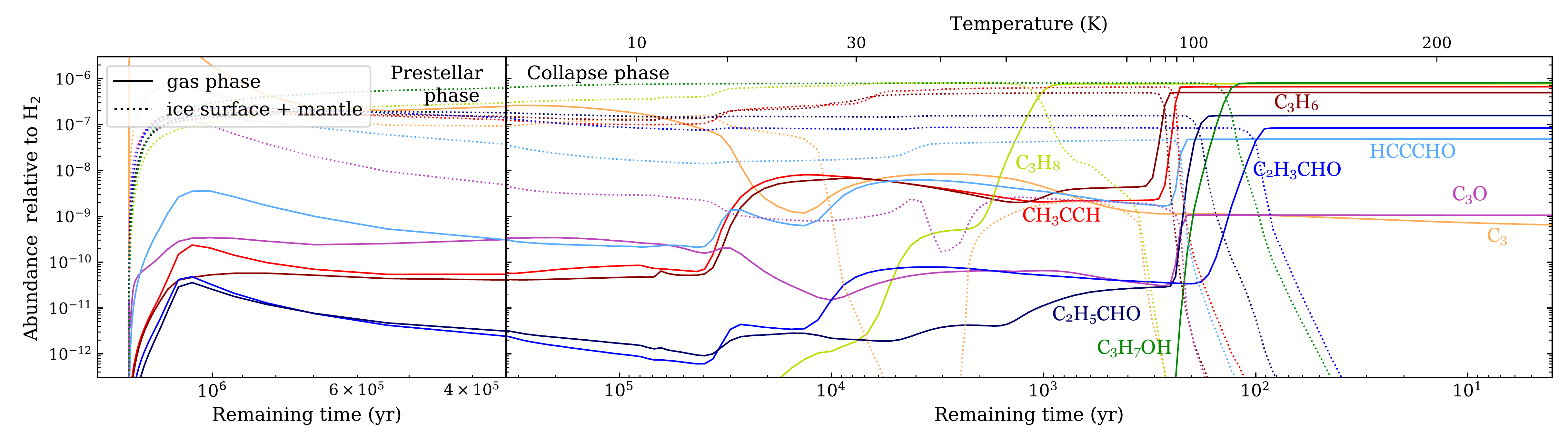}}\\
\adjustbox{trim=0 {0.03\height} 0 {0.03\height}, clip=true}{\includegraphics[width=0.97\textwidth]{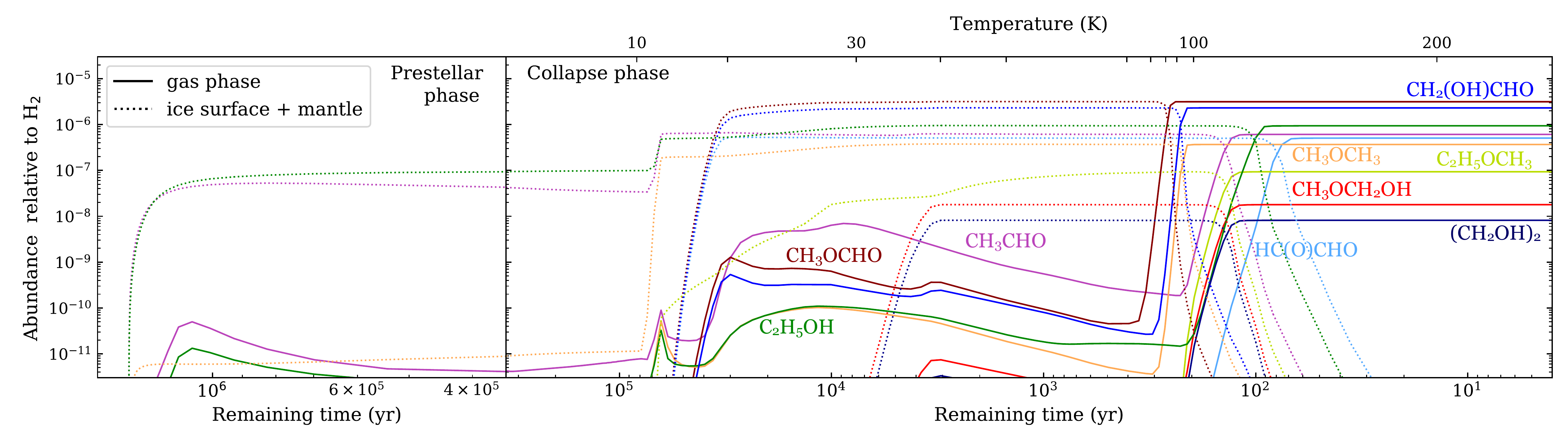}}
\caption{\label{fig:app:evol_C16_3}\small Evolution of the abundances of three-carbon species during the prestellar and the collapse phase of the simulation run D2. The time axis is reversed to better visualise the abundance evolution. Each color corresponds to a single species, with its abundance in the gas phase and on grain surfaces shown in solid and dotted lines, respectively.}
\end{figure}

\begin{figure}[h]\centering
\adjustbox{trim=0 {0.03\height} 0 {0.03\height}, clip=true}{\includegraphics[width=0.97\textwidth]{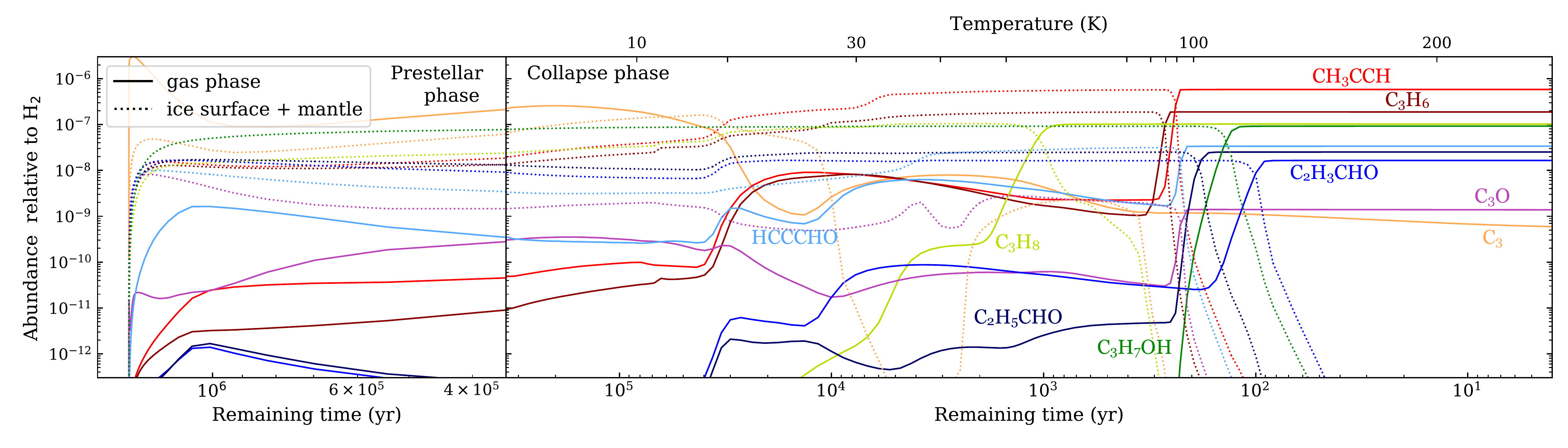}}\\
\adjustbox{trim=0 {0.03\height} 0 {0.03\height}, clip=true}{\includegraphics[width=0.97\textwidth]{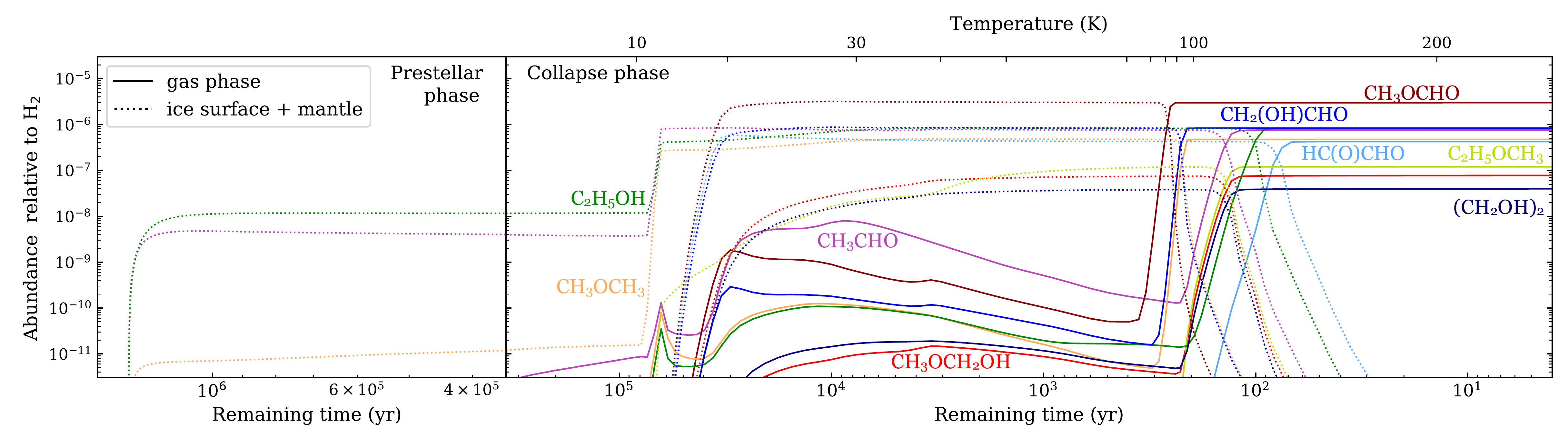}}
\caption{\label{fig:app:evol_C12_4}\small Evolution of the abundances of three-carbon species during the prestellar and the collapse phase of the simulation run E1. The time axis is reversed to better visualise the abundance evolution. Each color corresponds to a single species, with its abundance in the gas phase and on grain surfaces shown in solid and dotted lines, respectively.}
\end{figure}

\begin{figure}[h]\centering
\adjustbox{trim=0 {0.03\height} 0 {0.03\height}, clip=true}{\includegraphics[width=0.97\textwidth]{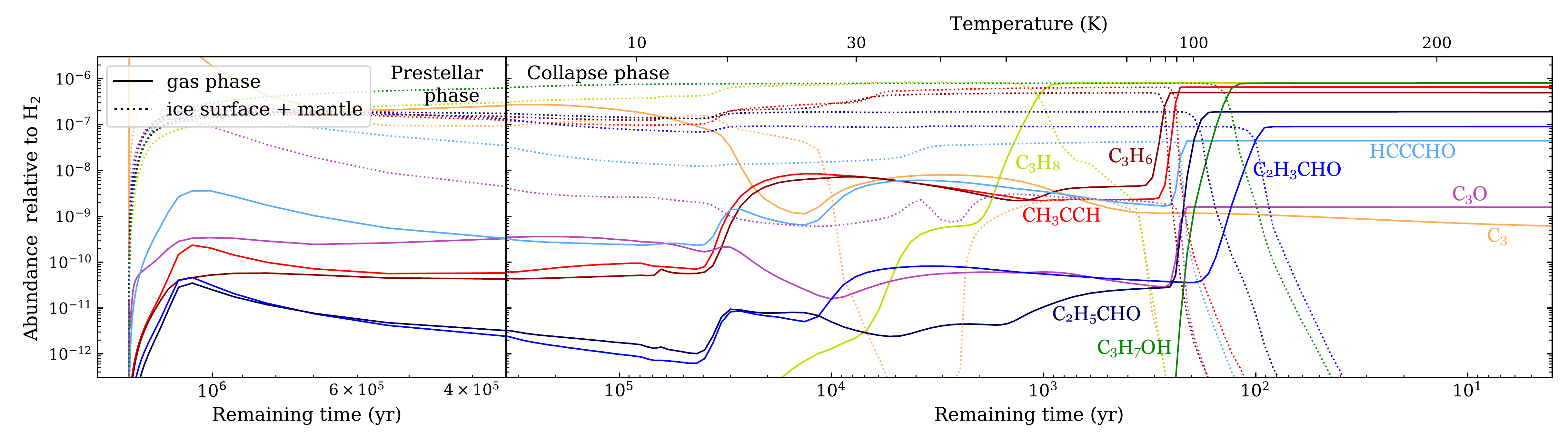}}\\
\adjustbox{trim=0 {0.03\height} 0 {0.03\height}, clip=true}{\includegraphics[width=0.97\textwidth]{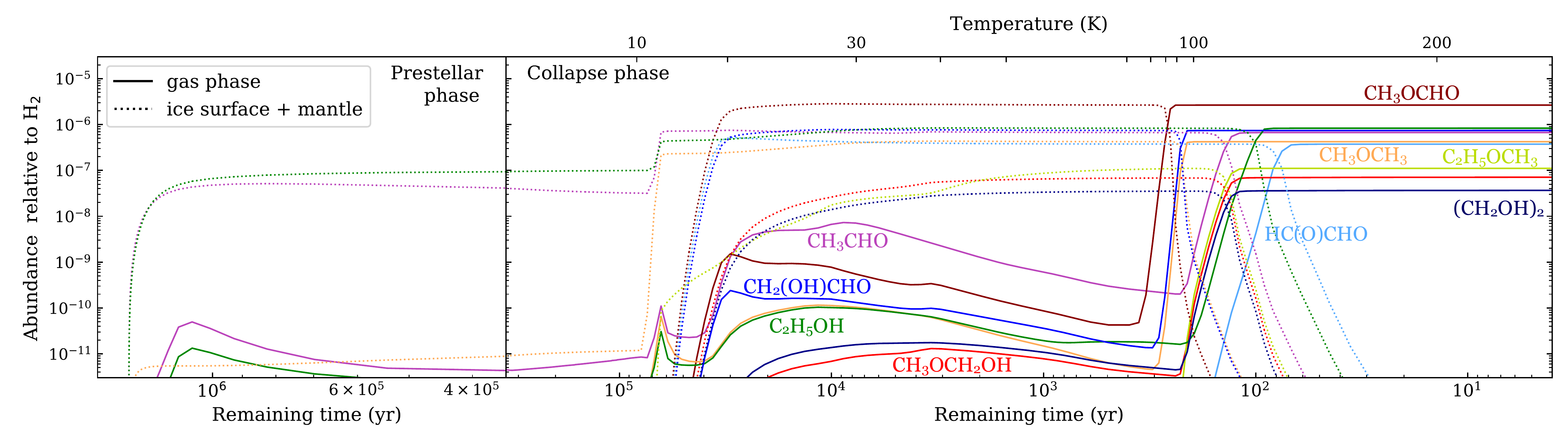}}
\caption{\label{fig:app:evol_C16_4}\small Evolution of the abundances of three-carbon species during the prestellar and the collapse phase of the simulation run E2. The time axis is reversed to better visualise the abundance evolution. Each color corresponds to a single species, with its abundance in the gas phase and on grain surfaces shown in solid and dotted lines, respectively.}
\end{figure}

This appendix section presents the results of the simulations runs for all the relevant species of this study. Table \ref{app-tab-Runs} shows the final abundances of the runs while Figures \ref{fig:app:evol_C12_0} to \ref{fig:app:evol_C16_4} represent the abundance evolution during each simulation run for the studied species.

\begin{table*}[h]
\caption{\small\label{app-tab-Runs} Species abundances, relative to CH$_3$OH, at the end of the collapse phase for all the runs.}
\begin{tabular}{l cccccccc}
\hline\hline
Runs & \multicolumn{8}{c}{Species abundance (relative to CH$_3$OH)} \\
/ Obs. & CH$_3$CCH & C$_3$H$_6$ & C$_3$H$_8$ & C$_3$O & HCCCHO & C$_2$H$_3$CHO & C$_2$H$_5$CHO & C$_3$H$_7$OH\tablefootmark{a} \\
\hline
PILS & $1.1\times10^{-3}$ & $4.2\times10^{-3}$ & $8.0\times10^{-3}$ & $2.0\times10^{-6}$ & $5.0\times10^{-5}$ & $3.4\times10^{-5}$ & $2.2\times10^{-4}$ & $6.0\times10^{-4}$ \\
\hline
A1 & $9.5\times10^{-3}$ & $2.8\times10^{-3}$ & $1.5\times10^{-3}$ & $1.4\times10^{-5}$ & $5.8\times10^{-4}$ & $2.7\times10^{-4}$ & $4.0\times10^{-4}$ & $1.5\times10^{-3}$ \\
B1 & $9.5\times10^{-3}$ & $2.8\times10^{-3}$ & $1.5\times10^{-3}$ & $1.5\times10^{-5}$ & $5.8\times10^{-4}$ & $2.7\times10^{-4}$ & $4.0\times10^{-4}$ & $1.6\times10^{-3}$ \\
C1 & $1.0\times10^{-2}$ & $7.6\times10^{-3}$ & $1.1\times10^{-2}$ & $2.1\times10^{-7}$ & $3.4\times10^{-4}$ & $3.1\times10^{-4}$ & $7.6\times10^{-4}$ & $2.3\times10^{-3}$ \\
D1 & $9.7\times10^{-3}$ & $2.9\times10^{-3}$ & $1.6\times10^{-3}$ & $1.5\times10^{-5}$ & $5.9\times10^{-4}$ & $1.6\times10^{-4}$ & $2.3\times10^{-4}$ & $1.6\times10^{-3}$ \\
E1 & $9.3\times10^{-3}$ & $3.0\times10^{-3}$ & $1.6\times10^{-3}$ & $2.2\times10^{-5}$ & $5.4\times10^{-4}$ & $2.6\times10^{-4}$ & $4.0\times10^{-4}$ & $1.5\times10^{-3}$ \\
A2 & $1.2\times10^{-2}$ & $9.0\times10^{-3}$ & $1.4\times10^{-2}$ & $1.8\times10^{-5}$ & $8.8\times10^{-4}$ & $1.8\times10^{-3}$ & $3.6\times10^{-3}$ & $1.5\times10^{-2}$ \\
B2 & $1.2\times10^{-2}$ & $9.0\times10^{-3}$ & $1.4\times10^{-2}$ & $1.9\times10^{-5}$ & $8.8\times10^{-4}$ & $1.8\times10^{-3}$ & $3.6\times10^{-3}$ & $1.5\times10^{-2}$ \\
C2 & $1.0\times10^{-2}$ & $8.7\times10^{-3}$ & $1.5\times10^{-2}$ & $2.1\times10^{-7}$ & $3.5\times10^{-4}$ & $3.1\times10^{-4}$ & $1.1\times10^{-3}$ & $1.6\times10^{-2}$ \\
D2 & $1.2\times10^{-2}$ & $9.3\times10^{-3}$ & $1.4\times10^{-2}$ & $1.8\times10^{-5}$ & $8.9\times10^{-4}$ & $1.6\times10^{-3}$ & $2.9\times10^{-3}$ & $1.5\times10^{-2}$ \\
E2 & $1.2\times10^{-2}$ & $8.7\times10^{-3}$ & $1.4\times10^{-2}$ & $2.7\times10^{-5}$ & $7.8\times10^{-4}$ & $1.6\times10^{-3}$ & $3.3\times10^{-3}$ & $1.4\times10^{-2}$ \\
\hline\hline
\textcolor{white}{\LARGE L} & C$_2$H$_5$OH & C$_2$H$_5$OCH$_3$ & CH$_3$COCH$_3$ & CH$_3$OCHO & CH$_3$OCH$_2$OH & HC(O)CHO & CH$_2$(OH)CHO & (CH$_2$OH)$_2$ \\
\hline
PILS & $2.3\times10^{-2}$ & $1.8\times10^{-3}$ & $1.7\times10^{-3}$ & $2.6\times10^{-2}$ & $1.4\times10^{-2}$ & $5.0\times10^{5}$ & $6.8\times10^{-3}$ & $2.0\times10^{-3}$ \\
\hline
A1 & $1.6\times10^{-2}$ & $1.6\times10^{-3}$ & $1.4\times10^{-4}$ & $5.8\times10^{-2}$ & $3.9\times10^{-4}$ & $9.5\times10^{-3}$ & $2.1\times10^{-2}$ & $1.8\times10^{-4}$ \\
B1 & $1.6\times10^{-2}$ & $1.6\times10^{-3}$ & $1.5\times10^{-4}$ & $4.6\times10^{-2}$ & $1.2\times10^{-3}$ & $7.0\times10^{-3}$ & $1.7\times10^{-2}$ & $7.7\times10^{-4}$ \\
C1 & $3.1\times10^{-2}$ & $3.2\times10^{-3}$ & $1.8\times10^{-4}$ & $2.1\times10^{-2}$ & $2.0\times10^{-3}$ & $2.1\times10^{-3}$ & $1.7\times10^{-2}$ & $2.1\times10^{-3}$ \\
D1 & $1.6\times10^{-2}$ & $1.6\times10^{-3}$ & $1.5\times10^{-4}$ & $5.9\times10^{-2}$ & $3.9\times10^{-4}$ & $9.7\times10^{-3}$ & $4.3\times10^{-2}$ & $1.8\times10^{-4}$ \\
E1 & $1.3\times10^{-2}$ & $1.9\times10^{-3}$ & $1.9\times10^{-4}$ & $4.8\times10^{-2}$ & $1.2\times10^{-3}$ & $6.8\times10^{-3}$ & $1.3\times10^{-2}$ & $6.4\times10^{-4}$ \\
A2 & $1.7\times10^{-2}$ & $1.6\times10^{-3}$ & $4.5\times10^{-4}$ & $5.7\times10^{-2}$ & $3.5\times10^{-4}$ & $9.2\times10^{-3}$ & $2.1\times10^{-2}$ & $1.6\times10^{-4}$ \\
B2 & $1.7\times10^{-2}$ & $1.6\times10^{-3}$ & $4.6\times10^{-4}$ & $4.4\times10^{-2}$ & $1.2\times10^{-3}$ & $6.7\times10^{-3}$ & $1.7\times10^{-2}$ & $7.8\times10^{-4}$ \\
C2 & $3.2\times10^{-2}$ & $3.3\times10^{-3}$ & $2.1\times10^{-4}$ & $2.0\times10^{-2}$ & $1.9\times10^{-3}$ & $2.0\times10^{-3}$ & $1.7\times10^{-2}$ & $2.0\times10^{-3}$ \\
D2 & $1.7\times10^{-2}$ & $1.7\times10^{-3}$ & $4.6\times10^{-4}$ & $5.9\times10^{-2}$ & $3.3\times10^{-4}$ & $9.4\times10^{-3}$ & $4.3\times10^{-2}$ & $1.5\times10^{-4}$ \\
E2 & $1.5\times10^{-2}$ & $1.9\times10^{-3}$ & $4.6\times10^{-4}$ & $4.7\times10^{-2}$ & $1.2\times10^{-3}$ & $6.6\times10^{-3}$ & $1.3\times10^{-2}$ & $6.4\times10^{-4}$ \\
\hline
\end{tabular}
\tablefoot{
\tablefoottext{a}{C$_3$H$_7$OH abundance for the PILS observations is the sum of the abundance of all the isomers.}
}
\end{table*}

\newpage

\newpage

\newpage

\newpage

\newpage

\newpage

\section{ C$_2$H$_3$CHO line list}

This section lists, in Table \ref{app-linelist-C2H3CHO}, the C$_2$H$_3$CHO transitions corresponding to the lines having an intensity above 3$\sigma$ in the observations. The firmly detected transitions, which are optically thin and unblended in the conditions of the best fit, are annotated with the letter `Y'.

\begin{longtable}{cccccc}
\caption{\label{app-linelist-C2H3CHO}Line list of the observed C$_2$H$_3$CHO transitions above 3 $\sigma$.}\\
\hline\hline
Transition & Fitted & $\mathrm{\nu}$ & $g_\mathrm{u}$ & $E_{\mathrm{u}}$ & $A_{\mathrm{ul}}$ \\
\multicolumn{2}{l}{$J(K_{\mathrm{a}},\ K_{\mathrm{c}})$ -- $J'(K_{\mathrm{a}}',\ K_{\mathrm{c}}')$}  & (GHz) & & (K) & (s$^{-1}$) \\
\hline
\endfirsthead
\caption{continued.}\\
\hline\hline
Transition & Fitted & $\mathrm{\nu}$ & $g_\mathrm{u}$ & $E_{\mathrm{u}}$ & $A_{\mathrm{ul}}$ \\
\multicolumn{2}{l}{$J(K_{\mathrm{a}},\ K_{\mathrm{c}})$ -- $J'(K_{\mathrm{a}}',\ K_{\mathrm{c}}')$}  & (GHz) & & (K) & (s$^{-1}$) \\
\hline
\endhead
\hline
\endfoot

25( 1,24) -- 24( 1,23) & Y & 225.1917 & 51 & 143 & $6.054\times10^{-4}$ \\
27( 0,27) -- 26( 0,26) & Y & 233.7577 & 55 & 159 & $6.788\times10^{-4}$ \\
26( 1,25) -- 25( 1,24) &  & 233.9402 & 53 & 154 & $6.793\times10^{-4}$ \\
26( 2,24) -- 25( 2,23) & Y & 235.9421 & 53 & 160 & $6.947\times10^{-4}$ \\
27(10,17) -- 26(10,16) &  & 240.4387 & 55 & 367 & $6.384\times10^{-4}$ \\
27(10,18) -- 26(10,17) &  & 240.4387 & 55 & 367 & $6.384\times10^{-4}$ \\
27( 9,18) -- 26( 9,17) &  & 240.4537 & 55 & 328 & $6.578\times10^{-4}$ \\
27( 9,19) -- 26( 9,18) &  & 240.4537 & 55 & 328 & $6.578\times10^{-4}$ \\
27( 8,19) -- 26( 8,18) & Y & 240.4793 & 55 & 293 & $6.752\times10^{-4}$ \\
27( 8,20) -- 26( 8,19) & Y & 240.4793 & 55 & 293 & $6.752\times10^{-4}$ \\
27( 7,21) -- 26( 7,20) & Y & 240.5217 & 55 & 262 & $6.908\times10^{-4}$ \\
27( 7,20) -- 26( 7,19) & Y & 240.5217 & 55 & 262 & $6.908\times10^{-4}$ \\
27( 6,22) -- 26( 6,21) &  & 240.5929 & 55 & 236 & $7.047\times10^{-4}$ \\
27( 6,21) -- 26( 6,20) &  & 240.5929 & 55 & 236 & $7.047\times10^{-4}$ \\
28( 2,27) -- 27( 2,26) &  & 247.4024 & 57 & 181 & $8.021\times10^{-4}$ \\
29( 0,29) -- 28( 0,28) & Y & 250.6283 & 59 & 182 & $8.377\times10^{-4}$ \\
37( 3,35) -- 36( 3,34) &  & 329.5826 & 75 & 319 & $1.902\times10^{-3}$ \\
37( 8,30) -- 36( 8,29) &  & 329.6375 & 75 & 432 & $1.826\times10^{-3}$ \\
37( 8,29) -- 36( 8,28) &  & 329.6375 & 75 & 432 & $1.826\times10^{-3}$ \\
37( 7,31) -- 36( 7,30) & Y & 329.7554 & 75 & 401 & $1.850\times10^{-3}$ \\
37( 7,30) -- 36( 7,29) & Y & 329.7554 & 75 & 401 & $1.850\times10^{-3}$ \\
37( 5,33) -- 36( 5,32) &  & 330.2566 & 75 & 352 & $1.892\times10^{-3}$ \\
37( 5,32) -- 36( 5,31) &  & 330.3155 & 75 & 352 & $1.893\times10^{-3}$ \\
37( 4,34) -- 36( 4,33) & Y & 330.5505 & 75 & 334 & $1.909\times10^{-3}$ \\
37( 4,33) -- 36( 4,32) &  & 331.4060 & 75 & 334 & $1.924\times10^{-3}$ \\
38( 2,37) -- 37( 2,36) &  & 333.8975 & 77 & 323 & $1.985\times10^{-3}$ \\
37( 3,34) -- 36( 3,33) & Y & 334.7764 & 75 & 321 & $1.994\times10^{-3}$ \\
39( 1,39) -- 38( 1,38) &  & 334.9111 & 79 & 325 & $2.008\times10^{-3}$ \\
39( 0,39) -- 38( 0,38) & Y & 335.0926 & 79 & 325 & $2.012\times10^{-3}$ \\
37( 2,35) -- 36( 2,34) &  & 335.4273 & 75 & 313 & $2.012\times10^{-3}$ \\
38( 1,37) -- 37( 1,36) & Y & 336.4730 & 77 & 322 & $2.032\times10^{-3}$ \\
38( 3,36) -- 37( 3,35) &  & 338.4007 & 77 & 336 & $2.060\times10^{-3}$ \\
38( 7,32) -- 37( 7,31) &  & 338.6860 & 77 & 418 & $2.008\times10^{-3}$ \\
38( 7,31) -- 37( 7,30) &  & 338.6860 & 77 & 418 & $2.008\times10^{-3}$ \\
38( 4,35) -- 37( 4,34) &  & 339.5117 & 77 & 350 & $2.071\times10^{-3}$ \\
38( 4,34) -- 37( 4,33) &  & 340.5275 & 77 & 351 & $2.089\times10^{-3}$ \\
39( 2,38) -- 38( 2,37) &  & 342.4848 & 79 & 339 & $2.143\times10^{-3}$ \\
40( 1,40) -- 39( 1,39) &  & 343.3917 & 81 & 342 & $2.166\times10^{-3}$ \\
40( 0,40) -- 39( 0,39) & Y & 343.5483 & 81 & 342 & $2.169\times10^{-3}$ \\
38( 3,35) -- 37( 3,34) &  & 344.0665 & 77 & 338 & $2.166\times10^{-3}$ \\
38( 2,36) -- 37( 2,35) &  & 344.2952 & 77 & 330 & $2.177\times10^{-3}$ \\
39( 1,38) -- 38( 1,37) &  & 344.8722 & 79 & 338 & $2.190\times10^{-3}$ \\
39( 3,37) -- 38( 3,36) &  & 347.2066 & 79 & 352 & $2.226\times10^{-3}$ \\
39( 8,32) -- 38( 8,31) &  & 347.4788 & 79 & 465 & $2.151\times10^{-3}$ \\
39( 8,31) -- 38( 8,30) &  & 347.4788 & 79 & 465 & $2.151\times10^{-3}$ \\
39( 7,33) -- 38( 7,32) &  & 347.6181 & 79 & 434 & $2.176\times10^{-3}$ \\
39( 7,32) -- 38( 7,31) &  & 347.6181 & 79 & 434 & $2.176\times10^{-3}$ \\
41( 1,41) -- 40( 1,40) &  & 351.8700 & 83 & 358 & $2.331\times10^{-3}$ \\
41( 0,41) -- 40( 0,40) &  & 352.0051 & 83 & 358 & $2.334\times10^{-3}$ \\
39( 2,37) -- 38( 2,36) &  & 353.1277 & 79 & 347 & $2.350\times10^{-3}$ \\
40( 1,39) -- 39( 1,38) &  & 353.2630 & 81 & 355 & $2.354\times10^{-3}$ \\
39( 3,36) -- 38( 3,35) &  & 353.3494 & 79 & 355 & $2.348\times10^{-3}$ \\
40( 3,38) -- 39( 3,37) & Y & 355.9997 & 81 & 369 & $2.401\times10^{-3}$ \\
40( 7,34) -- 39( 7,33) &  & 356.5518 & 81 & 451 & $2.353\times10^{-3}$ \\
40( 7,33) -- 39( 7,32) &  & 356.5518 & 81 & 451 & $2.353\times10^{-3}$ \\
40( 4,37) -- 39( 4,36) & Y & 357.4249 & 81 & 384 & $2.420\times10^{-3}$ \\
41( 2,40) -- 40( 2,39) & Y & 359.6292 & 83 & 373 & $2.483\times10^{-3}$ \\
42( 1,42) -- 41( 1,41) &  & 360.3463 & 85 & 376 & $2.504\times10^{-3}$ \\
42( 0,42) -- 41( 0,41) &  & 360.4625 & 85 & 376 & $2.506\times10^{-3}$ \\
40( 2,38) -- 39( 2,37) &  & 361.9237 & 81 & 364 & $2.531\times10^{-3}$ \\
40( 3,37) -- 39( 3,36) &  & 362.6215 & 81 & 372 & $2.539\times10^{-3}$ 
\end{longtable}

\section{ C$_3$H$_6$ line list}

This section lists, in Table \ref{app-linelist-C3H6}, the C$_3$H$_6$ transitions corresponding to the lines having an intensity above 3$\sigma$ in the observations. The firmly detected transitions, which are optically thin and unblended in the conditions of the best fit, are annotated with the letter `Y'.

\begin{longtable}{cccccc}
\caption{\label{app-linelist-C3H6}Line list of the observed C$_3$H$_6$ transitions above 3 $\sigma$.}\\
\hline\hline
Transition & Fitted & $\mathrm{\nu}$ & $g_\mathrm{u}$ & $E_{\mathrm{u}}$ & $A_{\mathrm{ul}}$ \\
\multicolumn{2}{l}{$J(K_{\mathrm{a}},\ K_{\mathrm{c}})$ -- $J'(K_{\mathrm{a}}',\ K_{\mathrm{c}}')$}  & (GHz) & & (K) & (s$^{-1}$) \\
\hline
\endfirsthead
\caption{continued.}\\
\hline\hline
Transition & Fitted & $\mathrm{\nu}$ & $g_\mathrm{u}$ & $E_{\mathrm{u}}$ & $A_{\mathrm{ul}}$ \\
\multicolumn{2}{l}{$J(K_{\mathrm{a}},\ K_{\mathrm{c}})$ -- $J'(K_{\mathrm{a}}',\ K_{\mathrm{c}}')$}  & (GHz) & & (K) & (s$^{-1}$) \\
\hline
\endhead
\hline
\endfoot

13( 2,12) -- 12( 2,11) E &  & 225.1013 & 108 & 83 & $1.517\times10^{-5}$ \\
13( 2,12) -- 12( 2,11) A &  & 225.1024 & 108 & 83 & $1.517\times10^{-5}$ \\
13( 1,12) -- 12( 1,11) E &  & 231.2205 & 108 & 80 & $1.672\times10^{-5}$ \\
13( 1,12) -- 12( 1,11) A &  & 231.2233 & 108 & 80 & $1.672\times10^{-5}$ \\
13( 2,11) -- 12( 2,10) E &  & 232.5859 & 108 & 85 & $1.676\times10^{-5}$ \\
13( 2,11) -- 12( 2,10) A &  & 232.5869 & 108 & 85 & $1.676\times10^{-5}$ \\
14( 0,14) -- 13( 0,13) E & Y & 235.2697 & 116 & 86 & $1.772\times10^{-5}$ \\
14( 0,14) -- 13( 0,13) A & Y & 235.2714 & 116 & 86 & $1.772\times10^{-5}$ \\
19( 1,18) -- 18( 1,17) E & Y & 331.5709 & 156 & 164 & $4.986\times10^{-5}$ \\
19( 1,18) -- 18( 1,17) A & Y & 331.5751 & 156 & 164 & $4.986\times10^{-5}$ \\
20( 1,20) -- 19( 1,19) E &  & 331.7410 & 164 & 170 & $5.023\times10^{-5}$ \\
20( 1,20) -- 19( 1,19) A &  & 331.7421 & 164 & 170 & $5.024\times10^{-5}$ \\
19( 8,11) -- 18( 8,10) E &  & 331.8207 & 156 & 274 & $4.148\times10^{-5}$ \\
19( 8,12) -- 18( 8,11) A &  & 331.8207 & 156 & 274 & $4.148\times10^{-5}$ \\
19( 8,11) -- 18( 8,10) A &  & 331.8207 & 156 & 274 & $4.148\times10^{-5}$ \\
19( 7,13) -- 18( 7,12) A &  & 331.9592 & 156 & 248 & $4.363\times10^{-5}$ \\
19( 7,12) -- 18( 7,11) A &  & 331.9592 & 156 & 248 & $4.363\times10^{-5}$ \\
19( 3,17) -- 18( 3,16) E & Y & 332.1087 & 156 & 176 & $4.926\times10^{-5}$ \\
19( 3,17) -- 18( 3,16) A & Y & 332.1091 & 156 & 176 & $4.926\times10^{-5}$ \\
19( 6,14) -- 18( 6,13) A &  & 332.1877 & 156 & 224 & $4.554\times10^{-5}$ \\
19( 6,13) -- 18( 6,12) A &  & 332.1890 & 156 & 224 & $4.554\times10^{-5}$ \\
20( 0,20) -- 19( 0,19) E &  & 332.1893 & 164 & 170 & $5.045\times10^{-5}$ \\
20( 0,20) -- 19( 0,19) A &  & 332.1909 & 164 & 170 & $5.045\times10^{-5}$ \\
19( 6,14) -- 18( 6,13) E &  & 332.1909 & 156 & 224 & $4.548\times10^{-5}$ \\
19( 6,13) -- 18( 6,12) E &  & 332.1909 & 156 & 224 & $4.548\times10^{-5}$ \\
19( 5,15) -- 18( 5,14) A &  & 332.5731 & 156 & 204 & $4.724\times10^{-5}$ \\
19( 5,15) -- 18( 5,14) E &  & 332.5773 & 156 & 204 & $4.674\times10^{-5}$ \\
19( 5,14) -- 18( 5,13) A &  & 332.6150 & 156 & 204 & $4.726\times10^{-5}$ \\
19( 5,14) -- 18( 5,13) E &  & 332.6154 & 156 & 204 & $4.676\times10^{-5}$ \\
19( 4,16) -- 18( 4,15) A &  & 333.0380 & 156 & 188 & $4.871\times10^{-5}$ \\
19( 4,16) -- 18( 4,15) E &  & 333.0396 & 156 & 188 & $4.870\times10^{-5}$ \\
19( 4,15) -- 18( 4,14) A &  & 333.7945 & 156 & 188 & $4.904\times10^{-5}$ \\
19( 4,15) -- 18( 4,14) E &  & 333.7959 & 156 & 188 & $4.904\times10^{-5}$ \\
19( 3,16) -- 18( 3,15) A &  & 338.1148 & 156 & 177 & $5.202\times10^{-5}$ \\
19( 3,16) -- 18( 3,15) E &  & 338.1152 & 156 & 177 & $5.202\times10^{-5}$ \\
19( 2,17) -- 18( 2,16) E & Y & 340.8266 & 156 & 170 & $5.403\times10^{-5}$ \\
19( 2,17) -- 18( 2,16) A & Y & 340.8293 & 156 & 170 & $5.403\times10^{-5}$ \\
20( 2,19) -- 19( 2,18) E &  & 343.0971 & 164 & 181 & $5.512\times10^{-5}$ \\
20( 2,19) -- 19( 2,18) A &  & 343.0990 & 164 & 181 & $5.512\times10^{-5}$ \\
20( 1,19) -- 19( 1,18) E & Y & 347.7734 & 164 & 180 & $5.760\times10^{-5}$ \\
20( 1,19) -- 19( 1,18) A & Y & 347.7777 & 164 & 180 & $5.760\times10^{-5}$ \\
21( 1,21) -- 20( 1,20) E &  & 348.0323 & 172 & 187 & $5.809\times10^{-5}$ \\
21( 1,21) -- 20( 1,20) A &  & 348.0335 & 172 & 187 & $5.809\times10^{-5}$ \\
21( 0,21) -- 20( 0,20) E & Y & 348.3802 & 172 & 187 & $5.827\times10^{-5}$ \\
21( 0,21) -- 20( 0,20) A &  & 348.3815 & 172 & 187 & $5.827\times10^{-5}$ \\
20(10,11) -- 19(10,10) E & Y & 349.1524 & 164 & 356 & $4.411\times10^{-5}$ \\
20(10,11) -- 19(10,10) A & Y & 349.1526 & 164 & 356 & $4.412\times10^{-5}$ \\
20(10,10) -- 19(10, 9) A & Y & 349.1526 & 164 & 356 & $4.412\times10^{-5}$ \\
20( 8,13) -- 19( 8,12) A &  & 349.3206 & 164 & 291 & $4.948\times10^{-5}$ \\
20( 8,12) -- 19( 8,11) A &  & 349.3206 & 164 & 291 & $4.948\times10^{-5}$ \\
20( 8,12) -- 19( 8,11) E &  & 349.3206 & 164 & 291 & $4.948\times10^{-5}$ \\
20( 3,18) -- 19( 3,17) E &  & 349.4123 & 164 & 192 & $5.757\times10^{-5}$ \\
20( 3,18) -- 19( 3,17) A &  & 349.4129 & 164 & 192 & $5.757\times10^{-5}$ \\
20( 7,14) -- 19( 7,13) A & Y & 349.4842 & 164 & 264 & $5.176\times10^{-5}$ \\
20( 7,13) -- 19( 7,12) A & Y & 349.4843 & 164 & 264 & $5.176\times10^{-5}$ \\
20( 6,15) -- 19( 6,14) A &  & 349.7525 & 164 & 241 & $5.379\times10^{-5}$ \\
20( 6,14) -- 19( 6,13) A &  & 349.7549 & 164 & 241 & $5.379\times10^{-5}$ \\
20( 6,15) -- 19( 6,14) E &  & 349.7561 & 164 & 241 & $5.358\times10^{-5}$ \\
20( 6,14) -- 19( 6,13) E &  & 349.7568 & 164 & 241 & $5.358\times10^{-5}$ \\
20( 5,16) -- 19( 5,15) A &  & 350.1976 & 164 & 221 & $5.563\times10^{-5}$ \\
20( 5,16) -- 19( 5,15) E &  & 350.2011 & 164 & 221 & $5.542\times10^{-5}$ \\
20( 5,15) -- 19( 5,14) A & Y & 350.2639 & 164 & 221 & $5.567\times10^{-5}$ \\
20( 5,15) -- 19( 5,14) E & Y & 350.2653 & 164 & 221 & $5.545\times10^{-5}$ \\
20( 4,17) -- 19( 4,16) A &  & 350.6713 & 164 & 205 & $5.720\times10^{-5}$ \\
20( 4,17) -- 19( 4,16) E &  & 350.6728 & 164 & 205 & $5.720\times10^{-5}$ \\
20( 4,16) -- 19( 4,15) A & Y & 351.7358 & 164 & 205 & $5.772\times10^{-5}$ \\
20( 4,16) -- 19( 4,15) E & Y & 351.7372 & 164 & 205 & $5.772\times10^{-5}$ \\
20( 3,17) -- 19( 3,16) A & Y & 356.6568 & 164 & 194 & $6.130\times10^{-5}$ \\
20( 3,17) -- 19( 3,16) E & Y & 356.6571 & 164 & 194 & $6.130\times10^{-5}$ \\
20( 2,18) -- 19( 2,17) E &  & 358.4284 & 164 & 187 & $6.296\times10^{-5}$ \\
20( 2,18) -- 19( 2,17) A &  & 358.4315 & 164 & 187 & $6.298\times10^{-5}$ \\
21( 2,20) -- 20( 2,19) E &  & 359.7217 & 172 & 199 & $6.365\times10^{-5}$ \\
21( 2,20) -- 20( 2,19) A &  & 359.7237 & 172 & 199 & $6.365\times10^{-5}$ \\
\end{longtable}

\end{document}